\documentclass[preprint,12pt,a4paper]{elsarticle}

\usepackage{hyperref}
\usepackage{graphicx}
\usepackage{subcaption}
\usepackage{amssymb}
\usepackage{amsthm}
\usepackage{amsmath}
\usepackage{multirow}
\usepackage[utf8]{inputenc}
\usepackage{cleveref}
\usepackage{booktabs}
\usepackage{listings}
\usepackage{microtype}
\usepackage[section]{placeins}
\newcommand{\code}[1]{\lstinline{#1}}

\usepackage{xcolor}
\usepackage{xargs}
\usepackage[colorinlistoftodos,textsize=footnotesize]{todonotes}

\newcommandx{\unsure}[2][1=]{\todo[linecolor=red,backgroundcolor=red!25,bordercolor=red,#1]{#2}}
\newcommandx{\change}[2][1=]{\todo[linecolor=blue,backgroundcolor=blue!25,bordercolor=blue,#1]{#2}}
\newcommandx{\info}[2][1=]{\todo[linecolor=OliveGreen,backgroundcolor=OliveGreen!25,bordercolor=OliveGreen,#1]{#2}}

\newcommand{\ten}[1]{\ensuremath{\mathbf{#1}}}

\theoremstyle{definition}
\newtheorem*{remark}{Remark}

\usepackage[displaymath,mathlines]{lineno}

\journal{}

\begin{document}

\begin{frontmatter}

  \title{Efficient and Accurate Adaptive Resolution for
    Weakly-Compressible SPH}

  \author[IITB]{Abhinav Muta\fnref{fn1}\corref{cor1}}
  \ead{abhinavm@aero.iitb.ac.in}
  \author[IITB]{Prabhu Ramachandran}
  \ead{prabhu@aero.iitb.ac.in}
\address[IITB]{Department of Aerospace Engineering, Indian Institute of
  Technology Bombay, Powai, Mumbai 400076}

\cortext[cor1]{Corresponding author}
\fntext[fn1]{Author names listed alphabetically. Both authors contributed
  equally to the work.}

\begin{abstract}
  In this paper we propose an accurate, and computationally efficient method
  for incorporating adaptive spatial resolution into weakly-compressible
  Smoothed Particle Hydrodynamics (SPH) schemes. Particles are adaptively
  split and merged in an accurate manner. Critically, the method ensures that
  the number of neighbors of each particle is optimal, leading to an efficient
  algorithm. A set of background particles is used to specify either
  geometry-based spatial resolution, where the resolution is a function of
  distance to a solid body, or solution-based adaptive resolution, where the
  resolution is a function of the computed solution. This allows us to
  simulate problems using particles having length variations of the order of
  1:250 with much fewer particles than currently reported with other
  techniques. The method is designed to automatically adapt when any solid
  bodies move. The algorithms employed are fully parallel. We consider a suite
  of benchmark problems to demonstrate the accuracy of the approach. We then
  consider the classic problem of the flow past a circular cylinder at a range
  of Reynolds numbers and show that the proposed method produces accurate
  results with a significantly reduced number of particles. We provide an open
  source implementation and a fully reproducible manuscript.
\end{abstract}


\end{frontmatter}


\section{Introduction}%
\label{sec:intro}

It would appear that a meshless particle method would be naturally suited for
adaptive resolution. However, accurate adaptive resolution for Smoothed
Particle Hydrodynamics (SPH) in the context of weakly-compressible and
incompressible fluid flow is still a challenging area of current
research~\cite{vacondio_grand_2020}.

In the context of incompressible and weakly-compressible fluid flow, there
have been some valuable developments starting with the pioneering work of
\citet{feldman_dynamic_2007} where the particles are adaptively split in an
accurate manner. This work has been extended further to include particle
merging by \citet{vacondio_splitting_2013,vacondio_variable_2016} and applied
to the shallow water equations~\cite{vacondio_accurate_2012}, soil
simulation~\cite{reyeslopez2013}, fluid-structure
interaction~\cite{hu2019}. The method has been designed to be very accurate
and a great deal of care is taken when splitting and merging
particles. However, the accuracy comes at a significant cost since each coarse
particle splits into 7 particles in two dimensions and around 14 in three
dimensions. This leads to an enormous increase in the number of particles as
the regions are refined. The particle de-refining method merges particles
pair-wise and it is argued~\cite{chiron_apr_2018} that the method is
computationally expensive since the rate of splitting particles is
significantly larger than the rate of merging. While the method is designed to
be accurate, the resulting refined particles also employ a very large
smoothing radius in comparison to what would be expected in a fixed particle
size discretization of the problem with a similar number of particles. This
poses significant additional performance limitations on the method. Moreover,
the method relies on manual specification of the spatial regions where the
adaptation is desired. This is inconvenient in general and especially when the
bodies are moving.

\citet{barcarolo_adaptive_2014,sun_plus-sph_2017}, and
\citet{chiron_apr_2018} refine each coarse particle (also called parent
particle), in two dimensions, into 4 child particles but also retain the
coarse particle. The parent particles are passively advected in the refined
regions. This implies that each coarse particle effectively splits into five
particles. This reduces the number of refined particles when compared with
\cite{vacondio_splitting_2013}. The significant advantage with this approach
is that de-refining particles is simple to implement; the parent particles are
re-activated and the child particles are removed. This approach has also been
used for some impressive multi-resolution simulations using the $\delta^+$-SPH
scheme~\cite{sun_plus-sph_2017,sun_multi-resolution_2018}. Another significant
advantage is that the smoothing length chosen is much smaller than the typical
values chosen in the approaches of \cite{vacondio_splitting_2013}. In order to
handle the interactions between the child and parent particles, a particle
property $\gamma$, is added to each particle. In the transition regions this
value is between 0 and 1 whereas in regions with uniform particle smoothing
length, the value is either 0 or 1. When the value is zero for a particle, the
particle is effectively switched off and when it is one it is active.
Intermediate values allow for the use of both the parent and child particles.

\citet{chiron_apr_2018} further refined this method by taking inspiration from
traditional Adaptive Mesh Refinement techniques to create an Adaptive Particle
Refinement (APR) procedure. In the intermediate regions where larger particles
are refined into smaller particles, both the parent and child particles are
retained and only particles of the same size interact and the properties are
carefully interpolated between the parent and child particles. The difficulty
with the approaches of \citet{barcarolo_adaptive_2014,sun_plus-sph_2017}, and
\citet{chiron_apr_2018} is that coarse particles effectively split into five
particles in each level of refinement. Furthermore, there are additional
complications due to the special handling required for the parent and child
particles either by the use of the $\gamma$ parameter or by the use of
prolongation and restriction operations in the APR method. It is also not
entirely clear what would happen in high-strain fluid flows where the four
child particles would drift significantly apart away from the parent particle.

Recently, another approach for dynamic particle splitting and merging has been
proposed by \citet{yang_smoothed_2017,yang_adaptive_2019} and applied to
multi-phase fluid simulations. This approach is similar to that employed by
\citet{vacondio_splitting_2013} but each coarse particle is only split into
two child particles. The parent particle is removed. However, the placement of
the child particles is done carefully along the perpendicular bisector of the
line joining the parent particle to its nearest particle. This method will
only work in two dimensions and no procedure for the three dimensional case is
proposed. The advantage with this approach is that the particle refinement is
much more gradual without a very large increase in the number of particles.
Merging is done only between two particles and therefore there is no profusion
of particles. The proposed method also elegantly handles gradual refinement of
the resolution around an interface using a single parameter. This has been
demonstrated for multi-phase problems~\cite{yang_adaptive_2019}. It appears
that no detailed study of the accuracy of the method has been
performed. However, previous accuracy studies by \cite{feldman_dynamic_2007}
suggest that splitting particles into only two child particles would introduce
significant error into the solution. Moreover, the method has only been
demonstrated for two dimensional fluid flows. An alternative to the
distance-based spatial adaptation is the recovery-based a \emph{posteriori}
error estimator~\cite{hu2019}, where the error in the SPH velocity gradient is
measured and particles are adaptively refined in regions where the error
exceeds a tolerance.

In the area of computer graphics, \citet{desbrun1999} use splitting and
merging operators in the SPH method and applied it to highly deformable
substances. \citet{adams2007} use extended local feature size to adaptively
refine the particles in the regions of geometric interest.
\citet{solenthaler2011} use two-scales, a lower resolution and a higher
resolution, and couple the two with appropriate boundary conditions and
feedback forces. However, these works are designed more for computer graphics
applications and do not test the accuracy with any standard benchmark
problems.

\citet{spreng2014} have proposed the use of the method of
\citet{vacondio_splitting_2013} for performing adaptive particle resolution
for structural mechanics problems. They also propose a novel method to merge
multiple particles by considering neighboring particles which are identified
in two different ways.  It is not clear if the proposed algorithm is parallel
as the details of the implementation are not discussed.  In a more recent
work, \citet{spreng2020a} have proposed the use of adaptive refinement to
improve the discretization errors.

Recently, \citet{sun2021a} have employed the adaptive particle refinement and
de-refinement approach to strongly compressible, multi-phase fluid
flows. However, the main focus of the adaptive particle refinement is to
ensure a homogenous and isotropic particle distribution when the fluid is
highly compressible.

The idea of splitting and merging particles is not new and has been
successfully applied in the context of vortex methods~\cite{rossi_csm_1996}.
This technique has also been used by various researchers employing SPH for
computer graphics~\cite{desbrun1999,adams2007,solenthaler2011}. However, the
challenge in implementing this with the SPH method for incompressible and
weakly-compressible fluids is to have a method that is both accurate and
computationally efficient with a minimum of numerical parameters. This is a
significant challenge. It bears emphasis that none of the existing adaptive
resolution schemes for fluid flow problems with widely varying
scales~\cite{vacondio_splitting_2013,sun_multi-resolution_2018,chiron_apr_2018}
feature an automatic adaptation strategy, nor do they inherently support
complex moving geometries or provide any ability to introduce solution-based
adaptivity. The methods of \cite{hu2019,sun2021a,spreng2020a} do support
moving geometries and solution-adaptivity however, they do not seem to have
been applied to problems with widely-varying scales.

In this paper we propose a new approach which is automatically adaptive,
computationally efficient, accurate, supports moving bodies, and solution
adaptivity. The basic strategy is to split and merge particles carefully as
originally proposed by \citet{feldman_dynamic_2007} and
\citet{vacondio_splitting_2013}. However, we adaptively merge particles to
reduce the large particle counts. This is done in a computationally efficient
manner, in parallel, and our simulations suggest that this approach is also
accurate. We are thus able to control the particle refinement adaptively so as
to effectively only double the number of particles in each refinement region
while retaining accuracy. In addition, we carefully set the smoothing radius
of the refined particles to be optimal for the particular refinement region
thereby further improving performance in comparison to the approach of
\citet{vacondio_variable_2016}. We use ideas inspired from the work of
\citet{yang_adaptive_2019} to automatically set the refinement criterion. This
allows us to specify the geometry, a few parameters determining the maximum
and minimum length scales and the algorithm automatically refines the
particles as required. We discuss in some detail the algorithm proposed and
show how it can be used to (i) handle complex geometries, (ii) specify
user-specified refinement regions, (iii) handle moving geometries, and (iv) be
used for solution-based adaptivity. We do not extensively explore
solution-based adaptivity in this work but outline the basic ideas and
demonstrate this with some simulations. The algorithms employed in this work
are parallel and in principle may be executed on a General-Purpose Graphics
Processing Unit (GPGPU).

We only consider two dimensional problems in this manuscript but in principle
the ideas naturally extend to three-dimensional cases. Although we use a
modified EDAC-SPH~\cite{edac-sph:cf:2019} scheme for the SPH discretization
any similar method could be used. In the present work we do not consider any
free-surface problems, however, our adaptive refinement algorithm can be
easily extended to work with such problems. We consider several simple
benchmark problems to demonstrate the accuracy of the approach. We then
simulate the flow past a circular cylinder at a variety of Reynolds numbers in
the range 40 - 9500 and compare these with some very well established
simulations to show that the method is capable of resolving the necessary
details with a minimum of particles. This translates to a proportional
reduction in the computational time.  The new method allows us to perform such
computations with far fewer particles than reported elsewhere with the SPH
method. For the case of the flow past a circular cylinder the results we
present require at least an order of magnitude fewer particles than those
reported in \cite{sun_multi-resolution_2018} for a similar
resolution. Finally, we note that none of the existing methods for adaptive
SPH feature open source implementations. We provide a fully open source
implementation based on the PySPH
framework~\cite{PR:pysph:scipy16,pysph2020}. The source code can be obtained
from \url{https://gitlab.com/pypr/adaptive_sph}. Our manuscript is fully
reproducible and every figure is automatically generated through the use of an
automation framework~\cite{pr:automan:2018}.

\section{The SPH method}%
\label{sec:sph}

In this paper we deal specifically with weakly-compressible flows.  We use the
entropically damped artificial compressibility (EDAC)
method~\cite{edac-sph:cf:2019} to simulate the weakly-compressible flows. The
position update, pressure evolution, and momentum equations in the EDAC
formulation are,
\begin{equation}
  \label{eq:pos-ode}
  \frac{\mathrm{d} \ten{r}}{\mathrm{d} t} = \ten{u},
\end{equation}
\begin{equation}
  \label{eq:edac}
  \frac{\mathrm{d} p}{\mathrm{d} t} = -\rho c_s^2 \text{div}(\ten{u})
  + \nu_{\text{e}} \nabla^2 p,
\end{equation}
\begin{equation}
  \label{eq:mom}
  \frac{\text{d} \ten{u}}{\text{d}t} =
  -\frac{1}{\rho} \nabla p
  + \nu \nabla^2 \ten{u} + \ten{f},
\end{equation}
where $\ten{r}$, $\ten{u}$, $p$, and $t$ denotes the position, velocity,
pressure, and time respectively. $\rho$ is the density, $\nu$ is the kinematic
viscosity of the fluid, $c_s$ is the artificial speed of sound, $\ten{f}$ is
the external body force, and $\nu_{\text{e}}$ is the EDAC viscosity parameter.

In order to further enhance the uniformity of the particles we use the transport
velocity formulation~\cite{Adami2013}, with the corrections
incorporated~\cite{adepu2021}. Then the above equations are re-formulated as,
\begin{equation}
  \label{eq:pos-ode:tvf}
  \frac{\mathrm{d} \ten{r}}{\mathrm{d} t} = \tilde{\ten{u}}
\end{equation}
\begin{equation}
  \label{eq:edac-corr}
  \frac{\tilde{\mathrm{d}}p}{\mathrm{d}t} =
  -\rho c_s^2 \text{div}(\ten{u})
  + \nu_{\text{e}} \nabla^2 p
  + (\tilde{\ten{u}}
  - \ten{u}) \cdot \nabla p
\end{equation}
\begin{equation}
  \label{eq:mom-corr}
  \frac{\tilde{\text{d}} \ten{u}}{\text{d}t} =
  -\frac{1}{\rho} \nabla p
  + \nu \nabla^2 \ten{u} + \ten{f}
  + \frac{1}{\rho} \nabla \cdot \rho (\ten{u} \otimes (\tilde{\ten{u}} - \ten{u}))
  + \ten{u}\,\mathrm{div}(\tilde{\ten{u}})
\end{equation}
where $\tilde{\ten{u}}$ refers to the transport velocity, and
$\frac{\tilde{\mathrm{d}} (.)}{\mathrm{d} t} = \frac{\partial (.)}{\partial t} +
\tilde{\ten{u}} \cdot \text{grad} (.)$ is the material time derivative of a
particle advecting with the transport velocity $\tilde{\ten{u}}$. The
computation of the transport velocity is shown in \cref{sec:shift}.
\begin{remark}
  In our numerical experiments with the Taylor-Green problem we found that the
  addition of the divergence correction terms in the pressure evolution
  equation is crucial for accuracy. However, we find that the use of the last
  two terms in the momentum equation~\eqref{eq:mom-corr} introduces noise
  where the particles are merged or split. Consequently, we do not use them in
  this work.  We note that \citet{sun_consistent_2019} observes that the
  effect of these terms in the momentum equation is minor.
\end{remark}

We discretize the governing equations using variable-$h$ SPH.~The domain is
discretized into points whose spatial location is denoted by $\ten{r}_i$, where
the subscript $i$ denotes the index of an arbitrary particle. The mass of the
particle, which vary as a function of space, is denoted by $m_i$, and its
smoothing length by $h_i$. In the variable-$h$ SPH the density is approximated
by the summation density equation using a~\emph{gather
  formulation}~\cite{hernquist1989, vacondio_accurate_2012} written as,
\begin{equation}
  \label{eq:sd-gather}
  \rho(\ten{r}_i) = \sum_j m_j W(|\ten{r}_i - \ten{r}_j|, h_i),
\end{equation}
where, $W(|\ten{r}_i - \ten{r}_j|, h_i)$ is the kernel function. We use the
quintic spline kernel in all our simulations, the quintic spline kernel is given
by,
\begin{equation}
  \label{eq:kernel}
  W(q) =
  \begin{cases}
    \sigma_2 [{(3 - q)}^5 - 6{(2 - q)}^5 + 15{(1 - q)}^5] \quad%
    &\text{if}\ 0 \le{} q < 1, \\
     \sigma_2 [{(3 - q)}^5 - 6{(2 - q)}^5] \quad
    &\text{if}\ 1 \le{} q < 2, \\
     \sigma_2 {(3 - q)}^5 \quad &\text{if}\ 2 \le{} q < 3, \\
     0 \quad &\text{if}\ q \ge{} 3, \\
  \end{cases}
\end{equation}
where $\sigma_2 = 7/(478 \pi {h(\ten{r})}^2)$, and $q = |\ten{r}|/h$.

The EDAC pressure evolution equation in variable-$h$ SPH
(see~\cite{monaghan-review:2005,vacondio_accurate_2012}, for a derivation of the
terms in the R.H.S) is given by,
\begin{equation}
  \label{eq:edac:sph}
\begin{split}
  \frac{\tilde{\mathrm{d}}p}{\mathrm{d} t}(\ten{r}_i) =
  \quad&
     \frac{\rho_0 c_s^2}{\beta_i}\sum_j
     \frac{m_j}{\rho_j} \ten{u}_{ij} \cdot \nabla W(r_{ij}, h_i) \\
  +&
     \frac{1}{\beta_i}\sum_j \frac{m_j}{\rho_j} \nu_{\text{e}, ij}  (p_i - p_j)
     (
       \ten{r}_{ij} \cdot \nabla W(r_{ij}, h_{ij})
     ) \\
  +&
     \sum_j m_j
     [(\tilde{\ten{u}}_{i} - \ten{u}_{i})
     \cdot (P_i \nabla W(r_{ij}, h_i) + P_j \nabla W(r_{ij}, h_j))],
   \end{split}
 \end{equation}
 where $\rho_0$ is the reference density, $p_i$ is the pressure of particle $i$,
 $\rho_j$ is the density of the $j^{\text{th}}$ particle computed using
 summation density~\cref{eq:sd-gather},
 $\ten{u}_{ij} = (\ten{u}_i - \ten{u}_j)$,
 $r_{ij} =|\ten{r}_{ij}| = |\ten{r}_i - \ten{r}_j|$, $\beta_i$ is the
 variable-$h$ correction term~\cite{vacondio_accurate_2012}, which in $d$
 dimensions is given by,
\begin{equation}
  \label{eq:beta}
  \beta_i = - \frac{1}{\rho_i d} \sum_j m_j r_{ij}
  \frac{\mathrm{d} W(r_{ij}, h_i)}{\mathrm{d}r_{ij}},
\end{equation}
$P_i$ and $P_j$ are given by,
\begin{equation}
  \label{eq:mom:pre}
  P_i = \frac{(p_i - p_{\text{avg}, i})}{\rho_i^2 \beta_i}, \quad
  P_j = \frac{(p_j - p_{\text{avg}, j})}{\rho_j^2 \beta_j},
\end{equation}
here we employ the pressure reduction technique proposed
by~\citet{sph:basa-etal-2009}, where, the average pressure is computed as,
\begin{equation}
  \label{eq:pavg}
  p_{\text{avg}, i} = \frac{\sum_{j = 1}^{N_i} p_j}{N_i},
\end{equation}
where $N_i$ is the number of neighbours for a particle with index $i$, and
\begin{equation}
  \label{eq:dw-avg}
  \nabla W(r_{ij}, h_{ij}) =
  \left(\frac{\nabla W(r_{ij}, h_i) + \nabla W(r_{ij}, h_j)}{2}\right).
\end{equation}

The EDAC viscosity of the pressure diffusion term in the EDAC equation with the
SPH discretization is given by,
\begin{equation}
  \label{eq:edac-alpha}
  \nu_{\text{e}, i} = \frac{\alpha_{\text{e}} c_s h_i}{8},
\end{equation}
where $\alpha_{\text{e}} = 1.5$ is used in all our simulations. Since this is a
function of the smoothing length, which is varying in space, we use the approach
of~\citet{cleary1999} to model the pressure diffusion term where,
 \begin{equation}
   \label{eq:nu-cleary}
   \nu_{\text{e}, ij} = 4\frac{\nu_{\text{e}, i} \nu_{\text{e}, j}}
   {(\nu_{\text{e}, i} + \nu_{\text{e}, j})}.
 \end{equation}

The momentum equation in the variable-$h$ SPH discretization is given by,
\begin{equation}
  \label{eq:mom-sph}
  \begin{split}
  \frac{\tilde{\mathrm{d}}\ten{u}}{\mathrm{d} t}(\ten{r}_i, t) =
  -&\sum_j m_j
    \left((P_i + A_i) \nabla W(r_{ij}, h_i)
    + (P_j + A_j) \nabla W(r_{ij}, h_j)\right) \\
    +& \frac{1}{\beta_i}\sum_j m_j \frac{4 \nu}{(\rho_i + \rho_j)}
    \frac{\ten{r}_{ij} \cdot \nabla W(r_{ij}, h_{ij})}
    {(|\ten{r}_{ij}|^{2} + \eta)} \ten{u}_{ij} \\
    -&
    \frac{1}{\beta_i}\sum_{j} \frac{m_j}{\rho_j}
  [(\tilde{\ten{u}}_{ij} - \ten{u}_{ij}) \cdot \nabla W(r_{ij}, h_i)] \ten{u}_i
  \end{split}
\end{equation}
where,
\begin{equation}
  \label{eq:astress}
{A}_{i} = \frac{1}{\rho_{i} \beta_i} \ten{u}_{i} \otimes(\tilde{\ten{u}}_{i} -
\ten{u}_{i}),
\quad
{A}_{j} = \frac{1}{\rho_{j} \beta_j} \ten{u}_{j} \otimes(\tilde{\ten{u}}_{j} - \ten{u}_{j}),
\end{equation}
and $\eta = 0.001 h_i^2$ is a small number added to ensure a non-zero
denominator in case when $i = j$.

\begin{remark}
  We do not employ any artificial viscosity in our benchmark cases. We note that
  the proposed scheme is not conservative due to shifting, the adaptive-h
  correction terms, and the non-standard form of the pressure gradient.
\end{remark}

\subsection{Particle shifting}%
\label{sec:shift}

We use a limited form of the particle shifting technique
of~\citet{diff_smoothing_sph:lind:jcp:2009} which is based on evaluating the
gradient of the kernel function. A particle with an index $i$ at a current
position $\ten{r}_i$ is shifted to a new position $\ten{r}'_i$ as,
\begin{equation}%
  \label{eq:shift}
  \ten{r}'_{i} = \ten{r}_{i} + \theta \delta \ten{r}_{i},
\end{equation}
where,
\begin{equation}%
  \label{eq:shift-deltar}
  \delta \ten{r}_{i} = - \frac{h^2_i}{2} \sum_{j}
  \frac{m_j}{\rho_0} \left (
    1 + 0.24 {\left(
        \frac{W(r_{ij}, h_{ij})}{W(\Delta x, \xi h_{ij})}
      \right)}^{4}
  \right)
  \nabla W_{ij},
\end{equation}
where $\xi$ is the point of inflection of the kernel~\cite{crespo2008}, and
$h_{ij} = (h_i + h_j)/2$. For quintic spline the point of inflection is
$ \xi = 0.759298480738450$. We found that using $\rho_j$ in the volume
approximation makes the shifting less effective and hence have used $\rho_0$.  We
limit the shifting by restricting the movement of particle which is shifted by
more than 25\% of its smoothing length:
\begin{equation}
  \label{eq:shift-limit}
  \theta =
  \begin{cases}
     \frac{0.25 h_i}{|\delta \ten{r}_{i}|}\quad &\text{if}\ |\delta \ten{r}_{i}|
     > 0.25 h_i, \\
     1 \quad &\text{otherwise}.
  \end{cases}
\end{equation}

We employ shifting while solving the fluid equations and also after our adaptive
refinement procedure. Since we use the transport velocity scheme which already
accounts for the shifting no additional correction is necessary. However, after
the adaptive refinement procedure and subsequent shifting we correct the fluid
properties by using a Taylor series approximation. Consider a fluid property
$\varphi_i$ the corrected value $\varphi'_i$ is obtained by,
\begin{equation}%
  \label{eq:shift-correct}
  \varphi'_{i} = \varphi_i + {(\nabla \varphi)}_i\cdot \delta \ten{r}_{i}.
\end{equation}
The transport velocity is computed using the shifting as,
\begin{equation}
  \label{eq:shift-tvf}
  \tilde{\ten{u}}_i = \ten{u}_i + \frac{\delta \ten{r}_i}{\Delta t}.
\end{equation}
\subsection{Boundary conditions}%
\label{sec:bc}

We employ periodic, no-slip, free-slip, no-penetration and the inlet-outlet
boundary conditions in our test cases. We enforce periodic boundary conditions
by the use of ghost particles onto which the properties are directly copied from
the particles exiting the domain through a periodic boundary.

For the no-slip, free-slip and no-penetration boundary conditions we use the
dummy particle technique of~\citet{Adami2012}. Dummy particles placed in uniform
layers are used to discretize the wall. The no-penetration is implicitly
enforced by using the wall velocity in the EDAC
equation~\cite{Adami2012}. For the no-slip or free-slip we extrapolate the
values of velocity of the fluid onto the dummy wall particles by,
\begin{equation}
  \label{eq:bc-wall}
  u_w = 2\ten{u}_i - \hat{\ten{u}}_i,
\end{equation}
where the subscript $w$ denotes the dummy wall particles, $\ten{u}_i$ is the
prescribed wall velocity, and
\begin{equation}
  \label{eq:shepard}
  \hat{\ten{u}}_i = \frac{\sum_j \ten{u}_j W(r_{ij}, h_{ij})}
  {\sum_j W(r_{ij}, h_{ij})}
\end{equation}
is the Shepard extrapolated velocity of the fluid particles indexed by $j$ onto
the dummy wall particles $i$. The pressure on the wall is calculated from the
fluid, to accurately impose the pressure gradient, by,
\begin{equation}
  \label{eq:bc-pre}
  p_w = \frac{\sum_f p_f W(r_{wf}, h_{wf}) +
    (\ten{g} - \ten{a}_w)\cdot
    \sum_f \rho_f \ten{r}_{wf} W(r_{wf}, h_{wf})}
  {\sum_f W(r_{wf}, h_{wf})},
\end{equation}
where the subscript $f$ denotes the fluid particles, $\ten{a}_w$ is the
acceleration of the wall, $r_{wf} = |\ten{r}_w - \ten{r}_f|$, and
$h_{wf} = (h_w + h_f)/2$.

For the inlet and outlet we use the non-reflecting boundary condition of
\citet{lastiwka2009}. First we compute the characteristic properties, referred
to as $J_1, J_2,$ and $J_3$ in aforementioned article, of the fluid. Then, we
extrapolate the characteristic variables of the fluid onto the inlet and outlet
particles using Shepard interpolation. Finally we determine the fluid dynamical
properties from the characteristic variables.

\subsection{Force computation}%
\label{sec:force-comp}

We compute the forces on the circular cylinder in the flow past a circular
cylinder simulation and evaluate the coefficients of lift and
drag. Specifically, we compute the forces due to the pressure and the
skin-friction on the cylinder by evaluating,
\begin{equation}
  \label{eq:force}
  \ten{f}^{\text{solid}} = m^{\text{solid}}\left(-\frac{1}{\rho}\nabla p
  + \nu \nabla \cdot \nabla \ten{u}\right),
\end{equation}
which in the variable-$h$ SPH discretization is written as,
\begin{equation}
  \label{eq:force-sph}
  \begin{split}
  \ten{f}^{\text{solid}}_{i} =
  \ &\underbrace{-m^{\text{solid}}_{i} \sum_j m_j
  \left(P_i \nabla W(r_{ij}, h_i)
    + P_j \nabla W(r_{ij}, h_j)\right)}_{\ten{f}_{i, \text{p}}} \\
  &+ \underbrace{m^{\text{solid}}_{i} \frac{1}{\beta_i}
  \sum_j m_j \frac{4 \nu}{(\rho_i + \rho_j)}
  \frac{\ten{r}_{ij} \cdot \nabla W(r_{ij}, h_{ij})}
  {(|\ten{r}_{ij}|^{2} + \eta)} \ten{u}_{ij}}_{\ten{f}_{i, \text{visc}}}
  \end{split}
\end{equation}
where the summation index $j$ is over all the fluid particles in the
neighborhood of a solid particle indexed by $i$. We compute the coefficient of
pressure drag $c_{d, \text{pressure}}$ and skin-friction drag
$c_{d, \text{skin-friction}}$, and coefficient of lift $c_l$ due to pressure by,
\begin{equation}
  \label{eq:cdcl}
  c_{d, \text{pressure}} = \frac{\ten{f}_{\text{p}} \cdot \ten{e}_x}
  {\frac{1}{2} \rho_0 U^2_{\infty} L}, \quad
  c_{d, \text{skin-friction}} = \frac{\ten{f}_{\text{visc}} \cdot \ten{e}_x}
  {\frac{1}{2} \rho_0 U^2_{\infty} L}, \quad
  c_{l} = \frac{\ten{f}_{\text{p}} \cdot \ten{e}_y}
  {\frac{1}{2} \rho_0 U^2_{\infty} L},
\end{equation}
where $L$ is the characteristic length of the simulation, $U_{\infty}$ is the
free stream velocity, $\ten{f}_{\text{visc}} = \sum_j \ten{f}_{j, \text{visc}}$
and $\ten{f}_{\text{p}} = \sum_j \ten{f}_{j, \text{p}}$ is the sum over all the
dummy wall particles, and $\ten{e}_x$ and $\ten{e}_y$ are the unit vectors in
the $x$ and $y$ directions respectively.

\subsection{Time integration}%
\label{sec:integration}

We use Predict-Evaluate-Correct (PEC) integrator to integrate the position
$\ten{r}_i$, velocity $\ten{u}_i$, and pressure $p_i$. The integrator is as
follows: We first predict the properties at an intermediate time value
$n + \frac{1}{2}$,
\begin{align}
  \label{eq:int-pre}
  \ten{u}^{n + \frac{1}{2}}_i &= \ten{u}^{n}_i + \frac{\Delta t}{2}
  \frac{\tilde{\mathrm{d}} \ten{u}_i^{n}}{\mathrm{d} t}, \\
  \tilde{\ten{u}}^{n + \frac{1}{2}}_i &= \ten{u}^{n+\frac{1}{2}}_i
                                        + \frac{\delta \ten{r}^n_i}{\Delta t}, \\
  \ten{r}^{n + \frac{1}{2}}_i &= \ten{r}^{n}_i + \frac{\Delta t}{2}
  \tilde{\ten{u}}_i^{n+\frac{1}{2}}, \\
  p^{n + \frac{1}{2}}_i &= p^{n}_i + \frac{\Delta t}{2}
  \frac{\tilde{\mathrm{d}}\ten{p}_i^{n}}{\mathrm{d} t},
\end{align}
next we estimate the new accelerations at $n + \frac{1}{2}$. We then correct the
properties to get the corresponding values at the new time $n + 1$,
\begin{align}
  \label{eq:int-cor}
  \ten{u}^{n + 1}_i &= \ten{u}^{n}_i + \Delta t
  \frac{\tilde{\mathrm{d}} \ten{u}_i^{n + \frac{1}{2}}}{\mathrm{d} t}, \\
  \tilde{\ten{u}}^{n + 1}_i &= \ten{u}^{n+1}_i
                              + \frac{\delta \ten{r}^{n+\frac{1}{2}}_i}{\Delta t}, \\
  \ten{r}^{n + 1}_i &= \ten{r}^{n}_i + \Delta t
  \tilde{\ten{u}}_i^{n+1}, \\
  p^{n + 1}_i &= p^{n}_i + \Delta t
                \frac{\tilde{\mathrm{d}}\ten{p}_i^{n+\frac{1}{2}}}{\mathrm{d} t}.
\end{align}
The time-step is determined by the highest resolution used in the domain, and
the minimum of the CFL criterion and the viscous condition is taken:
\begin{align}
  \label{eq:timestep}
  \Delta t = \min\left(0.25 \left(\frac{h_{\min}}{U + c_s}\right),
  0.125 \left(\frac{h^2_{\min}}{\nu}\right)\right).
\end{align}

\section{Adaptive refinement}
\label{sec:adapt}

We first provide a broad overview of the method before delving into the
details. The adaptive refinement algorithm involves the following key ideas:

\begin{itemize}
\item A particle is split if its mass is greater than $m_{max}$. Note that
  $m_{max}$ is space varying. The splitting is performed using the approach of
  \citet{feldman_dynamic_2007} and \citet{vacondio_splitting_2013}. We
  normally split each particle into 7 child particles in two dimensions.
\item A particle $i$ is allowed to merge with another particle $j$ if
  $r_{ij} < (h_i + h_j)/2$ and $m_i + m_j < \max(m_{max}[i], m_{max}[j])$. The
  merging algorithm is fully parallel and only particles that are mutually
  closest to each other are merged. That is, only if particle $i$'s closest
  allowed merge particle is $j$ and $j$'s closest allowed merge particle is
  $i$, will particle $i$ and $j$ be merged. More details on the
  merging algorithm are provided below.
\item When the particles are split they are iteratively merged three times in
  order to merge any split particles with nearby particles.
\item The maximum mass and minimum mass at a particular location are set
  automatically using a reference $m_{ref}$ parameter that is automatically
  computed based on the minimum specified resolution and a ratio $C_r$ similar
  to what is done by \citet{yang_adaptive_2019}. $m_{max} = 1.05 m_{ref}$ and
  $m_{min} = 0.5 m_{ref}$.
\item A collection of ``background'' points is used to adaptively set the
  minimum and maximum mass values of the fluid particles to control the
  adaptive resolution. If the body moves, this background is also updated.
\end{itemize}

The global minimum and maximum size of the particles is specified. Any solid
bodies (barring the far-field slip walls) are assumed to be specified at the
smallest size. The reference mass increases by the ratio $C_r$ from the
smallest particle to the largest. This produces a smooth increase in the
number of particles in each region.

\subsection{Adaptive splitting}
\label{subsec:split}

\begin{figure}[h!]
  \centering
  \includegraphics[width=0.5\linewidth]{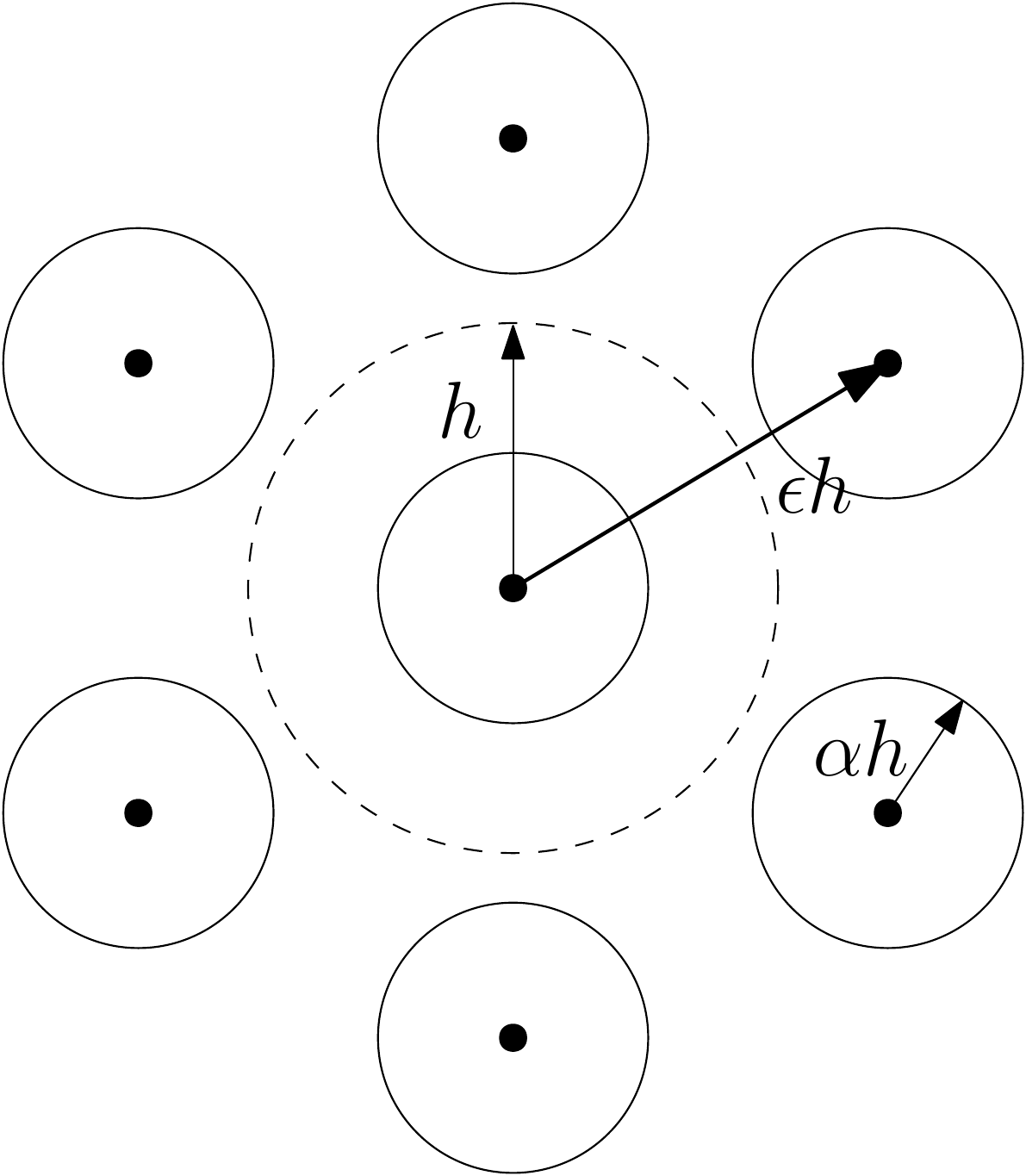}
  \caption{Sketch of particle splitting. The parent particle is shown as a
    dashed circle and has a smoothing length of $h$. Six child particles are
    placed in a hexagonal pattern with one child at the center.}
  \label{fig:split}
\end{figure}

The algorithm for splitting particles follows that of
\citet{feldman_dynamic_2007} and \citet{vacondio_accurate_2012}. If a
particle's mass is larger than the maximum allowed mass, $m_{max}$, then it is
split into 7 particles. The original particle is called the parent particle
and the split particles are called child particles. Six child particles are
placed in a hexagonal arrangement with one child particle at the center as
shown in Fig.~\ref{fig:split}. The parent particle has a smoothing length of
$h$, the child particles have a smoothing radius given by $\alpha h$ and they
are placed on a circle of radius $\epsilon h$. These parameters
$\alpha, \epsilon$ are normally computed so as to minimize the density error
as discussed in \cite{feldman_dynamic_2007,vacondio_accurate_2012}. We choose
the parameters $\alpha=0.9, \epsilon=0.4$ for the equal mass ratio case. We
note that in the present work these are only initial values of the distance
and the smoothing length factors. After we split the particles we perform
merging followed by shifting and a Taylor series correction. These corrections
are accurate due the choice of the values of $\alpha$ and
$\epsilon$. Subsequently, we use the optimal smoothing length as discussed in
the following. The mass of all the particles is the same and is equal to a
seventh of the parent's mass. This configuration produces very little
error. We use a quintic spline kernel for all computations in this work. The
density, velocity, and pressure values of the parent particle are copied to
the children. The children also copy the values of the $m_{min}$, and
$m_{max}$.

We note that in \cite{vacondio_accurate_2012}, the value of parameter $\alpha$
is $0.9$. This implies that the smoothing radius of the child particle is
0.9 times that of the parent despite it having a mass of around a seventh of
the parent. Normally in an SPH simulation one tends to choose $m = \rho \Delta
x^d$, where $d$ is the number of spatial dimensions and $\Delta x$ is the
inter-particle spacing. Furthermore, $h = k \Delta x$ and $k$ depends on the
choice of the kernel. Thus, the value of $\alpha=0.9$ is much larger than what
one would ordinarily expect. This makes the original approach computationally
inefficient and significantly increases the number of neighbors of each
particle. This also reduces the accuracy of the method since the
smoothing errors are larger. In the present work we find the average
mass of particles in the neighborhood of each particle and use this to set the
smoothing length using, $h = C (m/\rho)^{1/d}$, where $C$ is a constant. In
regions where the particle mass is uniform, this attains the ideal $h$ value
that would have been set without the use of adaptive resolution. This gives us
an optimal $h$ and is therefore computationally efficient. We test the
accuracy of our method with a suite of benchmark problems in
section~\ref{sec:results} and find that this does not affect the accuracy of
the method.

The implementation of particle splitting is relatively straightforward. The
adaptive splitting may be performed either every iteration or every
$n_{adapt}>1$ iterations. Any particles whose mass is greater than the
$m_{max}$ value are split. Once these particles are identified, the total
number of particles that need to be split can be identified. In addition, we
also identify the particles that are to be merged as discussed in the next
section. Hence, the total number of new particles that need to be created is
known. The new child particles are then stored over any unused merged
particles and new particles that have been created. Each of these steps are
easy to implement in parallel using a combination of elementwise and reduction
operations.

\subsection{Merging particles}
\label{subsec:merge}

The merging algorithm is in principle simple and we use essentially the same
approach as discussed in \cite{vacondio_splitting_2013}. We note that the
smoothing radius of the particles is initially set as discussed in
\cite{vacondio_splitting_2013}. If we have two particles at locations
$\ten{r}_a$, $\ten{r}_b$.  The location of the merged particle is at,
\begin{equation}
  \label{eq:merged-pos}
  \ten{r}_m = \frac{m_a \ten{r}_a + m_b \ten{r}_b}{m_m},
\end{equation}
where $m_m = m_a + m_b$ is the mass of the merged particle. The velocity is
set using the mass-weighted mean as,
\begin{equation}
  \label{eq:merged-prop}
  \ten{u}_m = \frac{m_a \ten{u}_a + m_b \ten{u}_b}{m_m}.
\end{equation}
A similar form is used for any scalar properties like pressure. The position
$\ten{r}_m$ and velocity $\ten{u}_m$ are obtained by ensuring the conservation
of momentum. The smoothing radius is set by minimizing the density error
\cite{vacondio_splitting_2013},
\begin{equation}
  \label{eq:merged-h}
  h = \left( \frac{m_m W(\ten{0}, 1)}
    {m_a W(\ten{r}_m - \ten{r}_a, h_a) + m_b W(\ten{r}_m - \ten{r}_b, h_b)}
  \right)^{1/d},
\end{equation}
where $W(\ten{x}, h)$ is the kernel function and $d$ is the number of spatial
dimensions.

Once the entire splitting and merging process is complete, the smoothing
length $h$ is set to an optimal value as discussed in the previous section
using the average mass of the neighboring particles.

The parallelization of the merge step is however, a non-trivial problem
which we discuss here. We wish to use a parallel algorithm that can identify
possible merge candidates in one loop over the particle neighbors. The
algorithm is designed so each particle can identify a suitable merge partner
in parallel. This is achieved using the following approach.

\begin{itemize}
\item A particle $i$ is allowed to merge with another particle $j$ if the
  particle $j$ has not been identified for splitting and $r_{ij} < (h_i + h_j)/2$
  and $m_i + m_j < \max(m_{max}[i], m_{max}[j])$. All neighbors of particle
  $i$ are searched and the closest particle index \code{closest_idx} that
  satisfies these criterion is identified. This is a completely parallel
  operation.
\item If the $i$'th particle's \code{closest_idx} is $j$, and if the closest
  index of the $j$'th particle is $i$, then the two particles may be merged.
  Otherwise the particles are not merged.
\item Once a pair of merging particles are identified, the particle with the
  smaller numerical index value is retained and the particle with the larger
  index is marked for deletion.
\end{itemize}

This algorithm is entirely parallel and can be implemented on a CPU or GPU
very easily. In fact, these computations may be implemented easily in the
context of a SPH calculation. After the identification is complete, one can
easily identify the particles that need to be deleted or merged.

The above algorithm may run into pathological particle configurations which
will not merge enough particles. However, we find that this does not happen in
practice and the algorithm works rather well especially since the particles
are constantly moving and are homegenized by the use of a particle shifting
procedure.

\begin{figure}[h!]
  \centering
  \includegraphics[width=0.8\linewidth]{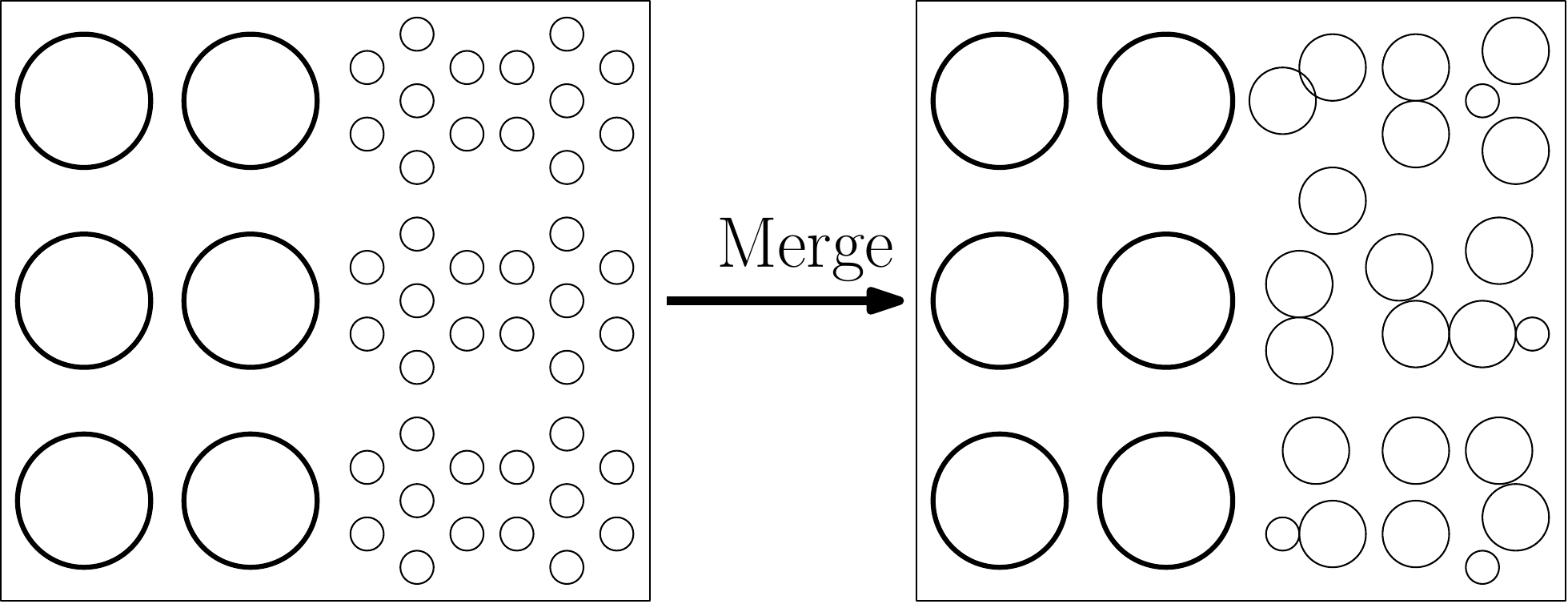}
  \caption{Sketch of particles splitting and then being merged. On the left
    side is a set of particles that split into 7 children each. On the right,
    these particles are merged to reduce the number of particles. Note that
    only one round of merging is complete at this stage.}%
  \label{fig:split-merge}
\end{figure}

It is important to note that when the particles are split, one particle is
split into 7 (as discussed in section~\ref{subsec:split}). In order to reduce
the number of particles we also iteratively perform merging using the same
algorithm as discussed above. The reason we choose to split particles into 7
and then merge is that this tends to produce much lower errors since the
particle distributions after splitting are more uniform and this makes it more
effective to find merge partners. Since the merging is performed pairwise and
we desire that on the average each particle be split into two particles, we
must have at least three merges. Increasing the number of merges is
computationally expensive so we limit it to three. In
Fig~\ref{fig:split-merge}, we show on the left two columns of parent particles
that are moving. As they move to the right, they split into 7 children
each. These are merged once to produce the particles on the right. With a
subsequent merge the remaining small particles are also merged into larger
particles depending on the allowed maximum and minimum masses. The figure
indicates that the particles are disordered. In order to correct these we
perform particle shifting iteratively three times using \cref{eq:shift} and
correct the properties of the fluid using \cref{eq:shift-correct}. In
subsection~\ref{sec:adapt} we show some particle plots (see
Fig.~\ref{fig:fluid-init-mass}) where one can clearly see that the particles
are uniformly distributed.

\subsection{Automatic adaptation}%
\label{sec:automatic-adapt}

The key part of the adaptive split and merge algorithm is in setting the
appropriate $m_{max}(\ten{r})$ and $m_{min}(\ten{r})$ spatially. In simple
cases, it is possible to manually assign the appropriate reference mass for
different spatial regions. On the other hand for more complex cases we may not
be able to set this manually. For example when simulating the flow past a
bluff body, we would like to prescribe the minimum and maximum resolutions and
automatically define the reference mass based on the distance from the solid
body. In addition when the solid body moves, the reference mass should be
suitably updated. Finally, the algorithm should also support solution
adaptivity. We first discuss the simpler case of geometry dependent spatial
adaptation and then discuss how solution adaptivity can be added.

We setup the discussion in the context of wind-tunnel-like problems where a
collection of stationary or moving solid bodies is placed in a stream of fluid
with a suitable inlet and outlet. In these class of problems, the solid body
typically defines the highest resolution since this is where the largest
gradients are observed.

We use the term \emph{size} of a particle to refer to the inter-particle
spacing $\Delta s$.  We determine a suitable reference mass in a region,
$m_{ref}$ and then set $m_{min} = 0.5 m_{ref}$ and $m_{max} = 1.05
m_{ref}$. The size of the particle immediately determines its $m_{ref}$, for
in two-dimensions, $m_{ref} = \rho {(\Delta s)}^d$ where $d$ is the number of
spatial dimensions. In order to smoothly vary the regions, we use a parameter
$1.05 \le C_r \le 1.2$. The sizes of particles in two adjacent regions are in
this ratio, i.e.\ $\Delta s_{k+1} = C_r \Delta s_{k}$, where $k$ indicates a
layer of particles with a similar resolution. We assume that the minimum
resolution $\Delta s_{min}$ and the maximum resolution $\Delta s_{max}$ for
the particles are known quantities. We note that $h = 3 \Delta s$ for the
simulations in this work.

The Lagrangian nature of the SPH method makes it difficult to use the fluid
particles to themselves define the reference mass. Instead, we employ a set of
stationary background particles. These background particles are not involved
in the computation of the governing equations of motion of the fluid or solid.
They are merely used to set $m_{ref}$ based on the requirements. The
background particles are initially setup with a constant size of
$\Delta s_{max}$.  The solid geometry of interest is discretized at a
resolution of $\Delta s_{min}$.  Given these, we initialize the background
when the simulation starts as follows.

\begin{enumerate}
\item Iterate over all the background particles. If a background particle has
  a solid particle as a neighbor, then the background particle is marked as
  being near a boundary. In our implementation, we have a simple integer mask
  which is set to the value 1. These particles are set to have the smallest
  size (the same as that of the solid particles). We call these the
  \emph{boundary background particles}.
\item Once the boundary background particles are identified, we iterate over
  the remaining particles and find the minimum ($\Delta s_{min}$), maximum
  ($\Delta s_{max}$), and geometric mean ($\Delta s_{avg}$) of the sizes of
  the neighboring particles. If the $\Delta s_{max}/\Delta s_{min} < C_r^3$,
  this suggests that the regions are near the ideal distribution, and we set
  the size of the particle to be equal to $C_r \Delta s_{min}$. If $\Delta
  s_{max}/\Delta s_{min} >= C_r^3$, then we set the size of the particle to
  $\Delta s_{avg}$. This allows the particle sizes to be refined rapidly in
  the initial stages when most of the particles are at the highest resolution.
  When the distribution is nearing the desired distribution we ensure that the
  nearby layers are such that $\Delta s_{k+1} = C_r \Delta s_{k}$, where $k$
  indicates a layer. We note that the size of the particle immediately
  determines $m_{ref}, m_{min}$, and $m_{max}$.
\item Once the reference mass of the particles is set, the particles are split
  if required using the same splitting algorithm as used for the fluid
  particles.
\item The smoothing length of the particles is now set such that the number of
  neighbors is roughly the same. Equation~\eqref{eq:h-smoothing} is used for
  this purpose and is discussed below. The background particles are also moved
  to distribute them uniformly using the same PST method as used for the
  fluid. Both the method of \cite{vacondio_splitting_2013} or
  \cite{diff_smoothing_sph:lind:jcp:2009} work well although we use that of
  \cite{diff_smoothing_sph:lind:jcp:2009} in this work since it is parameter
  free and works very well. These two operations of setting the smoothing
  length and using a PST are repeated three times.
\end{enumerate}

The equation used to iteratively set the smoothing length of the background
particles is the same as that used in \cite{yang_adaptive_2019} and is
reproduced here,
\begin{equation}
  \label{eq:h-smoothing}
  h_{i}^{n+1} = \frac{1}{2}\left(
    \frac{h_i^n}{2} \left( 1 + \sqrt{\frac{N_r}{N_i^n}} \right)
    + \frac{\sum_j h_j^n}{N_i^n}
  \right),
\end{equation}
where $N_r$ is a reference number of neighbor particles, $N_i^n$ is the number
of neighbors for particle $i$ at iteration level $n$. This approach ensures
that each particle has close to $N_r$ neighbors eventually. We reiterate that
this algorithm is only used for the background particles so that they smoothly
vary and the computations of the references masses are smooth. For the
two-dimensional flow problems considered here, we use $N_r=48$.

In order to initialize the background particles, the above steps are repeated
$3 \lceil \log(\Delta s_{max}/\Delta s_{min})/\log(C_r) \rceil $ times to
setup the initial background. The fluid particle resolution is set by finding
the minimum of the background particle reference mass in its
neighborhood. Thus the background particles only define the spatial resolution
for the fluid particles.

\begin{figure}[h!]
  \centering
  \includegraphics[width=0.7\linewidth]{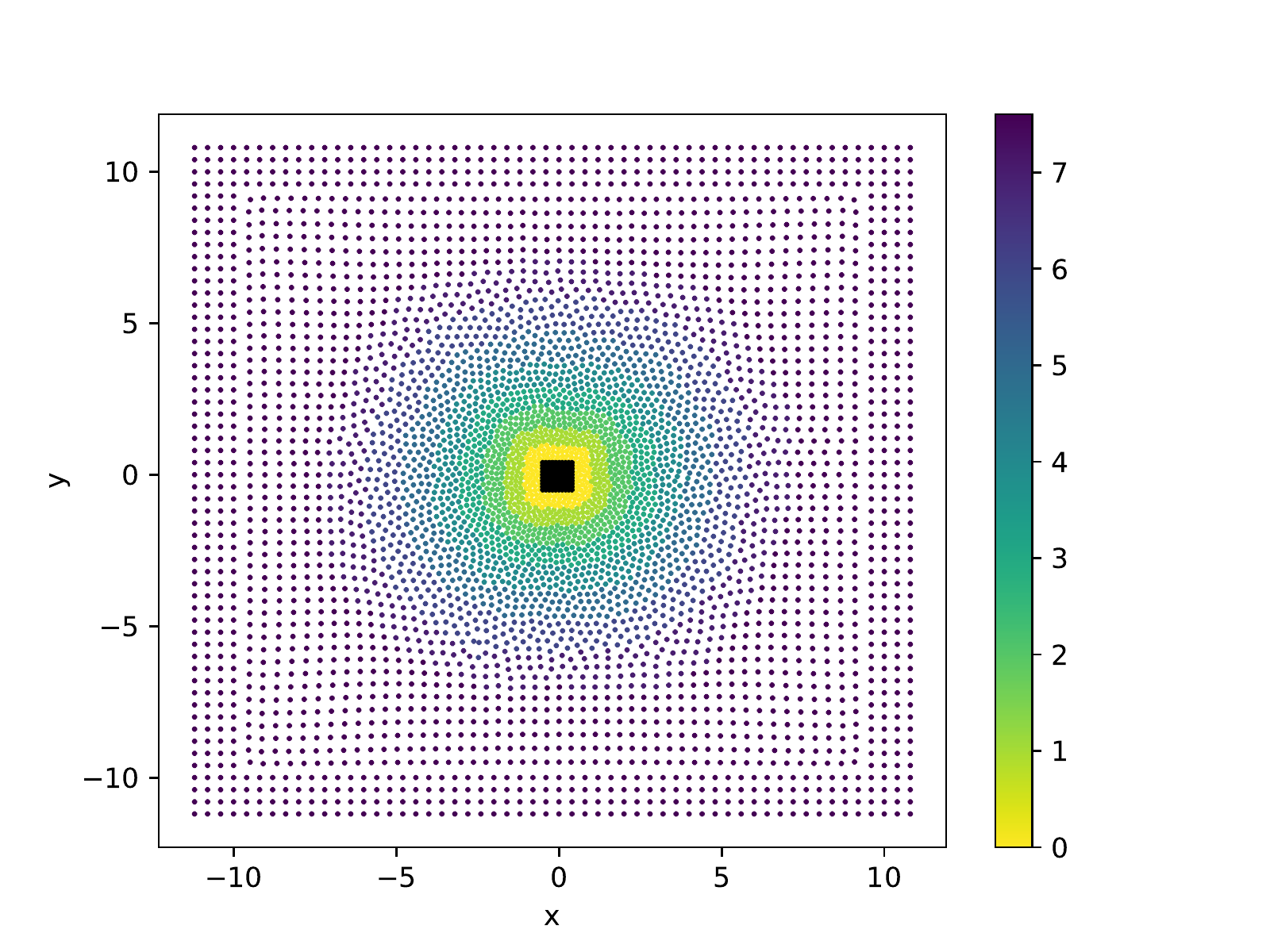}
  \caption{Background particle distribution for flow around a unit square
    shown in black.}
  \label{fig:bg-levels}
\end{figure}

\begin{figure}[h!]
  \centering
  \includegraphics[width=0.7\linewidth]{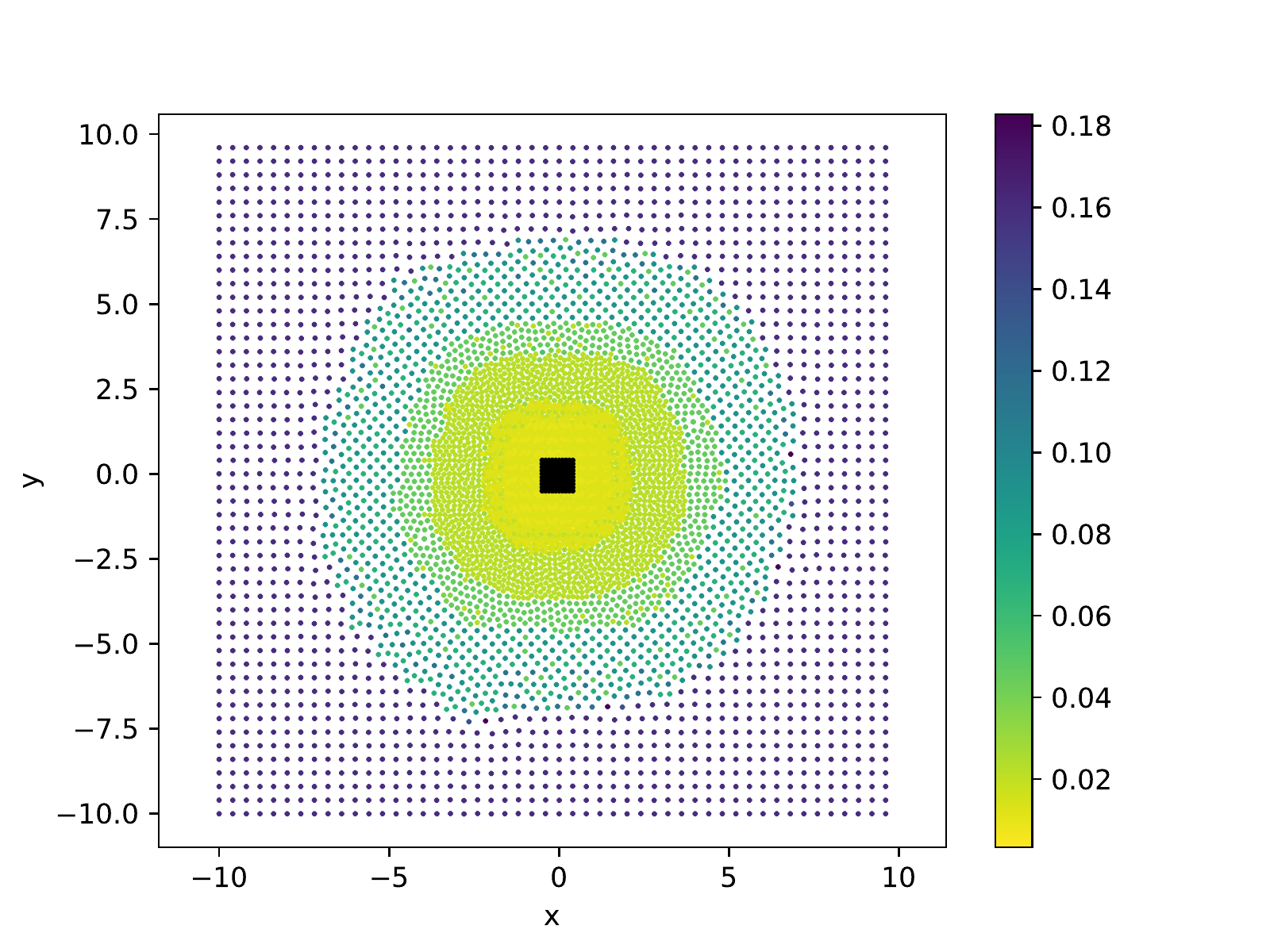}
  \caption{Corresponding fluid particles initialized using the background
    particles with the colors indicating the particle mass. The square solid
    is shown in black.}
  \label{fig:fluid-init-mass}
\end{figure}

In Fig.~\ref{fig:bg-levels} we show the background particles and the
corresponding number of split levels with 0 being the smallest size particles
and 7 being the largest. In this case the minimum particle spacing is 0.1 and
0.4 is the maximum spacing. We choose a $C_r=1.2$ and this generates roughly 8
regions with differing $m_{ref}$ values. The fluid particles created for this
distribution of particles is shown in Fig.~\ref{fig:fluid-init-mass}. Here we
can see that there are only 4 layers since when particles split they
effectively split when the mass from one layer to the next jumps by a factor
of two.

When the solid bodies move, the algorithm above is executed once every few
iterations. This automatically adapts the reference mass distribution in a
smooth fashion. Since the motion of the bodies in each time step is typically
quite small and a fraction of the local smoothing length, we only need to
perform one iteration of the above.

\begin{figure}[h!]
  \centering
  \includegraphics[width=0.7\linewidth]{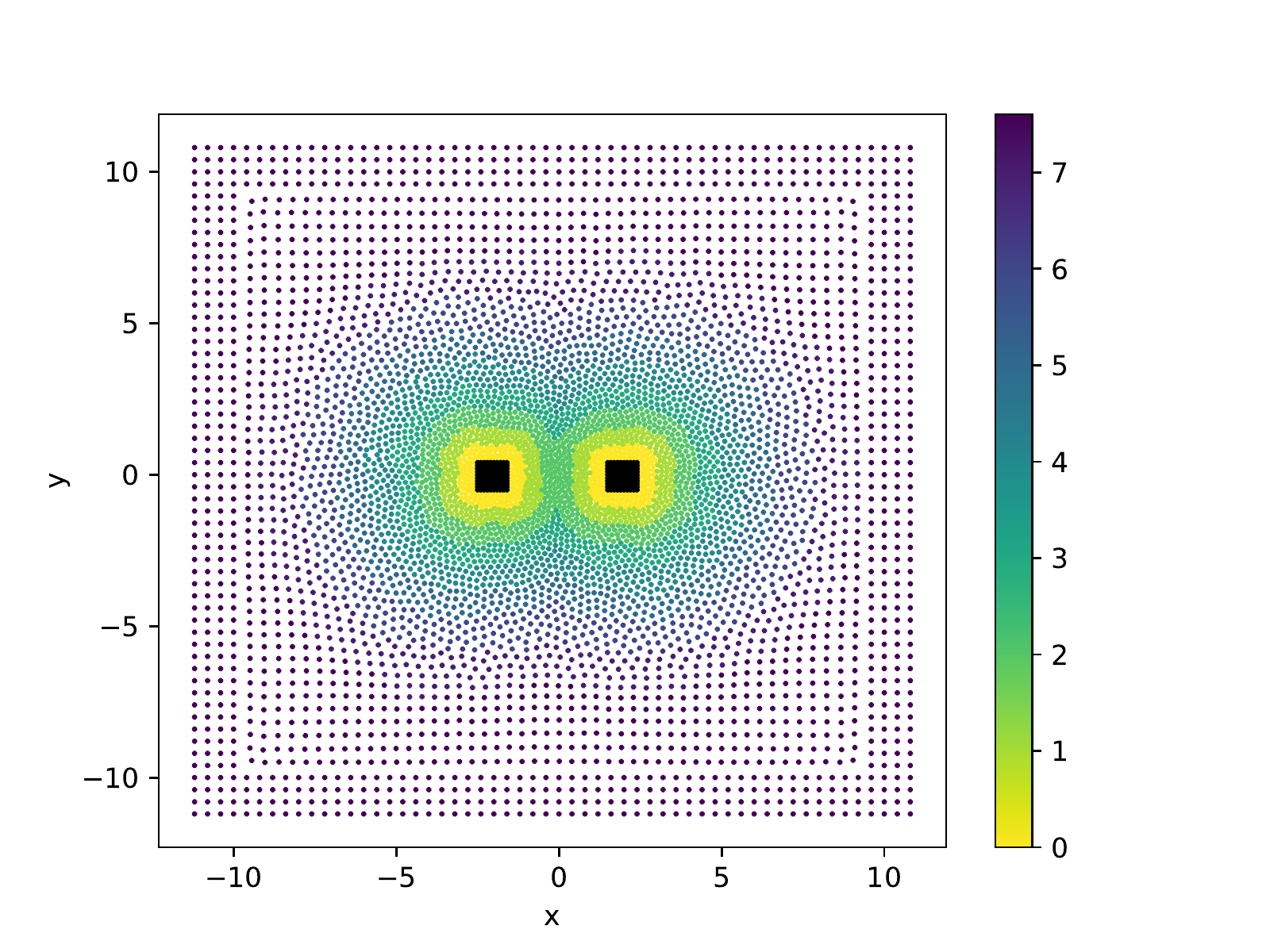}
  \caption{Background particle distribution with two square solid bodies shown
    in black.}
  \label{fig:two-body-bg}
\end{figure}

\begin{figure}[h!]
  \centering
  \includegraphics[width=0.7\linewidth]{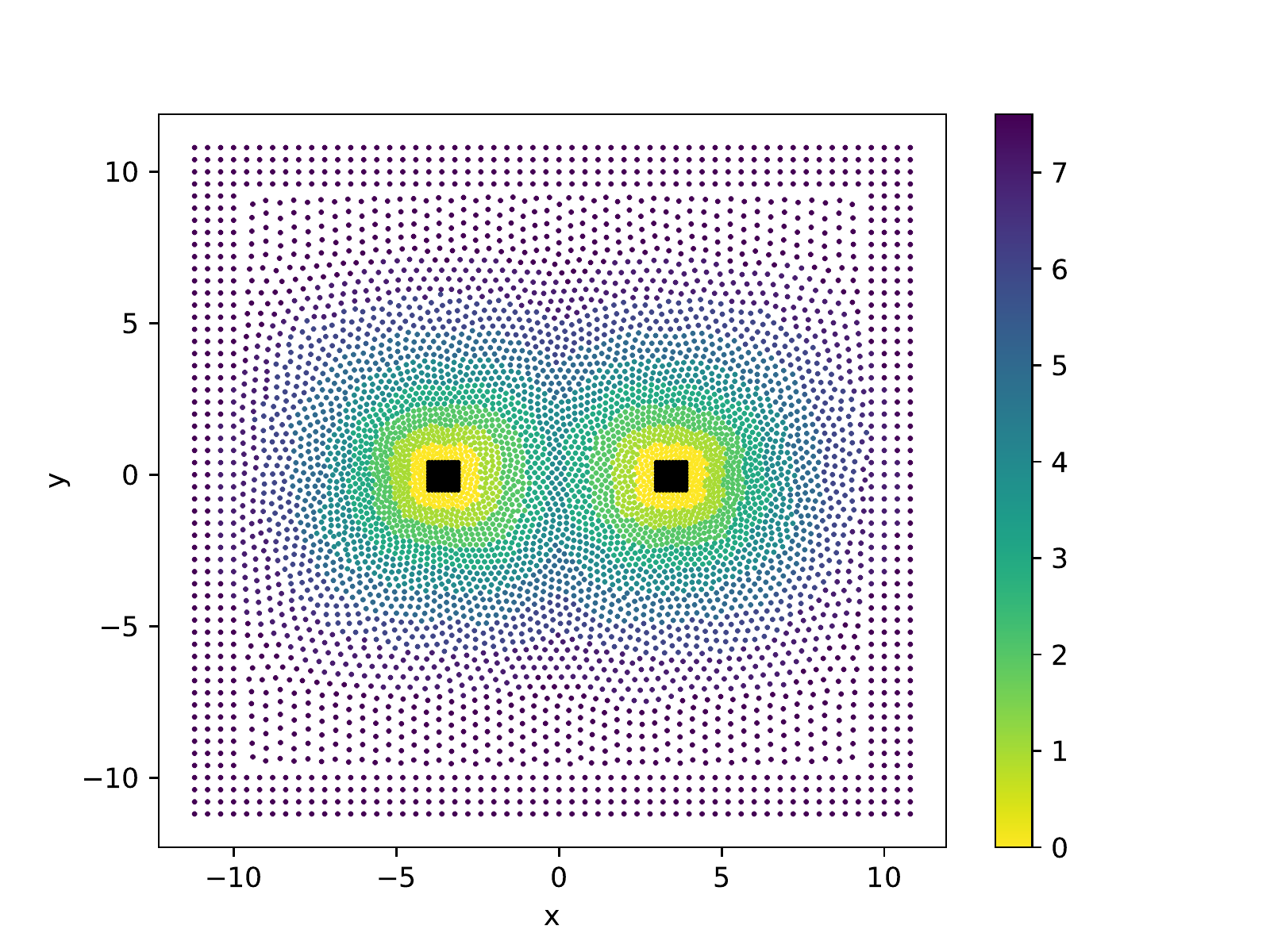}
  \caption{Background particle distribution with two square solid bodies after
    they move by a distance of 3 units in 60 steps.}
  \label{fig:two-body-moved-bg}
\end{figure}

In Fig.~\ref{fig:two-body-bg} we show the case of two unit square solids
placed in a fluid, the background particles are shown and the particle size is
smoothly decreasing towards the solid geometry. We move each square by 0.05
units away from each other in each step and update the background by
performing the steps discussed above once each time step. We do this 60 times
and the resulting background particles are shown in
Fig.~\ref{fig:two-body-moved-bg}. As can be clearly seen, the background mesh
adapts to the moving solid. This shows that the algorithm can
comfortably handle moving geometries.

We note that the current method may also be used to setup a specific
user-defined region with a desired resolution. This is done by creating a set
of particles that serve as a solid body but are only used to set the boundary
background particles. These particles do not participate in any fluid-solid
computations. Thus the approach offers a convenient way to define
user-specified regions with different resolutions.

While we do not explore this extensively in the current work, it is easy to
incorporate solution adaptivity using this framework. Let us assume that there
is some solution dependent scalar $\phi(\ten{r})$ that may be evaluated using
the fluid particles (like the magnitude of the vorticity) and are interpolated
onto the background particles. We can use a linear mapping between the range
of the values of $\phi$ to the minimum and maximum allowed resolution. Once
the boundary background particles are identified (step 1 in the algorithm for
the background particle) we use the $\phi$ value to appropriately set the
resolution. The rest of the algorithm then proceeds as before to update the
remaining particles.

We show examples of solution adaptivity based on the vorticity, $\omega$ in
section~\ref{sec:sol-adapt}. In this case, we compute the absolute magnitude
of the vorticity of the fluid particles and interpolate them onto the
background particles as the value of $\phi$. Any particles with a value of
$\phi > k \max(\omega)$, where $k$ is a user-specified value, are assigned the
highest resolution. This approach allows us to track the vorticity
adaptively. The approach may be easily extended to use different refinement
criterion if so desired. The proposed algorithm can thus handle a variety of
different forms of adaptivity.

We want to assess the errors due the splitting and merging. First, we estimate
the error in the global density due to merging of particles of different mass
ratios at varying separation distance. We consider a square domain $\Omega$ of
uniformly discretized points and two additional particles which are
merging. We compute the error in the global
density~\cite{vacondio_accurate_2012}, due to the merging, defined as,
\begin{equation}
  \label{eq:merge-error}
   \mathcal{E}_{m} (\ten{x}) = \int_{\Omega} \left[(m_a + m_b) W_m (x,
     h_m) - m_a  W_a (x, h_a) - m_b W_a (x, h_a)\right] \mathrm{d} \ten{x}
\end{equation}
where $h_m$ is set using \cref{eq:merged-h}, subscripts $a$ and $b$ denote
particles which are being merged, and subscript $m$ denotes the merged
particle. $\beta = \frac{m_a}{m_b}$ is the mass ratio of the particles being
merged. The integration is performed over the region of uniformly discretized
points and the particles have full kernel support.

\begin{figure}[h!]
  \centering
\begin{subfigure}{0.48\textwidth}
  \centering
  \includegraphics[width=\textwidth]{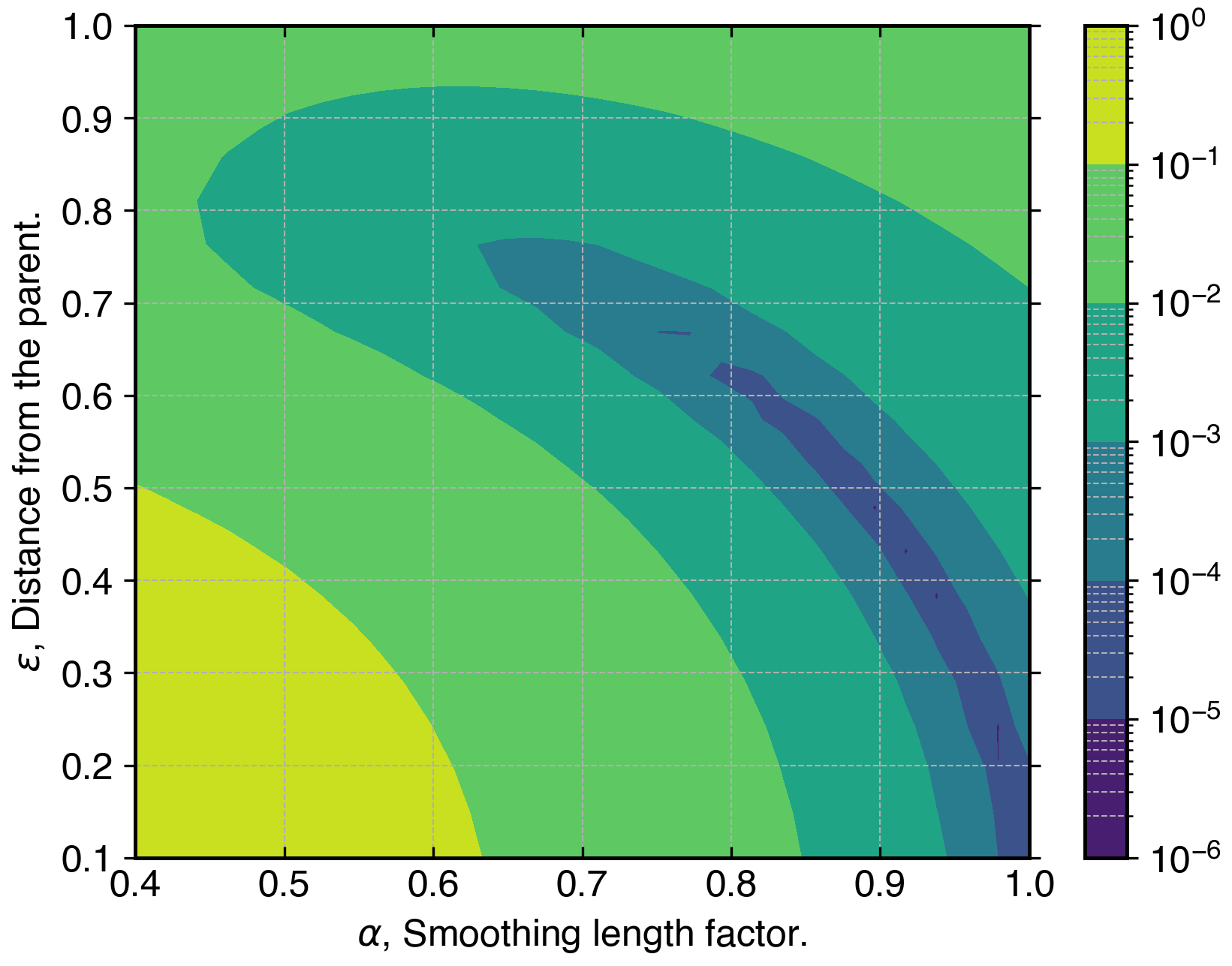}
  \subcaption{}%
  \label{fig:split-error}
\end{subfigure}
\begin{subfigure}{0.48\textwidth}
  \centering
  \includegraphics[width=\textwidth]{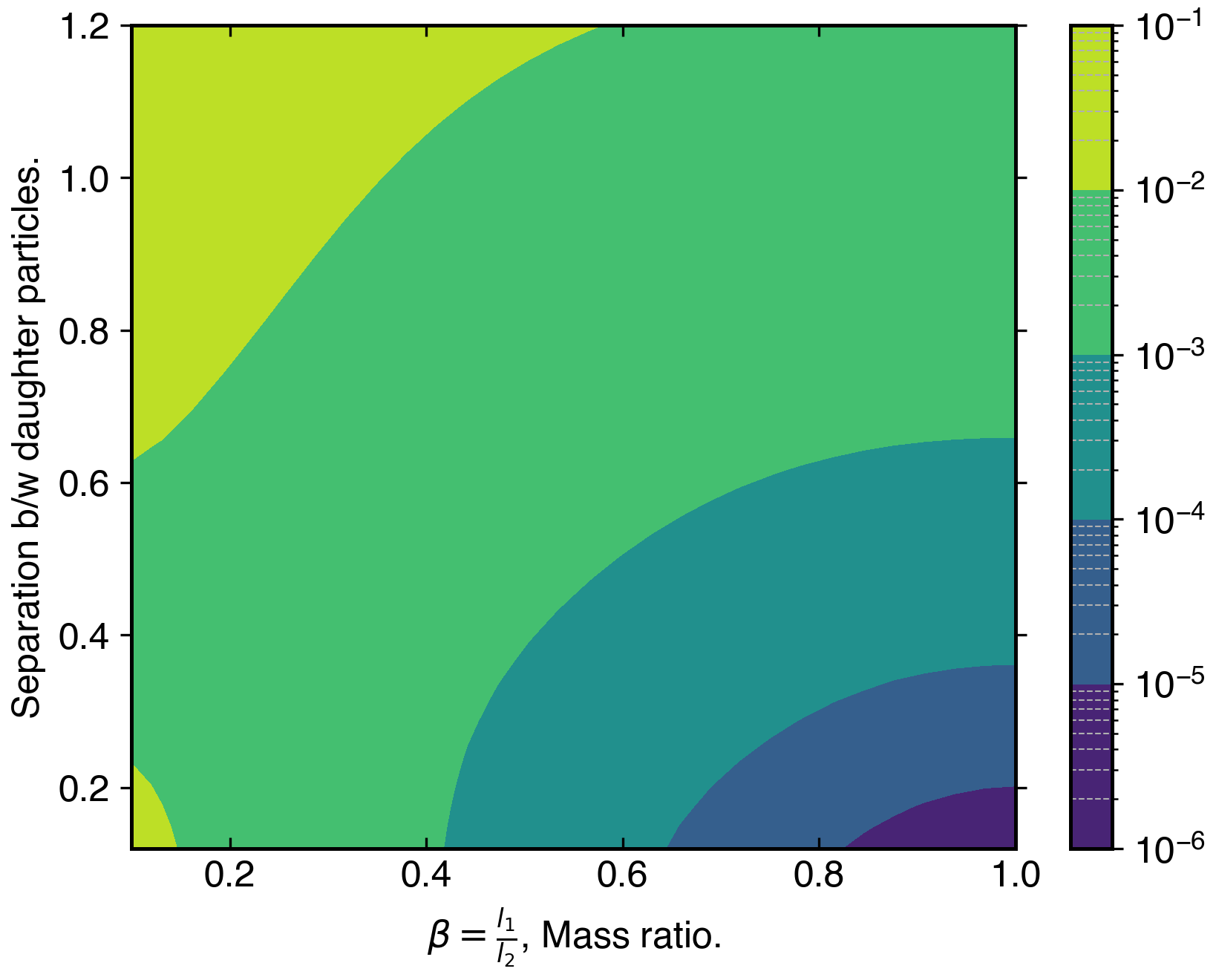}
  \subcaption{}%
  \label{fig:merge-error}
\end{subfigure}
\caption{(a) Error in density by splitting one particle into 7 daughter
  particles with equal mass. We vary the smoothing length factor $\alpha$ and
  the distance of the 6 children particles from the parent particle. (b) Error
  in density while merging two particles of varying mass ratios and separation
  distance.}%
\label{fig:split-merge-error}
\end{figure}

\Cref{fig:merge-error} shows the error in the global density for two particles
merging at varying separation distance and mass ratio. It is clear from the
figure that the merge errors are low when the particles are closer and have
nearly equal mass.

Next, following \citet{vacondio_accurate_2012} in order to assess the error
due to splitting one particle into 6+1 daughter particles, we compute the
error in global density. We consider a square domain $\Omega$ of uniformly
discretized points and a single SPH particle. We split a single SPH particle
into 7 daughter particles, where 6 daughter particles are placed on the
vertices of a hexagon centred around the parent particle and one daughter
particle at the location of the parent particle. The distance of the 6
daughter particles from the center is controlled by the parameter $\epsilon$,
where $r_p = \epsilon h_{p}$, and the smoothing length of the
k\textsuperscript{th} daughter particle, $h_k = \alpha h_{p}$, is controlled
by the parameter $\alpha$. The global density error is then computed by
evaluating,
\begin{equation}
  \label{eq:global-error}
\begin{split}
  \mathcal{E}_s (\ten{x}) = m^2_p \Bigl[ &\int_{\Omega} W_p^2(\ten{x}, h_p)
  \mathrm{d} \ten{x} - 2 \sum_{k=1}^7\int_{\Omega} \lambda_k W_p(\ten{x}, h_p)
  W_k(\ten{x}, h_k)
  \mathrm{d} \ten{x} \\
  &+ \sum_{k, l=1}^7 \int_{\Omega} \lambda_k \lambda_l W_k(\ten{x}, h_k)
  W_l(\ten{x}, h_l) \mathrm{d} \ten{x}\Bigr],
\end{split}
 \end{equation}
 where $\lambda_k$ is the mass ratio of the parent particle to the
 $k^{\text{th}}$ daughter particle, the subscript $p$ denote parent particle,
 the summation is over all the children particles. The kernel function is
 computed and the integration is performed over the uniformly discretized
 points. The parent and all the daughter particles always have full kernel
 support. We take $\lambda_k = 1/7$ in this work based on the results of the
 errors in merging shown above.

\Cref{fig:split-error} shows the error in the global density for a particle
split into 7 equal mass daughter particles. Based on the results of errors in
merging and splitting we choose the values of the daughters' smoothing length
factor $\alpha = 0.9$ and position from the center $\epsilon = 0.4$.

\subsection{Algorithm}
\label{subsec:algo}
In this section we summarize the adaptive resolution algorithm. We start with
a given solid body or multiple such bodies that are discretized at the highest
desired resolution, with particle spacing, $\Delta s_{min}$. For complex
geometries, we may use the particle packing method proposed in
\cite{packing_2021} to generate uniformly distributed particles for
discretizing the solid bodies. We prescribe a coarsest resolution $\Delta
s_{max}$ as well as the desired $C_r$ factor which is typically between the
values of 1.05 to 1.2. This effectively determines the width of each
refinement layer. One may also manually specify the constraints on the mass in
different spatial regions. Finally, we are given a domain of interest; in the
problems considered in this work, the domain size is fixed and known a priori.
Given this, the algorithm proceeds as follows.

\begin{enumerate}

\item We first initialize the background particles as discussed in
  section~\ref{sec:adapt}.

\item Using the background particles, we initialize the fluid particles at the
  initial time. This is done by splitting and merging the particles based on
  the reference mass of the background particles. This is discussed in detail
  in sections \ref{subsec:split} and \ref{subsec:merge}. The particle shifting
  algorithm of \citet{diff_smoothing_sph:lind:jcp:2009} is applied at each
  stage to get a smooth distribution of particles.

\item The initialized fluid particles along with the given solid particles are
  then used to simulate the governing equations using an appropriate scheme.
  In the present work we use a highly modified EDAC-SPH scheme as discussed in
  \cref{sec:sph}.

\item At the end of every iteration, we find the nearest fluid particles to
  the background particles and set the reference mass of the fluid particles.

\item The fluid particles are adaptively split and merged every $n_{adapt}$
  iterations as discussed in sections \ref{subsec:split} and
  \ref{subsec:merge}. Typically we choose $n_{adapt}$ to be between 1--10.
  Similarly, the background particles are updated (to accommodate moving
  bodies or solution adaptivity) every $n_{bg}$ iterations. When the
  adaptation is entirely spatial and the solid boundaries do not move, the
  background does not need to be updated at all; we usually set this to 100 or
  500 iterations. This parameter is adjustable depending on the requirements.

\item After merging, we shift the particles three times using \cref{eq:shift}
  and correct the properties of the fluid \cref{eq:shift-correct}.

\item The smoothing length of all the particles is set using the average mass
  of the neighbors as discussed in \cref{subsec:split}.
\end{enumerate}

After the initialization is complete, the time-marching procedure of the SPH
scheme is started. Before every iteration, the following is performed: (i) we
update the background particles every $n_{bg}$ iterations; (ii) thereafter,
every $n_{adapt}$ iterations, we split, merge, shift and correct the
properties of the particles; (iii) we update the smoothing length of the
particles based on average mass and update the nearest neighbor search
algorithm to fetch new neighbors; (iv) finally, we execute each time-step of
the time-marching scheme as discussed in the \cref{sec:integration}.

\begin{figure}[!htp]
  \centering
  \includegraphics[width=\textwidth]{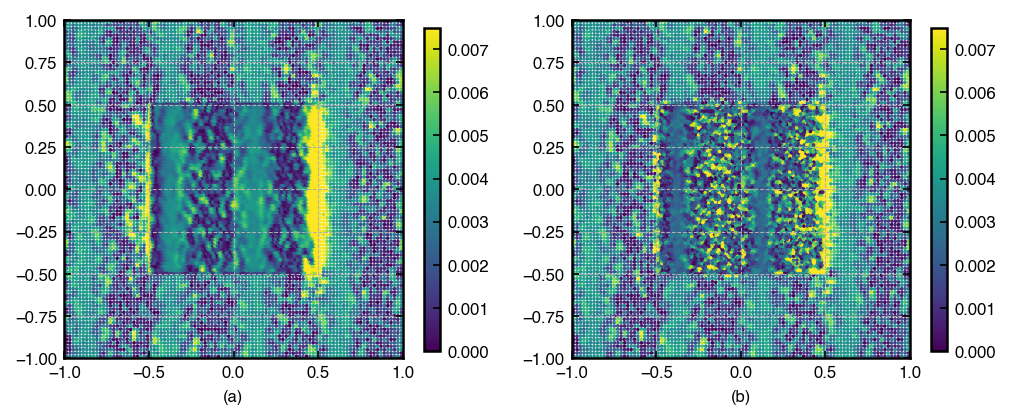}
  \caption{Error in approximation of a function $f(x) = sin(2 \pi x) + cos(2
    \pi x)$ using (a) \citet{vacondio_accurate_2012} formulation and (b) the
    proposed formulation.}%
  \label{fig:func-error}
\end{figure}

\begin{figure}[!htp]
  \centering
  \includegraphics[width=\textwidth]{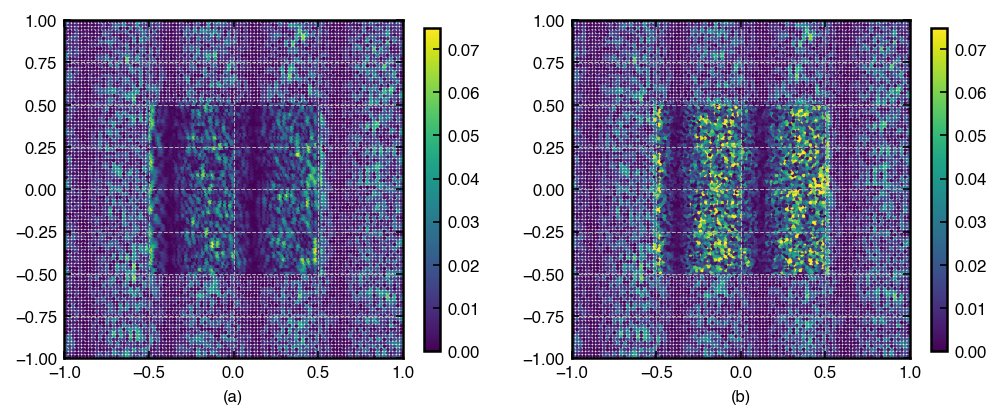}
  \caption{Error in approximation of the gradient of a function
    $f(x) = sin(2 \pi x) + cos(2 \pi x)$ using (a)
    \citet{vacondio_accurate_2012} formulation and (b) the proposed
    formulation.}%
  \label{fig:grad-error}
\end{figure}

In order to assess the accuracy of the algorithm, we consider the following
test case. We consider a square region of side two units on which we
discretize the function $f(x, y) = \sin(2 \pi x) + \cos(2 \pi x)$ using
particles of equal size. The inner square region of unit side is split into
particles. We consider two cases, the first using the formulation of
\citet{vacondio_accurate_2012} where each particle in the inner region is
split into 7 daughter particles, and the second using the proposed method. We
then compute the error in function and gradient approximation
\cref{fig:func-error,fig:grad-error}. The results are summarized in
\cref{table:func-grad-error}. As can be seen the proposed algorithm is
accurate and has far fewer neighbors than \citet{vacondio_accurate_2012}.
\begin{table}[ht]
  \centering
  \begin{tabular}[ht]{lll}
    \toprule
    & Vacondio et al. & Our Algorithm \\
    \midrule
    $L_{\infty}$ error in $\rho$& 2.00e-15 & 1.11e-15 \\
    $L_{\infty}$ error in $f$ & 1.0e-02 & 1.0e-02 \\
    $L_{1}$ error in $f$ &  2.5e-03 &  2.4e-03\\
    $L_{\infty}$ error in $\frac{\partial f}{\partial x}$ & 5.7e-02  & 1.2e-01 \\
    $L_{1}$ error in $\frac{\partial f}{\partial x}$ &  7.3e-03 & 8.0e-03 \\
    Average no.\ of neighbors & 116 & 36\\
    \bottomrule
  \end{tabular}
  \caption{Comparison of errors in $L_{\infty}$ and $L_1$ norm of the proposed
    formulation and the formulation of \citet{vacondio_accurate_2012}.}%
  \label{table:func-grad-error}
\end{table}

\section{Results and discussion}%
\label{sec:results}

We apply the adaptive resolution technique proposed in this work to the test
cases shown below. We first apply our method to the classical numerical test
cases with varying Reynolds numbers and compare with established results in the
literature. We then simulate the flow past a circular cylinder at Reynolds
numbers 40, 550, 1000, 3000, and 9500. We show the details that typically
require a large number of particles to capture accurately. We compare the
results of our method to the high resolution vortex method results
of~\citet{koumoutsakos1995}, and~\citet{ramachandran2004}. We also show the
results of solution-based dynamic particle resolution, where the particles are
adaptively resolved to the highest resolution based on the magnitude of
vorticity in the flow, for the flow past a circular cylinder and the flow past a
C-shape at $Re = 2000$. Every figure presented in this manuscript is
automatically generated by an automation framework~\cite{pr:automan:2018}. The
open-source code is available at~\url{https://gitlab.com/pypr/adaptive_sph}.

\subsection{Taylor-Green vortex}%
\label{sec:tgv}
The Taylor-Green problem is a widely used benchmark to study accuracy in
SPH~\cite{vacondio_accurate_2012,chiron_apr_2018}. The exact solution for the
Taylor-Green problem is given by,
\begin{align}
  \label{eq:tgv_sol}
  u^+ = -&U_{\infty} e^{bt} \cos(2 \pi x^+ / L) \sin(2 \pi y^+ / L), \\
  v^+ =\quad&U_{\infty} e^{bt}\sin(2 \pi x^+ / L) \cos(2 \pi y^+ / L), \\
  p^+ = -&U^2_{\infty} e^{2bt} (\cos(4 \pi x^+ / L) + \cos(4 \pi y^+ / L))/4,
\end{align}
where the superscript $(.)^+$ indicate dimensional quantity, $U_{\infty} = 1$
m/s, $b=-8\pi^2/Re$, $Re=U L /\nu$, and $L=1$ m. In the results we use the
dimensionless velocities $u = u^+/U_{\infty}$ and $v = v^+/U_{\infty}$, pressure
$p = p^+/\rho_0 U^2_{\infty}$, and distance $x = x/L$ and $y = y/L$. The domain
is a square region $[0, L] \times [0, L]$ with an inner refinement zone
$[0.1L, 0.5L] \times [0.1L, 0.5L]$ where the resolution is 2 times higher than
the outer resolution.  We simulate the problem using the parameters given
in~\cref{tab:tgv-params}.

\begin{table}[!ht]
  \centering
  \begin{tabular}[!ht]{ll}
    \toprule
    Quantity & Values\\
    \midrule
    $L$, length of the domain & 1 m \\
    Time of simulation & 2.5 s \\
    $c_s$ & 10 m/s \\
    $\rho_0$, reference density & 1 kg/m\textsuperscript{3} \\
    Reynolds number & 200 \& 1000 \\
    Resolution, $L/\Delta x_{\max} : L/\Delta x_{\min}$ & $[50:100]$ \& $[100:200]$ \& $[150:300]$\\
    Smoothing length factor, $h/\Delta x$ & 1.0\\
    \bottomrule
  \end{tabular}
  \caption{Parameters used for the Taylor-Green vortex problem.}%
  \label{tab:tgv-params}
\end{table}

We consider two Reynolds numbers $200$ and $1000$, and simulate the problem
using the adaptive algorithm proposed in this paper with three different minimum
resolutions $L/\Delta x_{\max}$ of 50, 100 and 150. We compare the results with the
exact solution and the non-adaptive simulation with a resolution that matches
the minimum resolution of the adaptive case. For the $Re = 200$ case we also
compare with the Adaptive Particle Refinement (APR) results
of~\cite{chiron_apr_2018}.

\begin{figure}[!ht]
  \centering
  \begin{subfigure}{0.48\textwidth}
    \centering
    \includegraphics[width=\textwidth]{%
      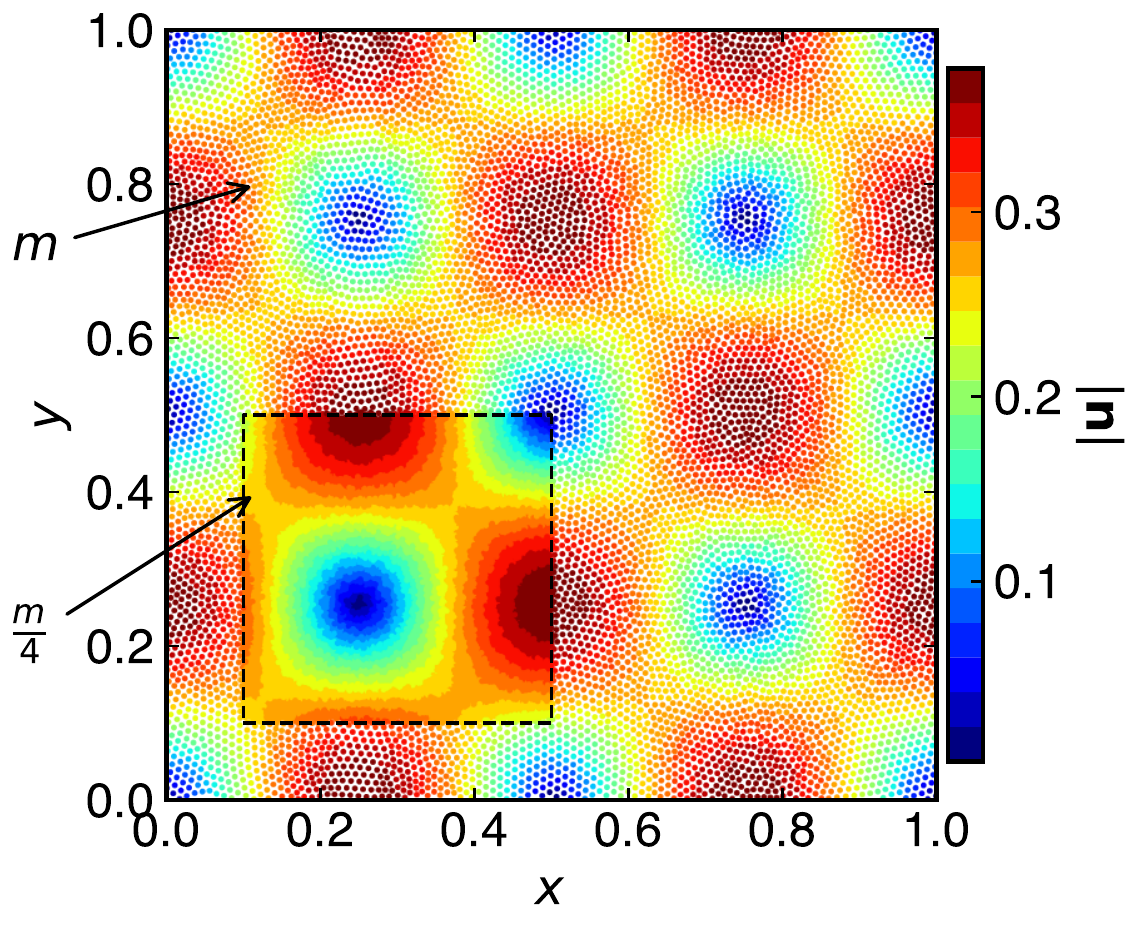%
    }%
    \label{fig:tg:pplot:vmag:200}
  \end{subfigure}
  \begin{subfigure}{0.48\textwidth}
    \centering
    \includegraphics[width=\textwidth]{%
      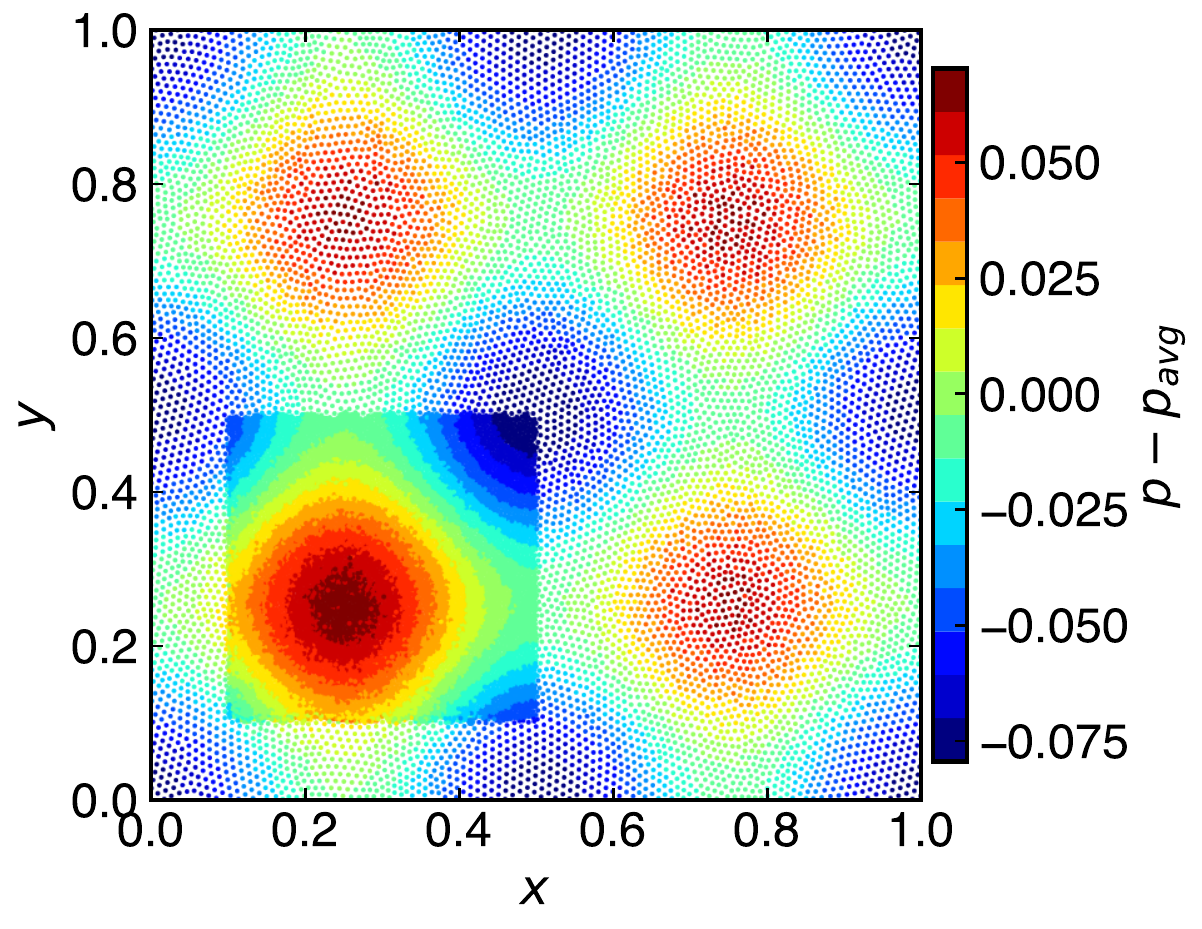%
    }%
    \label{fig:tg:pplot:p:200}
  \end{subfigure}
  \caption{Particle plots for the Taylor-Green vortex problem at $t = 2.5$ s and
    $L/\Delta x_{\max} = 100$. Reynolds number is $200$. The mass of the
    particles inside the dashed region is $1/4$ times the mass of the particles
    outside the region i.e., inside the region the resolution $L/\Delta x$ is
    200.}%
  \label{fig:tg:pplot:re200}
\end{figure}

\begin{figure}[!ht]
  \centering
  \begin{subfigure}{0.48\textwidth}
  \centering
  \includegraphics[width=\textwidth]{%
    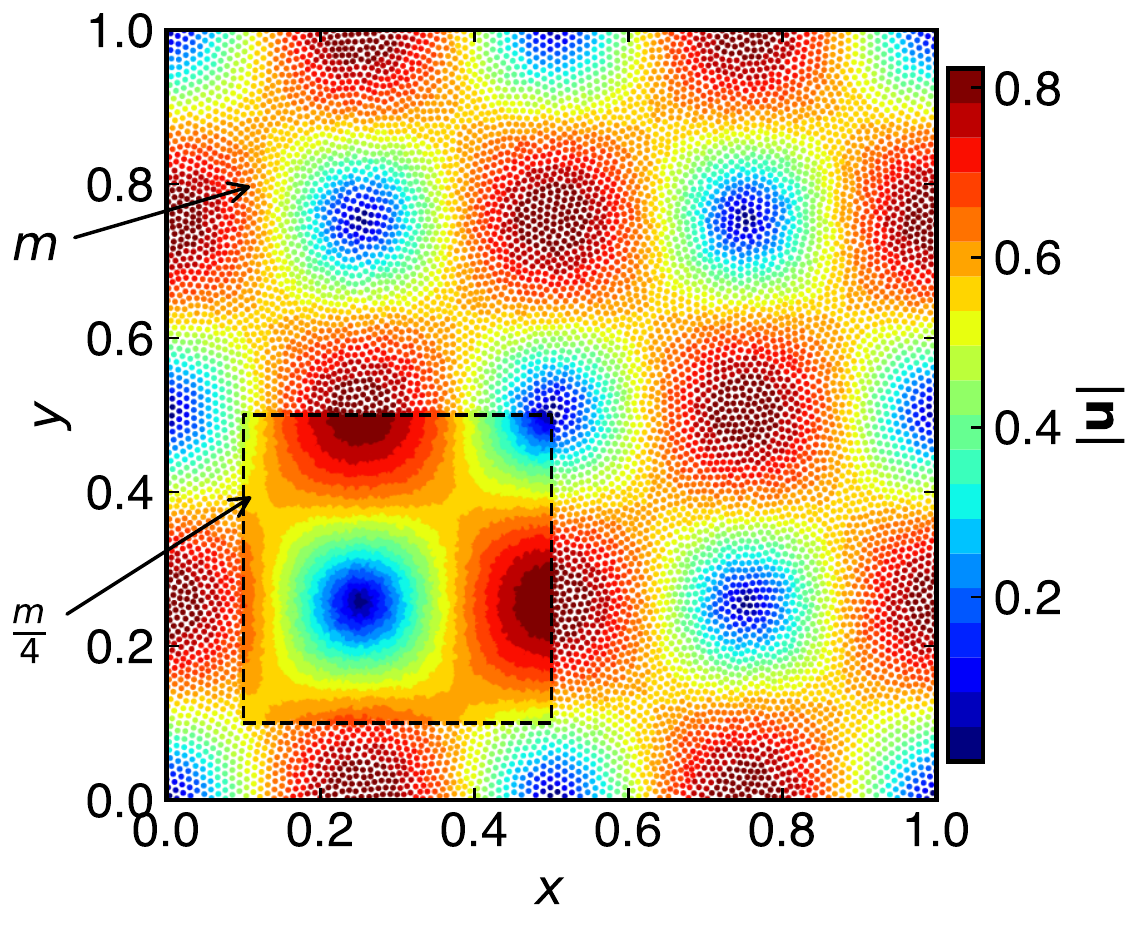%
  }%
  \label{fig:tg:pplot:vmag:1000}
\end{subfigure}
  \begin{subfigure}{0.48\textwidth}
  \centering
  \includegraphics[width=\textwidth]{%
    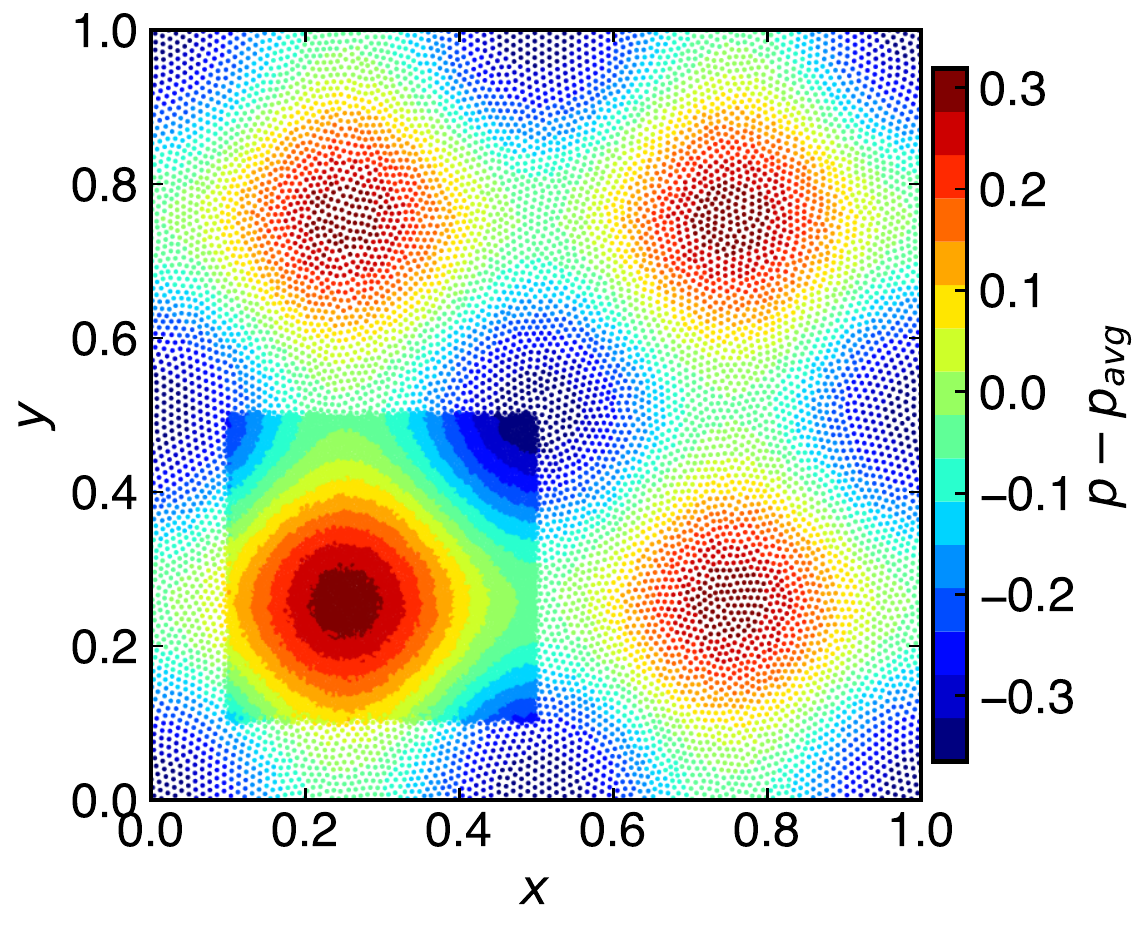%
  }%
  \label{fig:tg:pplot:p:1000}
  \end{subfigure}
  \caption{Particle plots for the Taylor-Green vortex problem at $t = 2.5$ s and
    $L/\Delta x_{\max} = 100$. Reynolds number is $1000$. The mass of the
    particles inside the dashed region is $1/4$ times the mass of the particles
    outside the region indicating the resolution inside corresponds to
    $L/\Delta x$ of 200.}%
  \label{fig:tg:pplot:re1000}
\end{figure}

\begin{figure}[!ht]
  \centering
  \begin{subfigure}{0.48\textwidth}
    \centering
    \includegraphics[width=\linewidth]{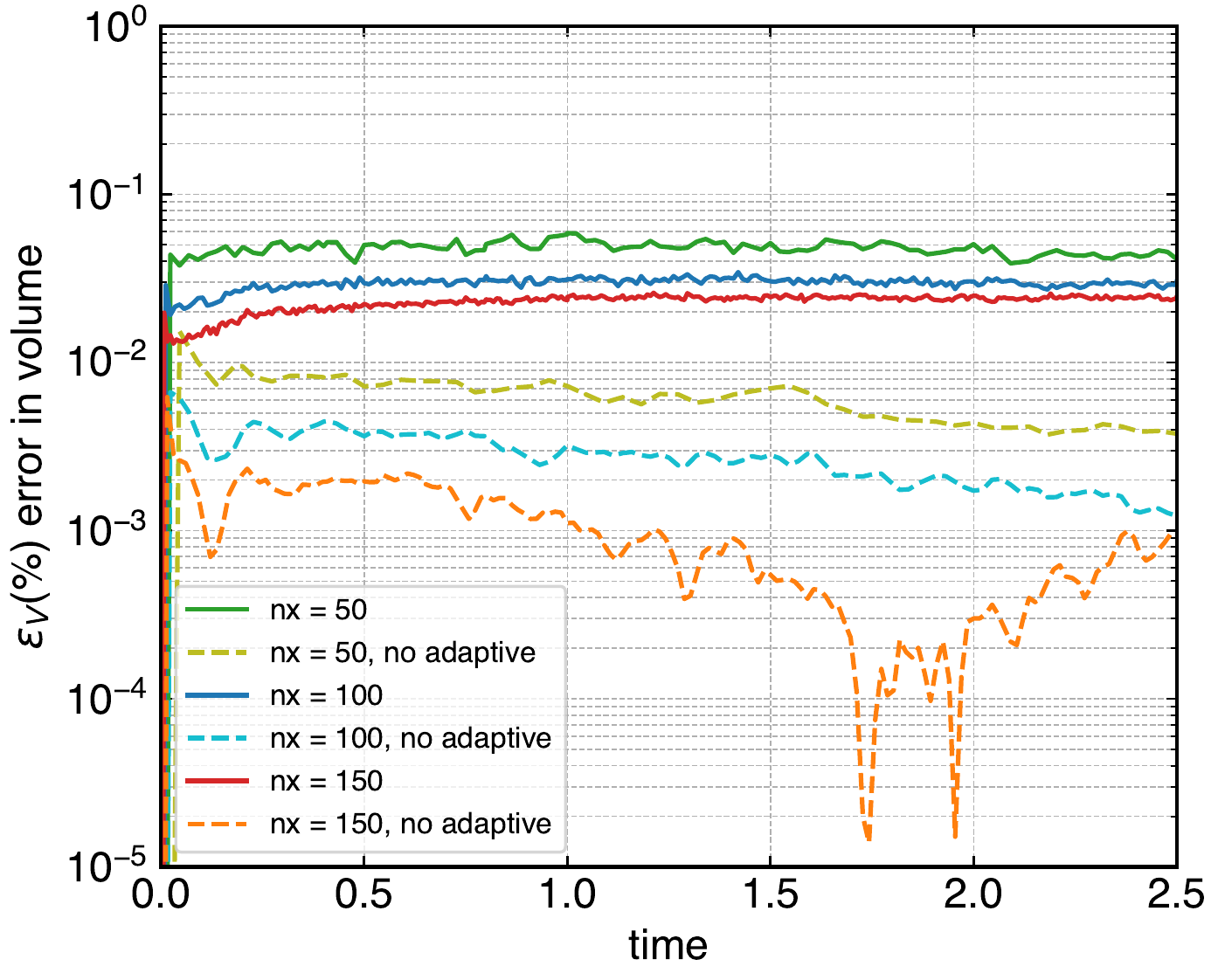}%
    \label{fig:tg:vol:consv:200}
  \end{subfigure}
  \begin{subfigure}{0.48\textwidth}
    \centering
    \includegraphics[width=\linewidth]{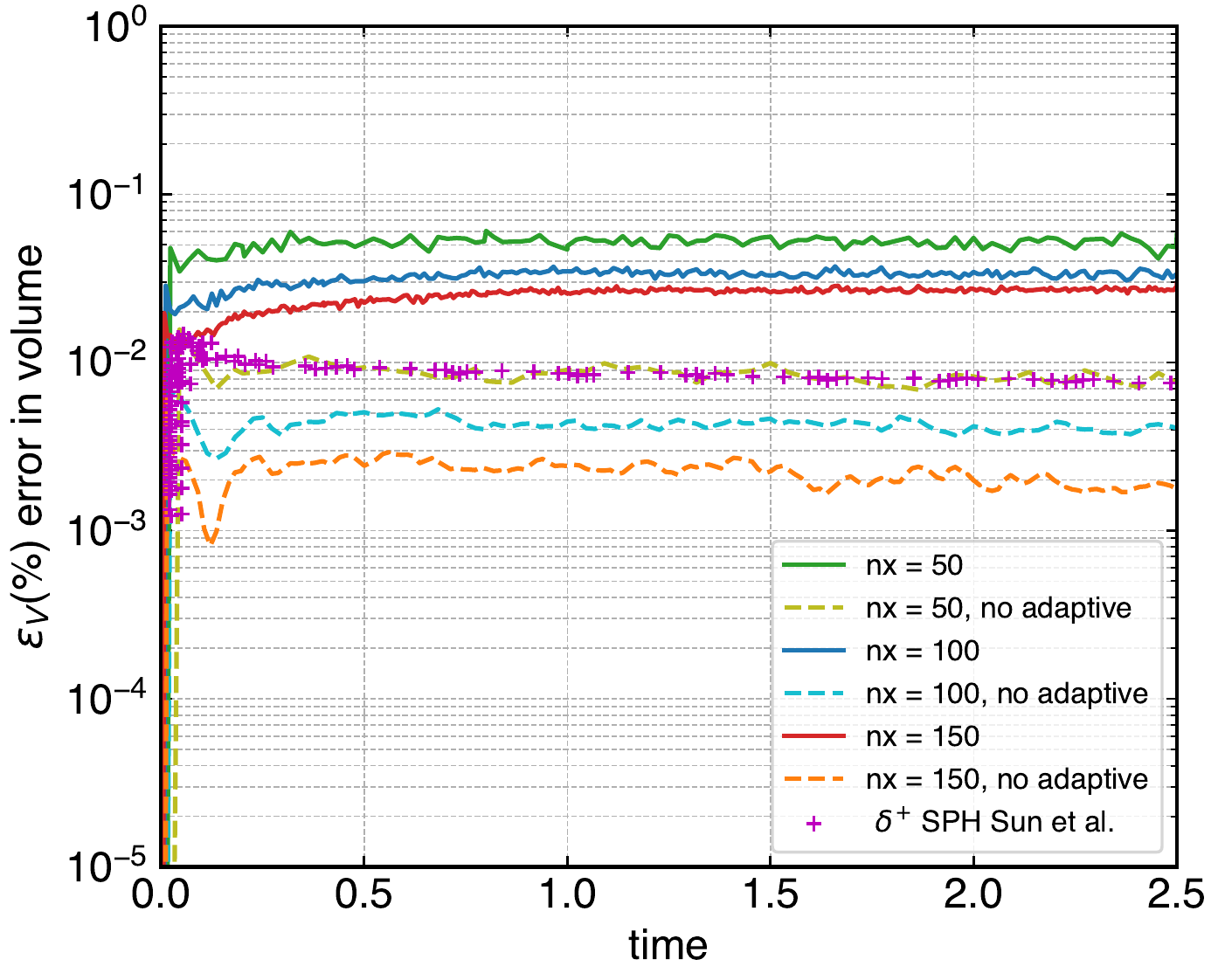}%
    \label{fig:tg:vol:consv:1000}
  \end{subfigure}
  \caption{Evolution of the percentage change in total volume occupied by the
    particles from the initial one, $\epsilon_V (\%)$, for the Reynolds number 200
    (left) and 1000 (right).}%
  \label{fig:tg:vol:consv}
\end{figure}

\Cref{fig:tg:pplot:re200} shows the velocity magnitude and the pressure particle
plots for $Re = 200$ at $t = 2.5$ s. The velocity and pressure contours show
less decay than~\cite{chiron_apr_2018}. Specifically, note that the high
velocity regions near the interface of the layers are well maintained in the
present work. We show the results for $Re = 1000$
in~\cref{fig:tg:pplot:re1000}. It can be seen that the results are as expected
and the contours do not show any decay or change in shape. We note that there is
a drift in the average pressure and subtract it from the pressure contour
plots. This drift in pressure is due to errors in the total volume conservation
\cite{sun_consistent_2019}, which is computed as follows,
\begin{equation}
  \label{eq:1}
  \epsilon_V(\%) = \Bigg|\frac{\sum_i m_i/\rho_i}{L^2} - 1\Bigg| \times 100.
\end{equation}
\Cref{fig:tg:vol:consv} shows the error in total volume. We compare the adaptive
and non-adaptive cases with the non-adaptive $\delta^{+}$-SPH results by
\cite{sun_consistent_2019} who simulated the problem with $Re = 1000$ at a
resolution $L/\Delta x = 400$. As can be seen in \cref{fig:tg:vol:consv:1000},
the errors in the present work for $L/\Delta x = 150$ are almost five times
smaller than the case with $L/\Delta x = 400$ of
\cite{sun_consistent_2019}. However, the adaptive cases have a larger errors
which drop as the resolution increases.

\begin{figure}[!ht]
  \centering
  \begin{subfigure}{0.48\textwidth}
  \centering
  \includegraphics[width=\textwidth]{%
    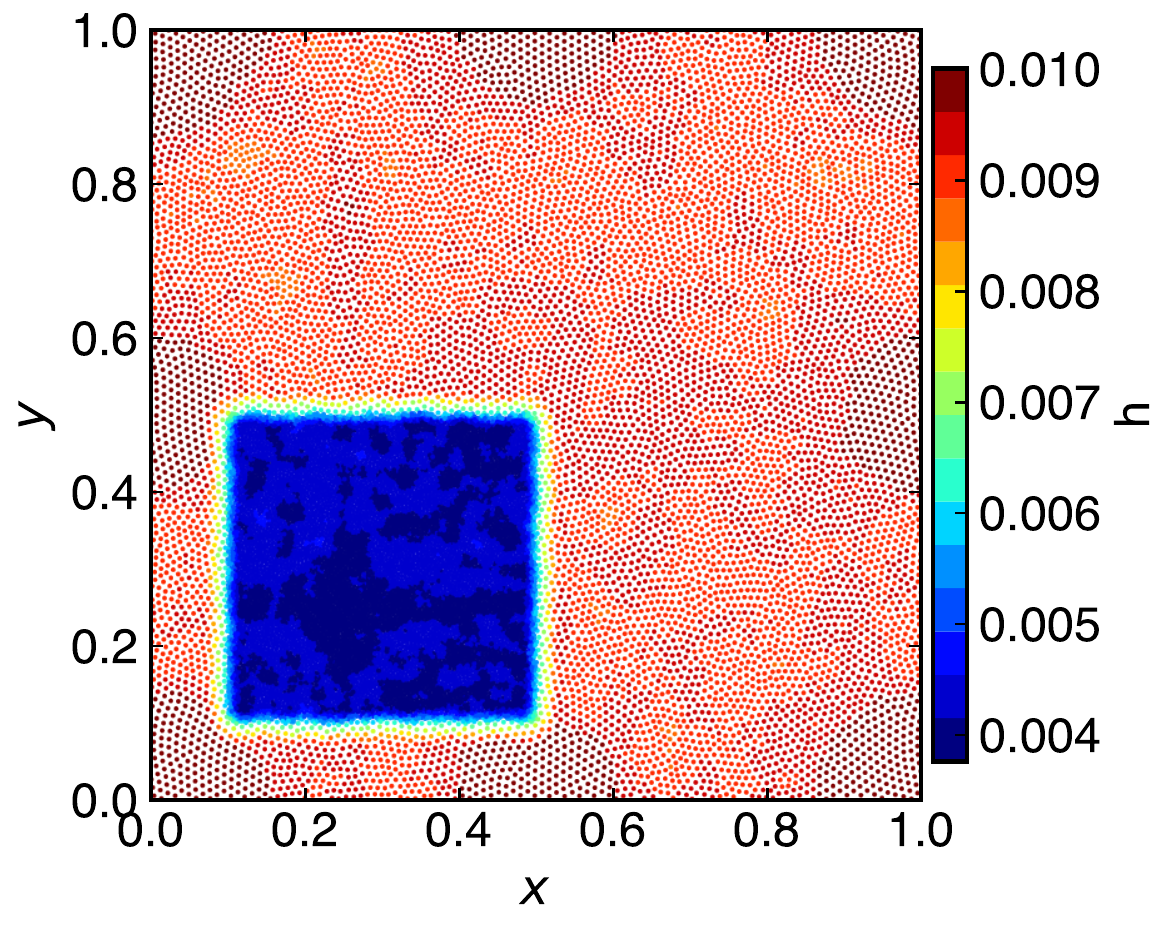%
  }%
  \subcaption{Smoothing length, $h$.}%
  \label{fig:tg:pplot:h}
  \end{subfigure}
  \begin{subfigure}{0.48\textwidth}
  \centering
  \includegraphics[width=0.9\textwidth]{%
    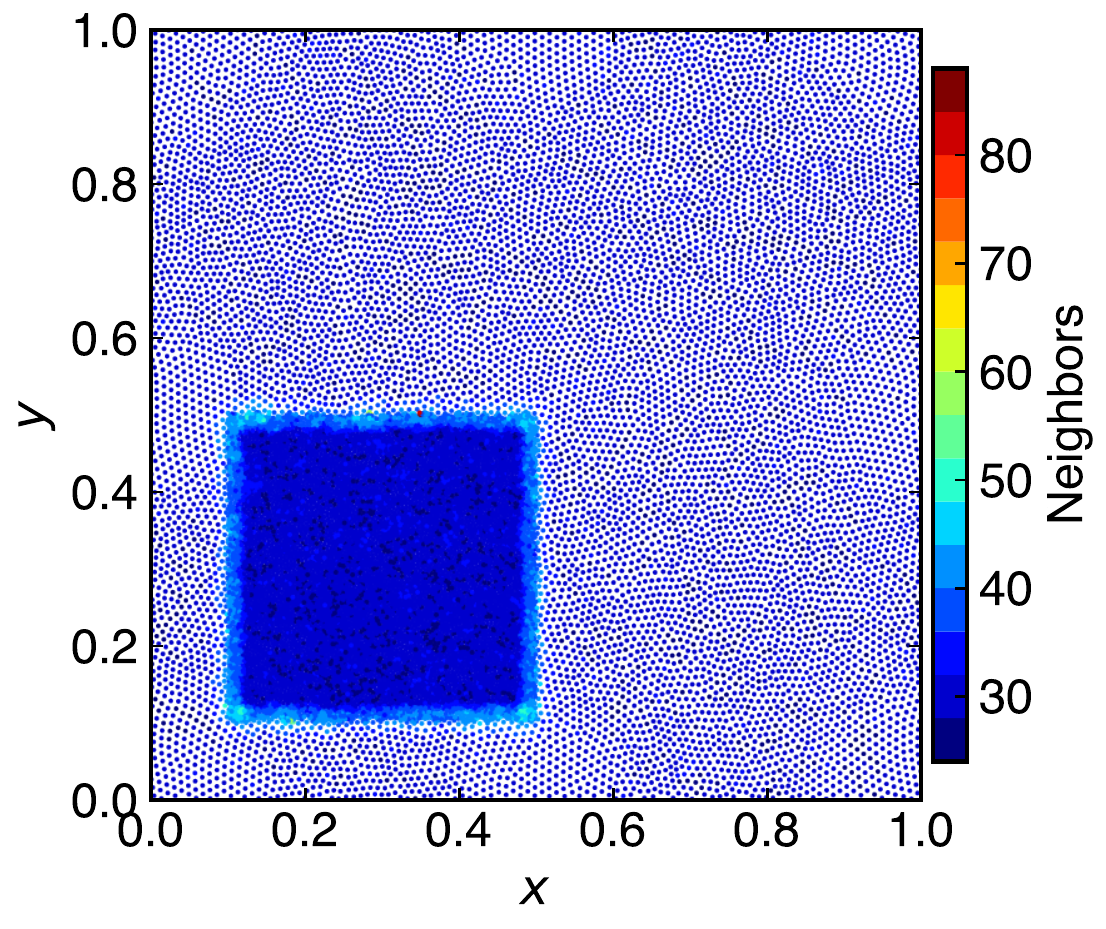%
  }%
  \subcaption{No.\,of neighbors.}%
  \label{fig:tg:pplot:nbrs}
  \end{subfigure}
  \caption{ Particle plots for the Taylor-Green vortex problem shown
    in~\cref{fig:tg:pplot:re1000}. In (a) the distribution of the smoothing
    length $h$ is shown, and in (b) the number of neighbors of each particle are
    shown.}%
  \label{fig:tg:pplot:h:nbrs}
\end{figure}

\Cref{fig:tg:pplot:h} shows the spatial distribution of the smoothing length
$h$. As can be seen the smoothing length is almost constant in the interior of
the respective regions. At the interface between the two regions having
different mass the value is changing gradually. \Cref{fig:tg:pplot:nbrs} show
the distribution of number of neighbors of each particle. It can be seen that in
the interior of the regions the value is around $30$. Whereas, in the interface
between the two regions it is larger as would be expected. Since bulk of the
particles have minimum number of neighbors the method is efficient.

\begin{figure}[!ht]
  \centering
  \includegraphics[width=1.0\textwidth]{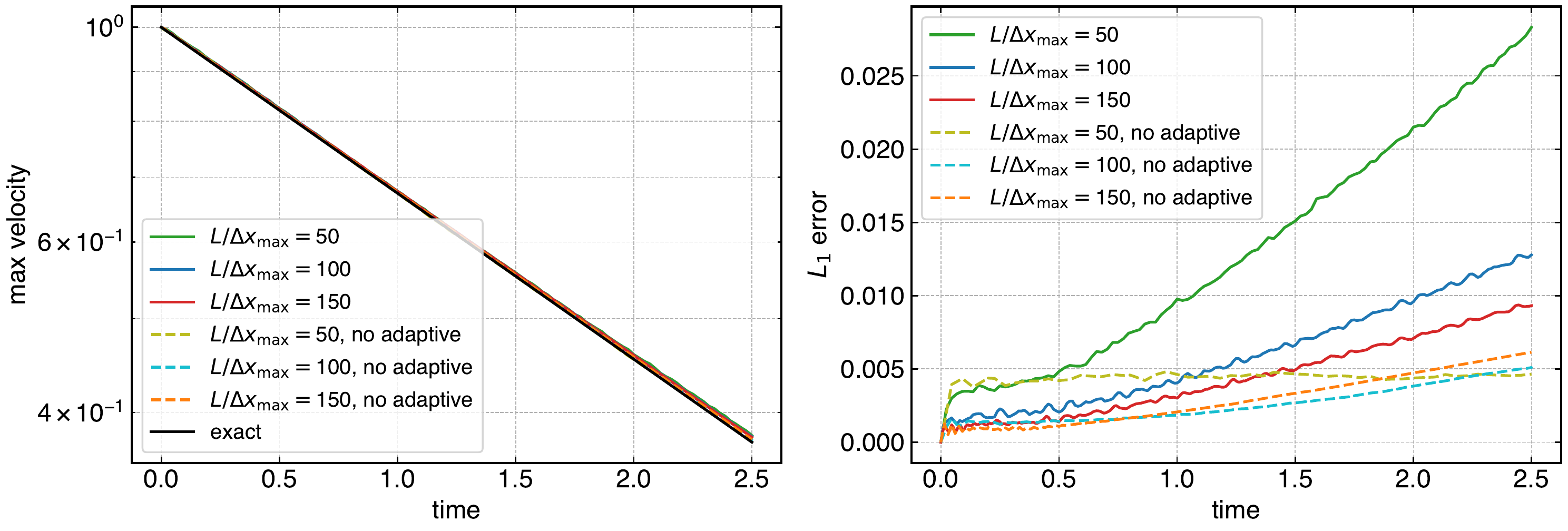}
  \caption{Decay of the maximum velocity (left) and $L_1$ error in the velocity
    (right) for the Taylor-Green vortex problem at $Re = 200$.}%
  \label{fig:tg:re:200}
\end{figure}

\begin{figure}[!ht]
  \centering
  \includegraphics[width=1.0\linewidth]{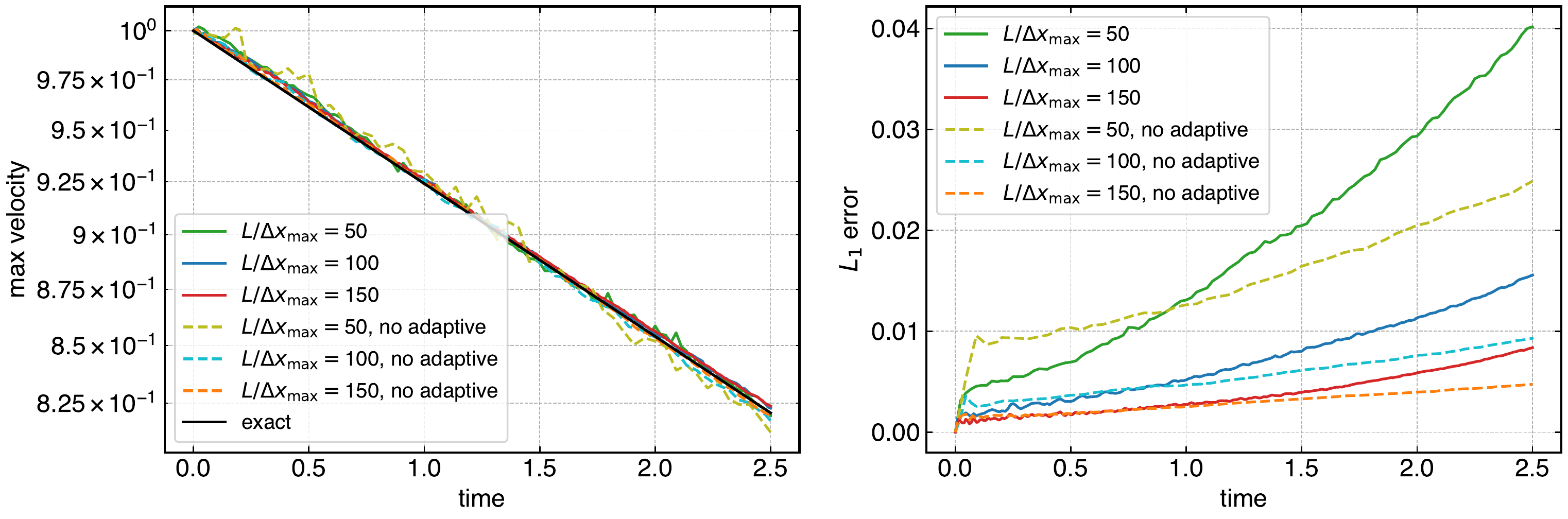}
  \caption{Decay of the maximum velocity (left) and $L_1$ error in the velocity
    (right) for the Taylor-Green vortex problem at $Re = 1000$.}%
  \label{fig:tg:re:1k}
\end{figure}

In~\cref{fig:tg:re:200} and~\cref{fig:tg:re:1k} we plot the maximum velocity
decay and the $L_1$ error in the velocity for $Re = 200$, and $Re = 1000$
respectively with different minimum resolutions. The maximum velocity decay
shows good agreement with the exact solution. We also compare with the
non-adaptive case at different resolutions. Although we do not expect greater
accuracy than the non-adaptive case due to the presence of lower resolution
regions, we expect the errors to be of the same order as that of the
non-adaptive case. The $L_1$ norms reveal that the errors in the adaptive case
are almost 2 times the errors in the non-adaptive case for $Re=200$, whereas for
the $Re=1000$ case, the errors are lower. The increase in the $L_1$ error for
the adaptive case at $Re = 200$ is due to the effect of having an approximately
constant number of neighbors in the adaptive region. Normally, the
discretization error due to the viscous operator reduces when the number of
neighbors is increased. In addition, this error is significant at low Reynolds
numbers and hence causes an increase in the $L_1$ error. This is consistent with
the findings of \cite{negi_convergence_2021} (figure 11 and table 5), and
\cite{negi2021a} (figure 15). A theoretical proof of this is available in the
work of \cite{fatehi2011}. At $Re = 200$ the viscous error dominates, however,
at $Re = 1000$ the viscous effect is not dominant and the $L_1$ error behaves as
expected; the adaptive method has a smaller $L_1$ error than the non-adaptive
case.

\begin{figure}[!ht]
  \centering
  \begin{subfigure}{0.48\textwidth}
    \centering
    \includegraphics[width=\linewidth]{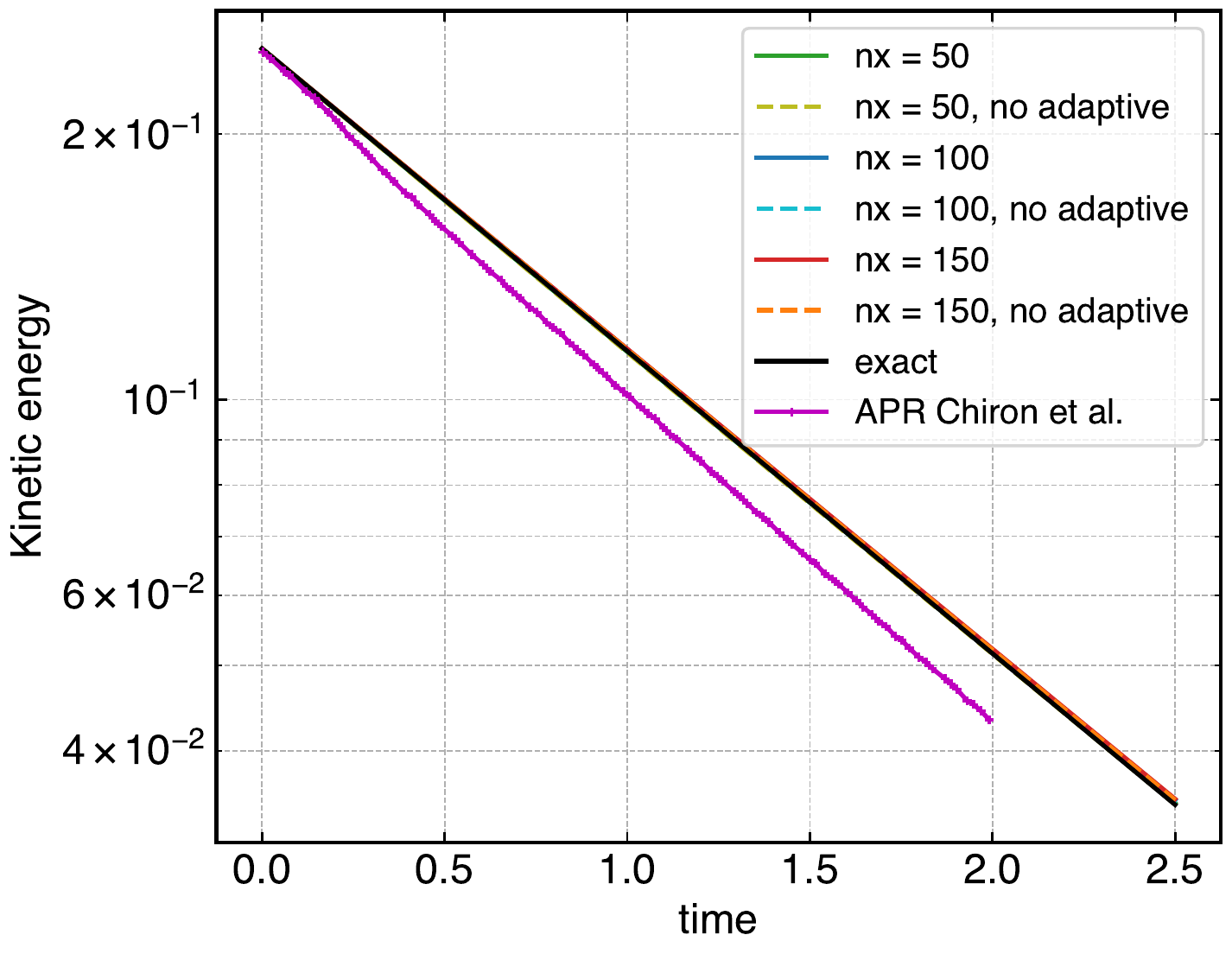}%
    \label{fig:tg:ke:200}
  \end{subfigure}
  \begin{subfigure}{0.48\textwidth}
    \centering
    \includegraphics[width=\linewidth]{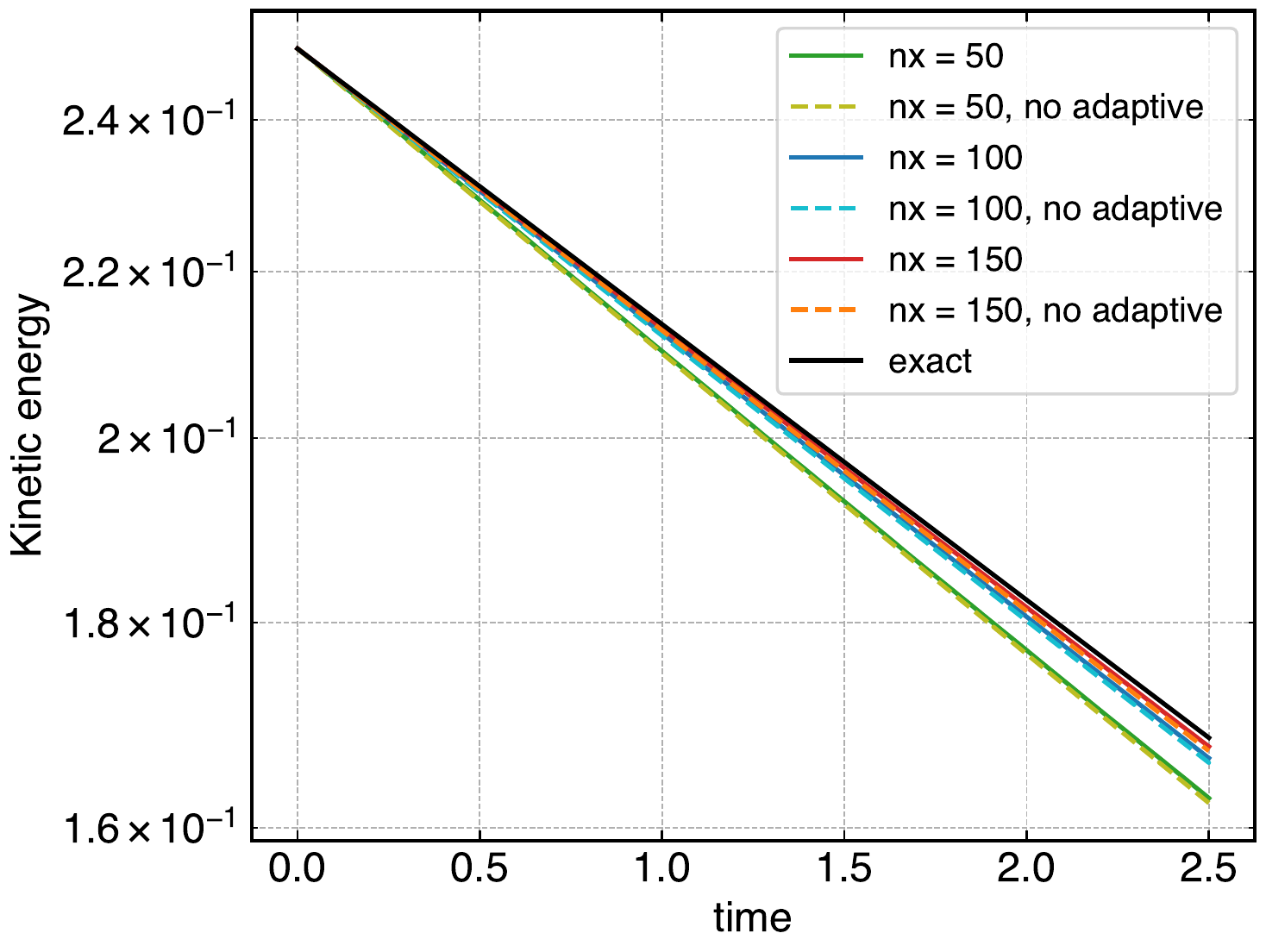}%
    \label{fig:tg:ke:1000}
  \end{subfigure}
  \caption{Kinetic energy decay of the Taylor-Green vortex problem at $Re = 200$
    (left) and $Re = 1000$ (right). We use the exact, non-adaptive, and APR
    scheme (only for $Re = 200$)~\cite{chiron_apr_2018} for comparison.}%
  \label{fig:tg:ke}
\end{figure}

\begin{figure}[!ht]
  \centering
  \begin{subfigure}{0.48\textwidth}
    \centering
    \includegraphics[width=\linewidth]{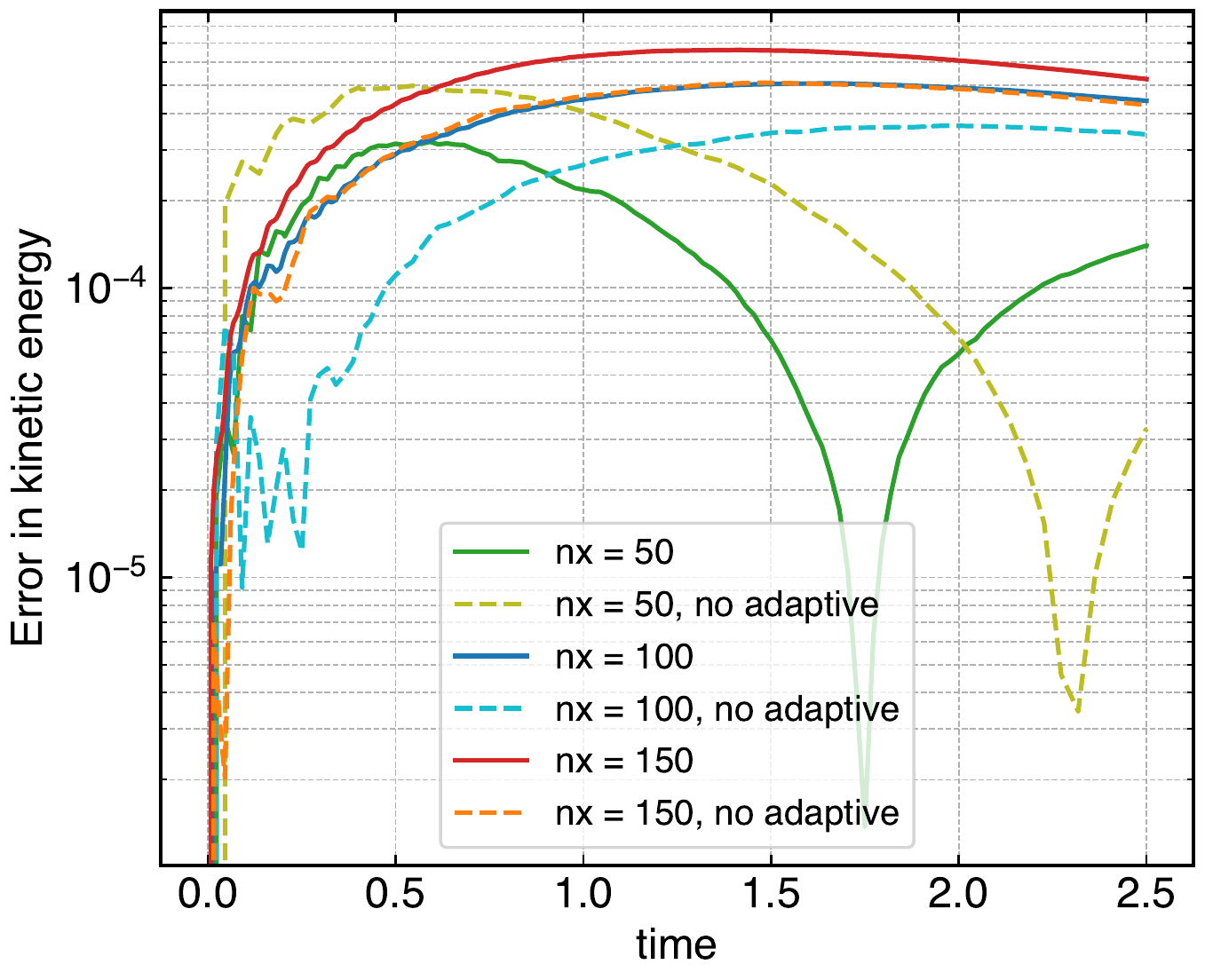}%
    \label{fig:tg:ke:error:200}
  \end{subfigure}
  \begin{subfigure}{0.48\textwidth}
    \centering
    \includegraphics[width=\linewidth]{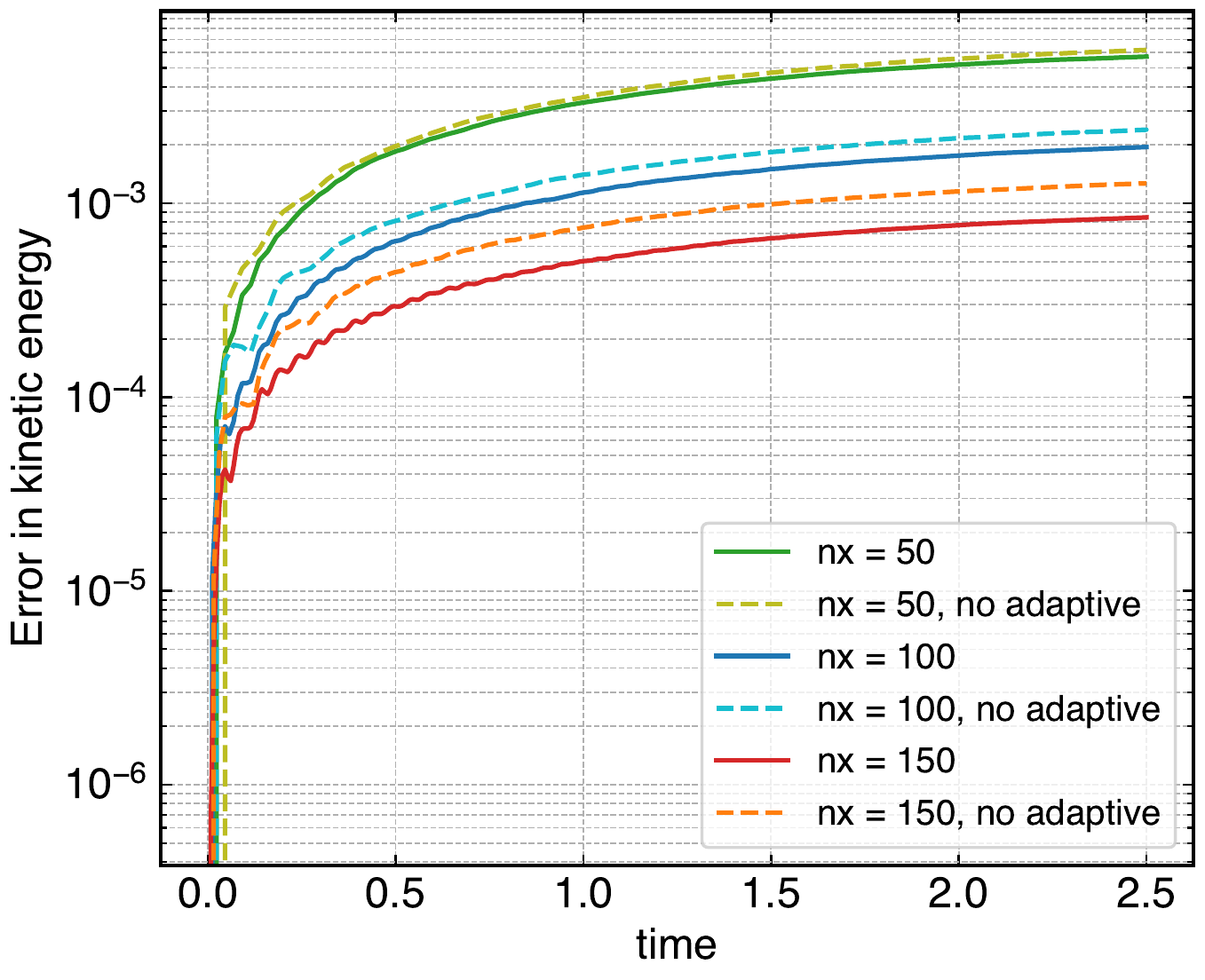}%
    \label{fig:tg:ke:error:1000}
  \end{subfigure}
  \caption{Error in kinetic energy decay of the Taylor-Green vortex problem at
    $Re = 200$ (left) and $Re = 1000$ (right).}%
  \label{fig:tg:ke:error}
\end{figure}

In~\cref{fig:tg:ke} we show the kinetic energy decay for $Re = 200$ and
$Re = 1000$ at different minimum resolutions. We compare with the exact,
non-adaptive, and the APR simulation of~\cite{chiron_apr_2018}. Our results
match the non-adaptive decay and barring a slight increase the exact decay,
whereas the APR scheme shows comparatively large decay. \Cref{fig:tg:ke:error}
shows the error in the kinetic energy versus time. For the $Re = 200$ case the
results show an anomalous behavior as seen from the $L_1$ error. The $Re = 1000$
cases behave as expected.

The above results show that the proposed method is accurate, displays less
dissipation than other recent techniques proposed for adaptive resolution, and
requires minimum number of neighbors for bulk of the particles. This makes the
proposed method both accurate and efficient.

\FloatBarrier%

\subsection{Gresho-Chan vortex}%
\label{sec:gc}

Gresho-Chan vortex~\cite{gresho1990} is a two-dimensional inviscid numerical
test case with periodic boundary conditions in both the $x$ and $y$
directions. This problem tests, to name a few, the numerical stability of the
method, and the conservation properties. Considered as a difficult test
case~\cite{rosswog2015a}, this test case is widely used by the astrophysical
community~\cite{liska2003,springel2010a,hopkins2015}. The problem is of a
rotating vortex inside a domain of unit length where the centrifugal force due
to azimuthal velocity balances the pressure gradient. The distances are
non-dimensionalized by the domain length $L$. The initial radial velocity is
zero, and the azimuthal velocity in non-dimensional form is given by,
\begin{equation}
  \label{eq:gc-az}
  u_{\varphi}(r) =
  \begin{cases}
    r/R \quad &\mathrm{for}\ 0 \le r < R,\\
    2 - r/R \quad &\mathrm{for}\ R \le r < 2R, \\
    0 \quad &\mathrm{for}\ r \ge 2R,
  \end{cases}
\end{equation}
where $R = 0.2L$, and $r$ is the distance from the centre of the vortex located
at the origin $(0, 0)$. The artificial speed of sound is $c_s = 10$ m/s, and the
smoothing length factor, $h/\Delta x = 1.0$. The non-dimensional pressure,
balanced by the centrifugal velocity, is given by,
\begin{equation}
  \label{eq:gc-p}
  p(r) = p_0 +
  \begin{cases}
    \frac{25}{2} r^2 \quad &\mathrm{for}\ 0 \le r < R, \\
    4 - 4 \log(0.2) + \frac{25}{2} r^2 - 20r + 4 \log(r)%
    \quad &\mathrm{for}\ R \le r < 2R, \\
    4 \log(2) - 4 \quad &\mathrm{for}\ r \ge 2R, \\
  \end{cases}
\end{equation}
where $p_0 = 5$ is the reference pressure.  We adaptively refine a semi-circular
region of radius $0.45$ around the origin with particles of mass around 0.5
times that of the outer particles. We simulate the problem for $t = 3$ s. We
compare our results with the exact solution and the non-adaptive cases. We
consider two different minimum resolutions $L/\Delta x_{\max}$ of 50 and
100. The particles are initially placed on a uniform Cartesian grid.

\begin{figure}[!ht]
  \centering
  \includegraphics[width=\textwidth]{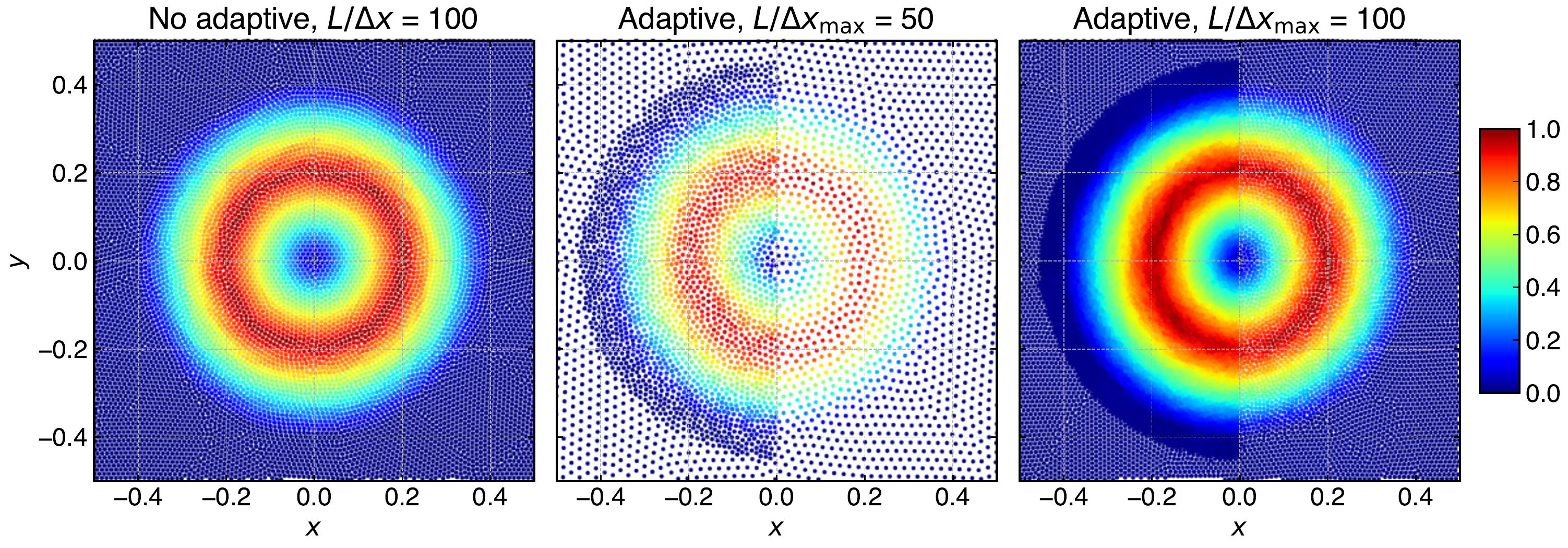}
  \caption{Velocity magnitude distribution for the Gresho-Chan vortex
    at $t=3$ s.}%
  \label{fig:gc-pplots}
\end{figure}

\begin{figure}[!ht]
  \centering
  \includegraphics[width=\textwidth]{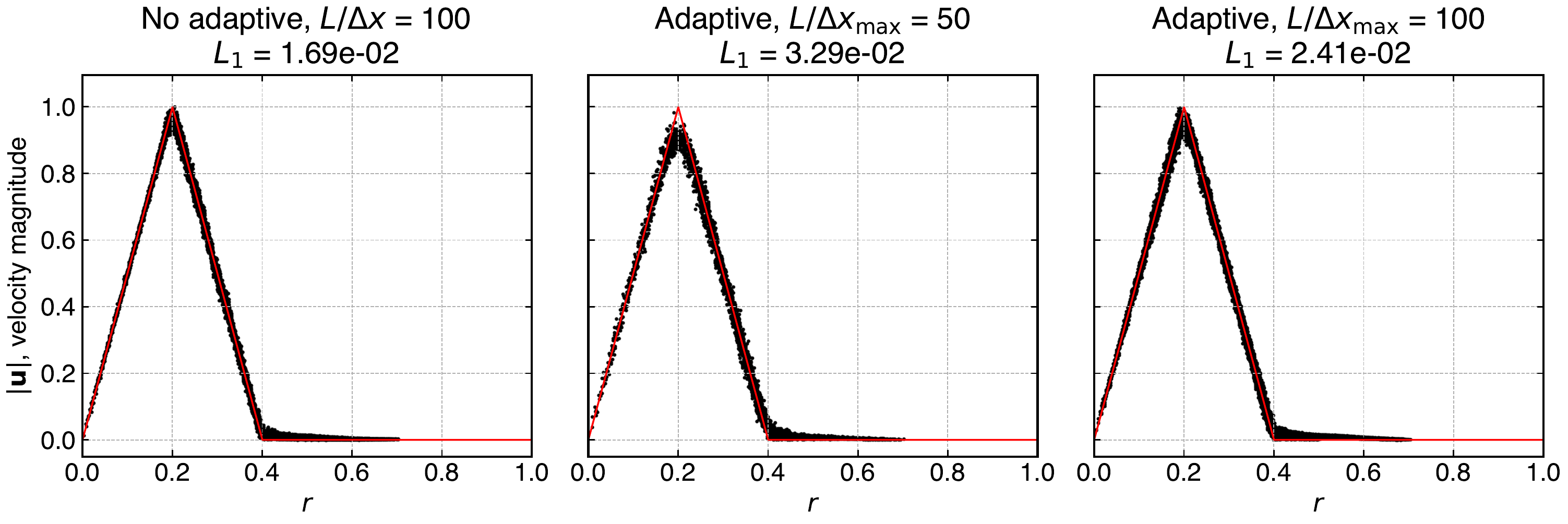}
  \caption{Comparison of the velocity magnitude as a function of $r$, the
    distance from the centre of the vortex, to the exact solution at $t = 3$ s.}%
  \label{fig:gc-vmag}
\end{figure}

\Cref{fig:gc-pplots} shows the particle positions at $t=3$ s.  It is difficult to
assess the difference between the simulations from this result.
\Cref{fig:gc-vmag} shows the magnitude of the velocity of all the particles in
the domain as a function of the distance $r$ from the centre of the vortex. The
red-line indicates the exact velocity magnitude. The figure also
indicates the $L_1$ norm of the error, which is computed as,
\begin{equation}
  \label{eq:gc-l1norm}
  L_1 = \frac{1}{N} \sum_{i=1}^{N}|\ten{u}_i - \ten{u}_{\mathrm{exact}}(\ten{r}_i)|.
\end{equation}
The plot shows decay and noise in the velocity magnitude.  This is an
inviscid problem and our simulation does not employ any artificial viscosity. It
is therefore highly sensitive to small perturbations. The results show that the
particle splitting and merging process introduce a small amount of noise in the
simulation. However, the results show that this is only slightly
dissipative. One can also see that these results are as good if not better than
the existing results~\cite{rosswog2015a,hopkins2015}.  Further,
\citet{hopkins2015} mentions that splitting and merging can be noisy and
diffusive. However, the present results show that careful splitting and merging
of particles as we have done produces acceptable results.

\begin{figure}[!ht]
  \centering
  \begin{subfigure}{0.48\textwidth}
    \centering
    \includegraphics[width=\textwidth]{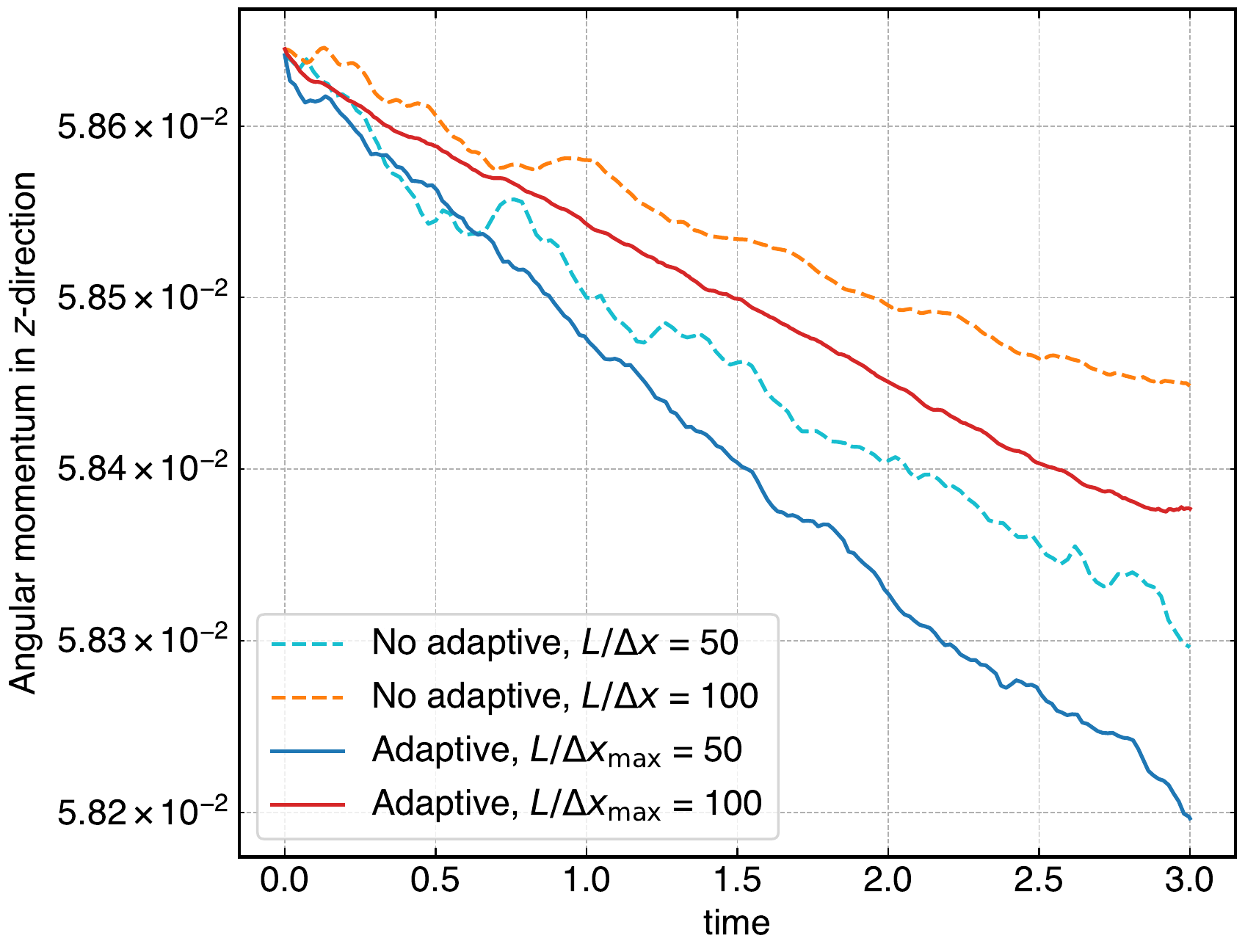}
    \subcaption{}%
    \label{fig:gc-conservation}
  \end{subfigure}
  \begin{subfigure}{0.48\textwidth}
    \centering
    \includegraphics[width=0.9\textwidth]{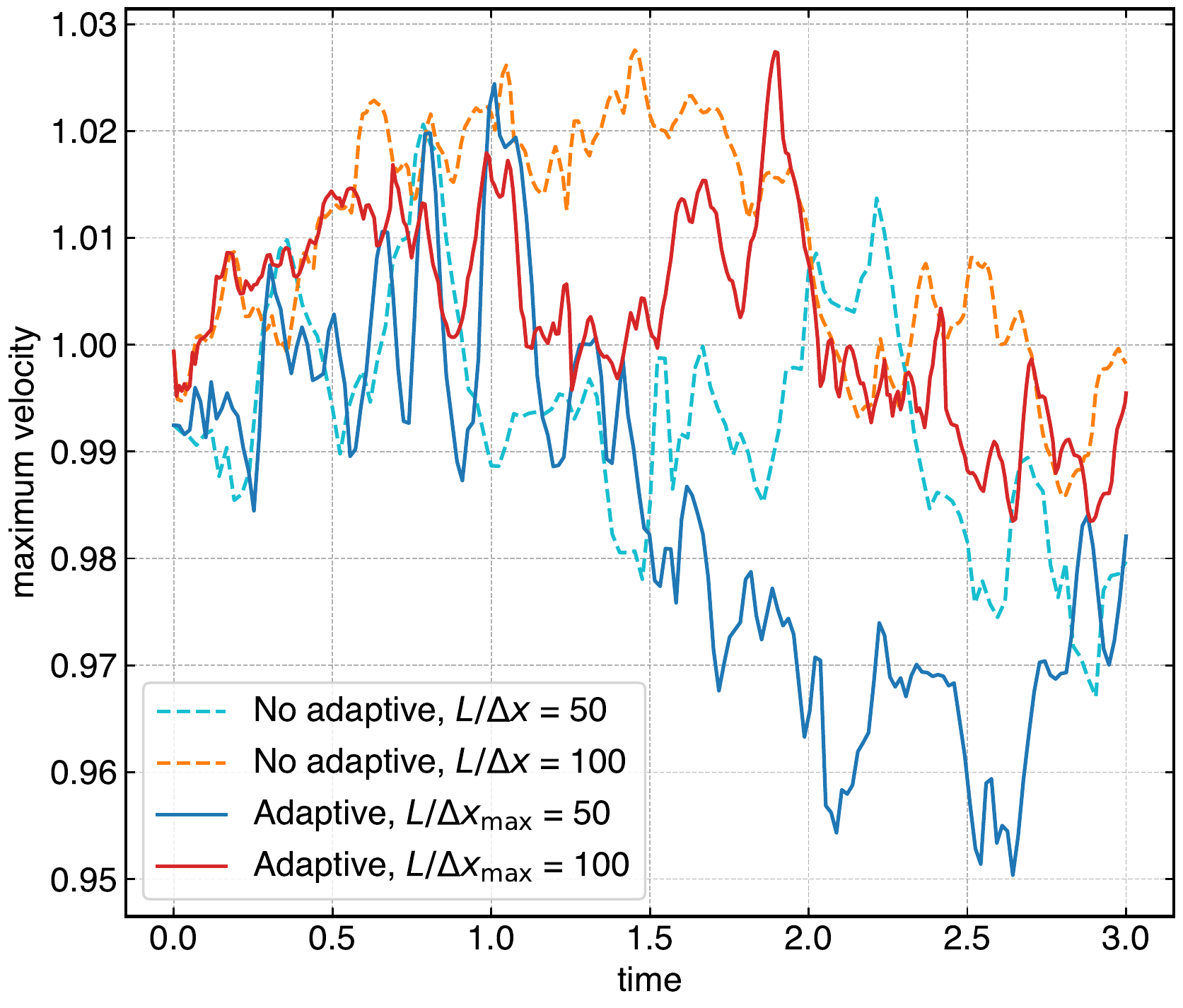}
    \subcaption{}%
    \label{fig:gc-decay}
  \end{subfigure}
  \caption{Evolution of the angular momentum (left), and the evolution of the
    maximum velocity (right) of the Gresho-Chan vortex problem.}%
  \label{fig:gc-conservation-decay}
\end{figure}

\begin{figure}[!ht]
  \centering
  \begin{subfigure}{0.48\textwidth}
    \centering
    \includegraphics[width=0.95\textwidth]{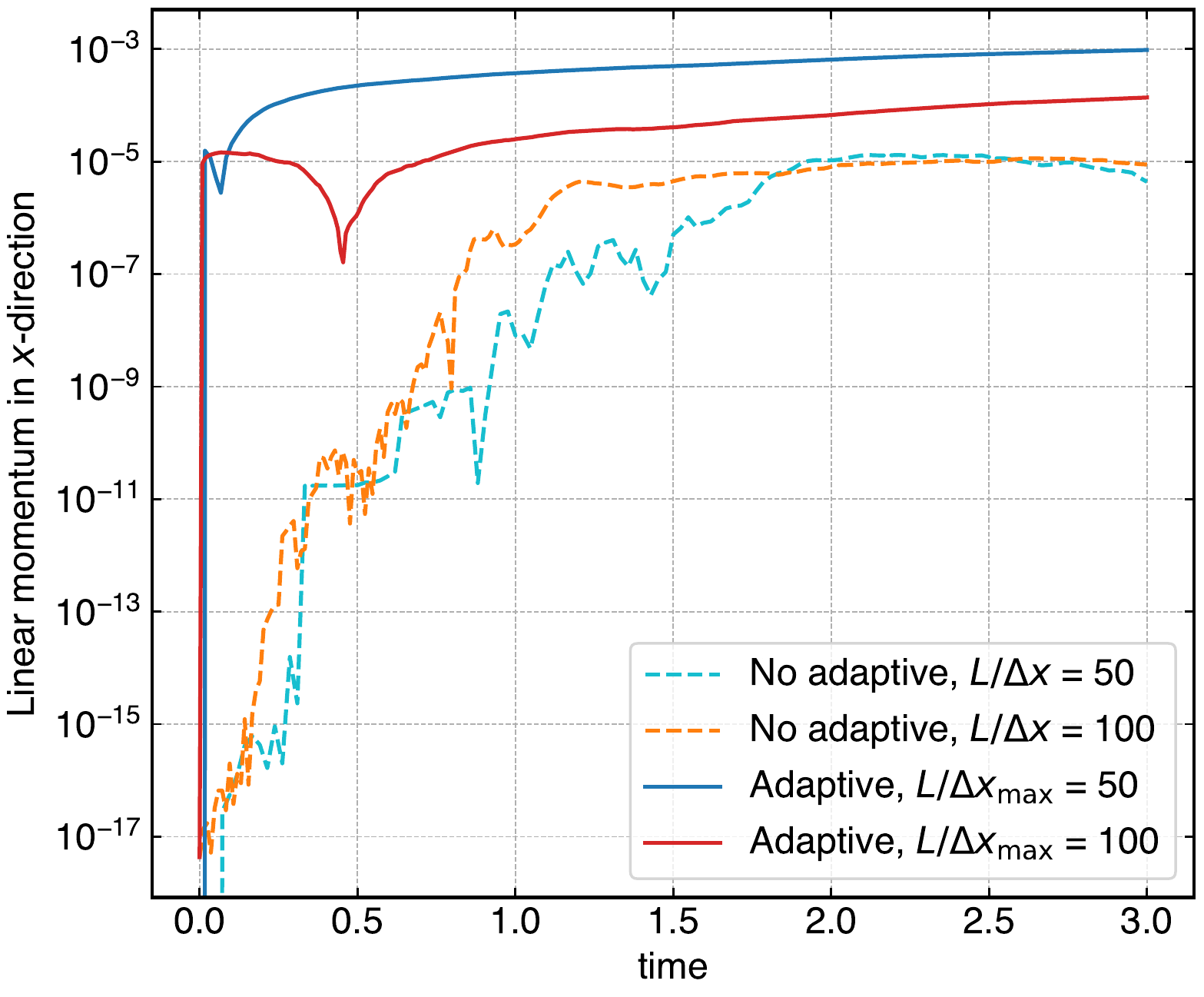}
    \subcaption{}%
    \label{fig:gc-lin-mom-x}
  \end{subfigure}
  \begin{subfigure}{0.48\textwidth}
    \centering
    \includegraphics[width=\textwidth]{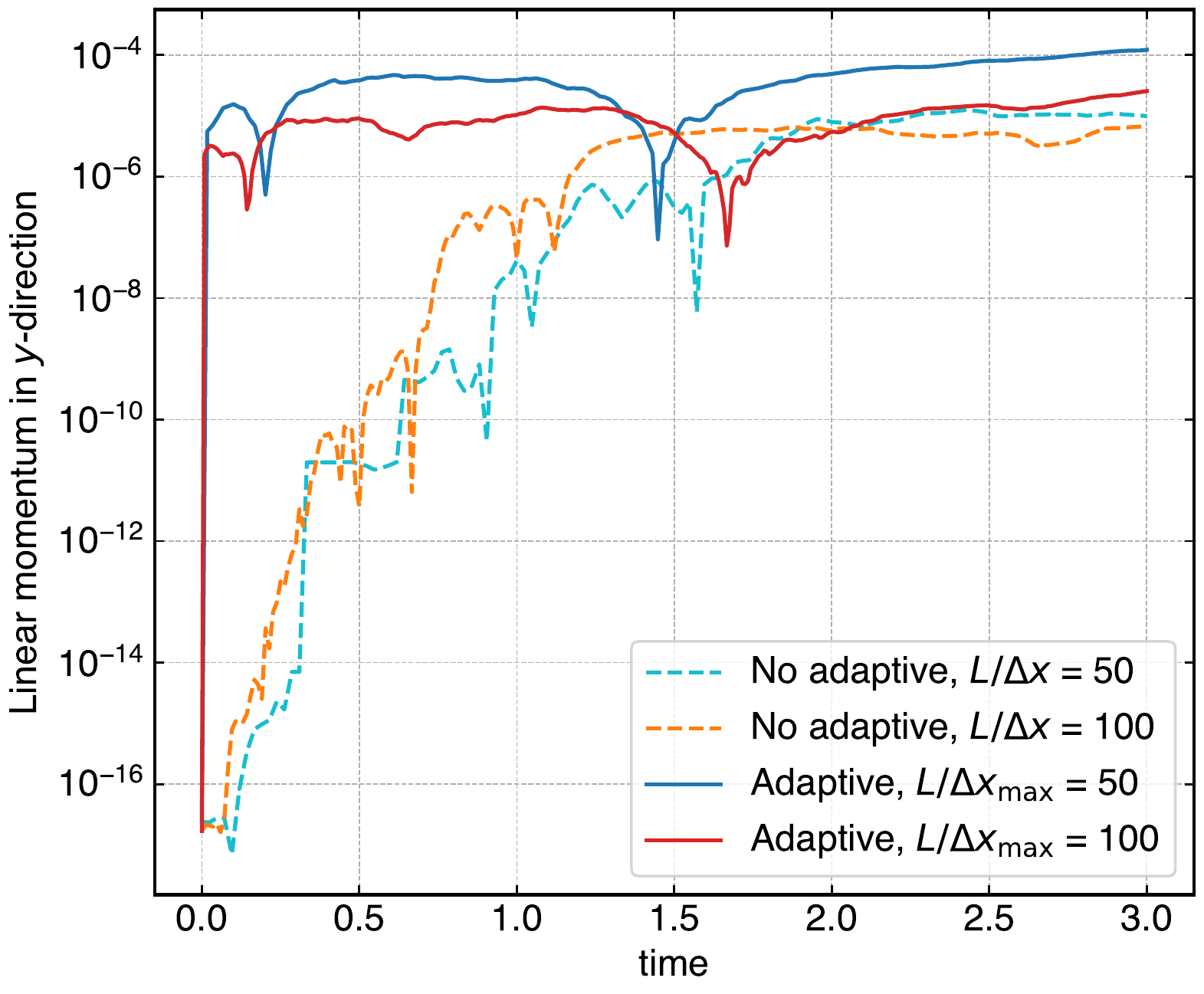}
    \subcaption{}%
    \label{fig:gc-lin-mom-y}
  \end{subfigure}
  \caption{Evolution of linear momentum in $x$ (left) and $y$ (right)
    directions of the Gresho-Chan vortex problem.}%
  \label{fig:gc-conservation-lin}
\end{figure}

\Cref{fig:gc-conservation} shows the angular momentum of the system as a
function of time. For the non-adaptive cases we can clearly see a small amount
of dissipation which reduces as we increase the resolution. A similar trend
follows for the adaptive cases. \Cref{fig:gc-conservation-lin} shows the
evolution of linear momentum in $x$ and $y$ directions where higher resolution
has better conservation. But due to the reasons mentioned in the remark of
\cref{sec:sph} the conservation of momentum does not hold. We note that the mass
is exactly conserved in all our cases. The adaptive $L/\Delta x_{\max} = 50$
case is more dissipative compared to the $L/\Delta x_{\max} = 100$ case. To
further study this, the maximum velocity evolution over time is shown
in~\cref{fig:gc-decay}. In this figure we see that for the adaptive
$L/\Delta x_{\max} = 100$ case, the maximum velocity does not decay
significantly. These results affirm the accuracy of our adaptive algorithm.

\FloatBarrier%

\subsection{Two dimensional lid-driven cavity}%
\label{sec:ldc}

\begin{figure}[!ht]
  \centering
  \includegraphics[width=0.48\textwidth]{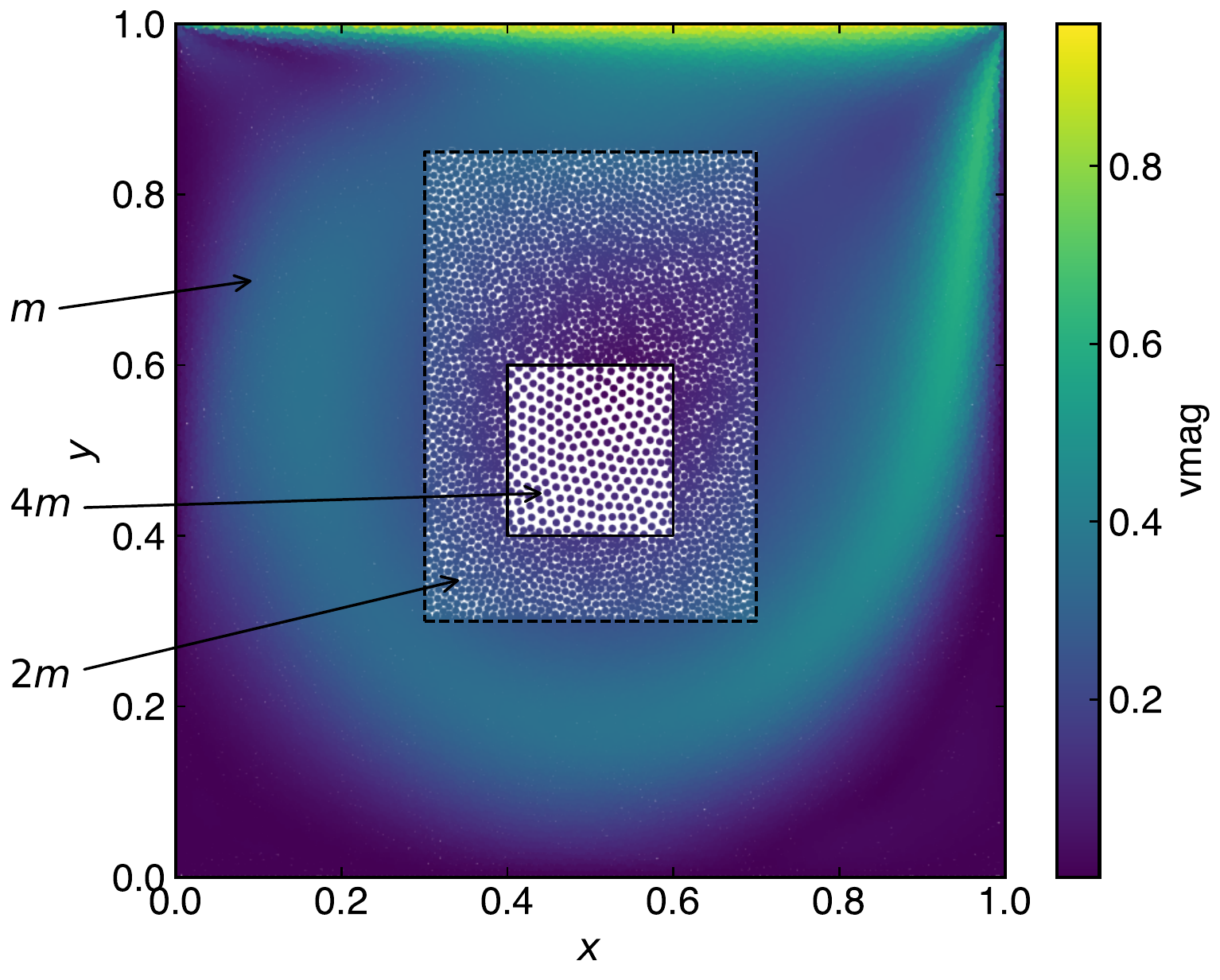}
  \includegraphics[width=0.48\textwidth]{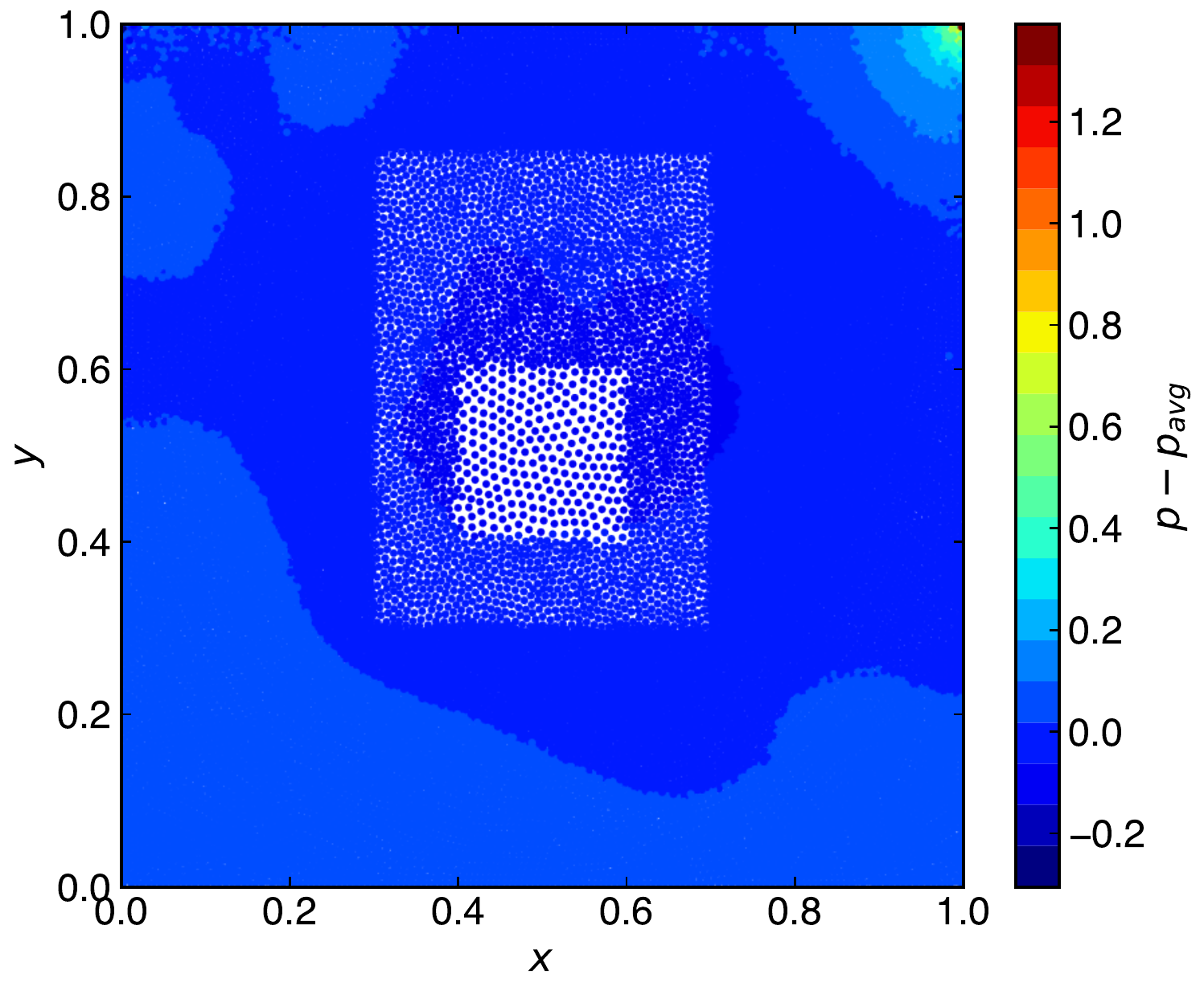}
  \caption{Particle plot with color indicating the velocity magnitude (left) and
    pressure $p - p_{avg}$ (right) for the lid-driven cavity problem simulated
    with $L/\Delta x_{\min} = 150$ at $Re = 1000$. The regions of discretization
    are also indicated.}%
  \label{fig:ldc:pplot}
\end{figure}

Lid-driven cavity is a two-dimensional viscous problem with solid boundaries. We
study this problem with two different Reynolds numbers $100$ and $1000$. We
compare our results to those of~\citet{ldc:ghia-1982}. The domain length $L$ is
1 m, and the top wall is moving with a velocity $U_{\text{wall}}$ of $1$ m/s. We
consider a square domain $[0, L] \times [0, L]$ with two refinement levels, the
intermediate refinement region
$[0.3L, 0.7L] \times [0.3L, 0.85L] - [0.4L, 0.6L] \times [0.4L, 0.6L]$ consists
of particles with mass twice that of the outer most region, and the inner most
refinement region $[0.4L, 0.6L] \times [0.4L, 0.6L]$ consists of particles with
mass four times that of the outer most and two times that of the intermediate
region.  We simulate with three different maximum resolutions, where
$L/\Delta x_{\min}$ is 50, 100 and 150. The adaptively refined regions are shown
in~\cref{fig:ldc:pplot}. The artificial speed of sound is $c_s = 10$ m/s, the
smoothing length factor, $h/\Delta x = 1.0$, and the reference density $\rho_0$
is 1 kg/m\textsuperscript{3}. We non-dimensionalize the velocity by the wall
velocity, $u = u^+/U_{\text{wall}}$ and $v = v^+/U_{\text{wall}}$, the pressure
as $p = p^+/\rho_{0} U^2_{\text{wall}}$, and the lengths by the domain length
$L$.

\begin{figure}[!ht]
  \centering
  \includegraphics[width=\textwidth]{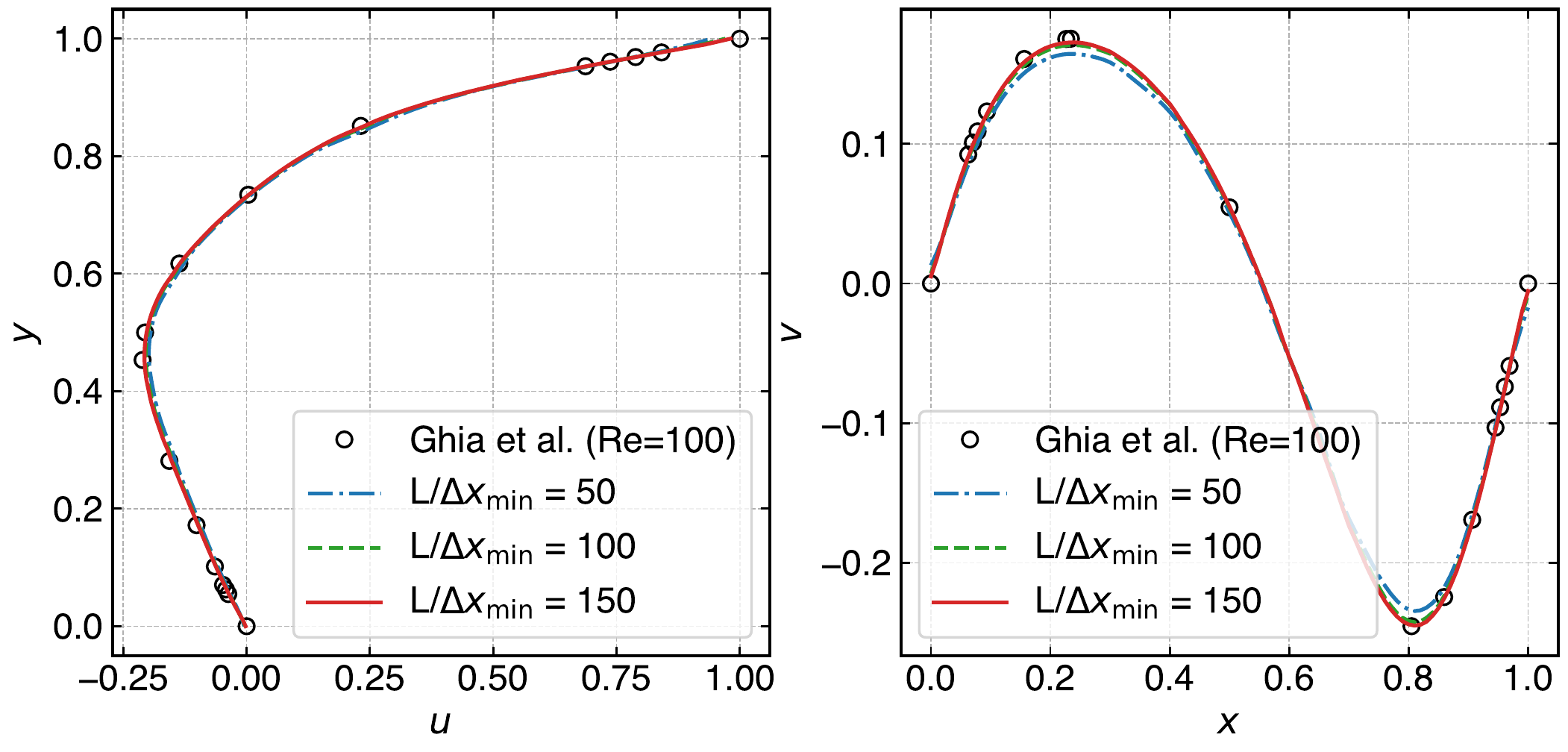}
  \caption{The horizontal (left) and the vertical (right) centerline velocity
    profiles for the lid-driven cavity at $Re = 100$ are compared with the
    results of~\citet{ldc:ghia-1982}.}%
  \label{fig:ldc:uv-re100}
\end{figure}

\begin{figure}[!htpb]
  \centering
  \includegraphics[width=\textwidth]{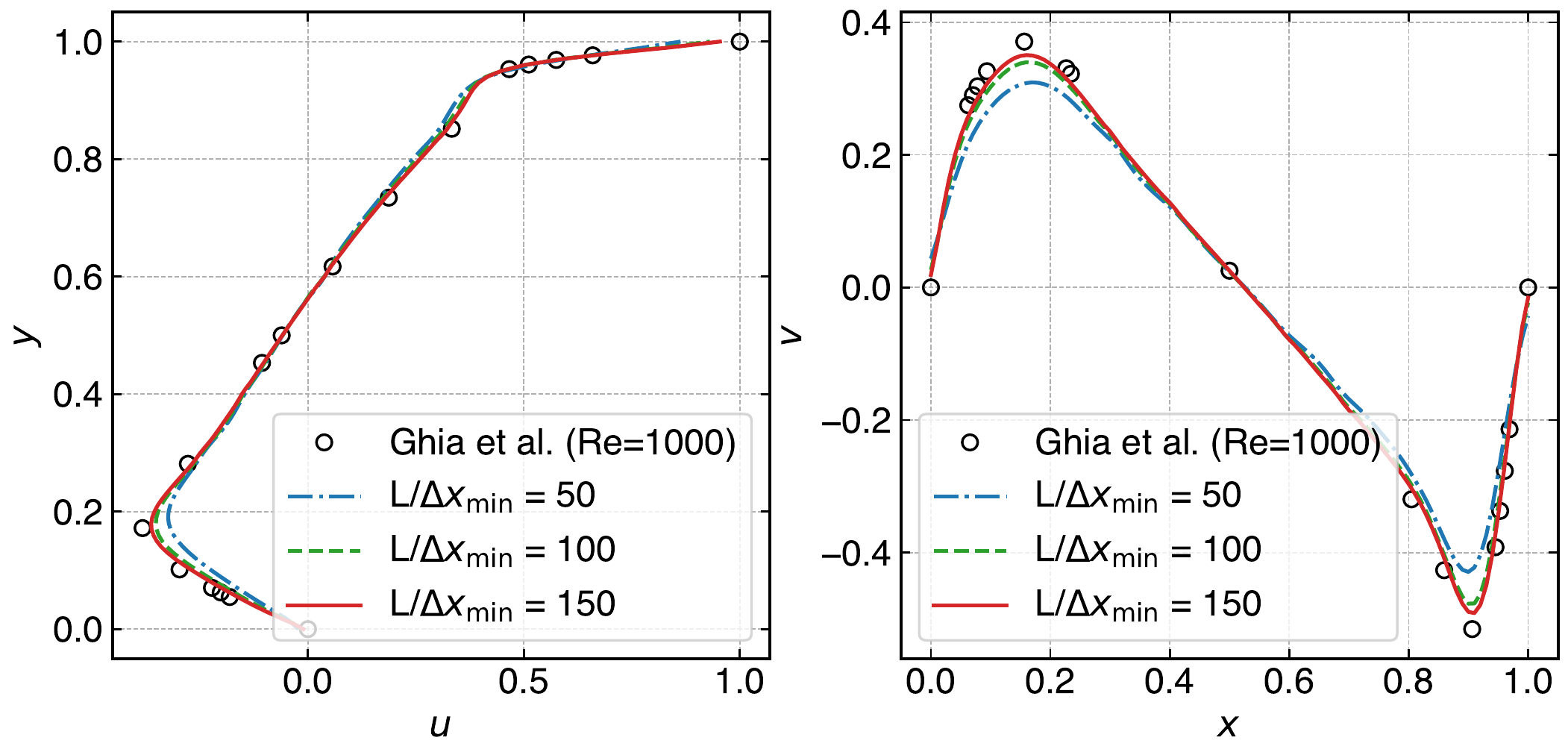}
  \caption{The horizontal (left) and the vertical (right) centerline velocity
    profiles for the lid-driven cavity at $Re = 1000$ are compared with the
    results of~\citet{ldc:ghia-1982}. }%
  \label{fig:ldc:uv-re1000}
\end{figure}

\Cref{fig:ldc:pplot} shows the velocity magnitude distribution and pressure
distribution for $Re = 1000$.  We use 3 layers to simulate this problem. The
outer layer of particles are at the highest resolution with the particle mass
corresponding to a resolution of $L/\Delta x_{\min} = 150$. The middle region is
at twice the mass of the outer region, this corresponds to a resolution of
$L/\Delta x = 100$. The inner most resolution is the coarsest of all with an
effective resolution of $L/\Delta x_{\max} = 75$. \Cref{fig:ldc:uv-re100} shows
the centerline velocity profiles at $Re = 100$. The results match well with the
results of~\cite{ldc:ghia-1982}. In~\cref{fig:ldc:uv-re1000} we show the
centerline velocity profiles for $Re = 1000$. It can be observed that as the
resolution is increased the centerline profiles show a good agreement
with~\cite{ldc:ghia-1982}.

\FloatBarrier%

\subsection{Flow past a circular cylinder}%
\label{sec:fpc}

\begin{figure}[ht]
  \centering
  \includegraphics[width=0.6\textwidth]{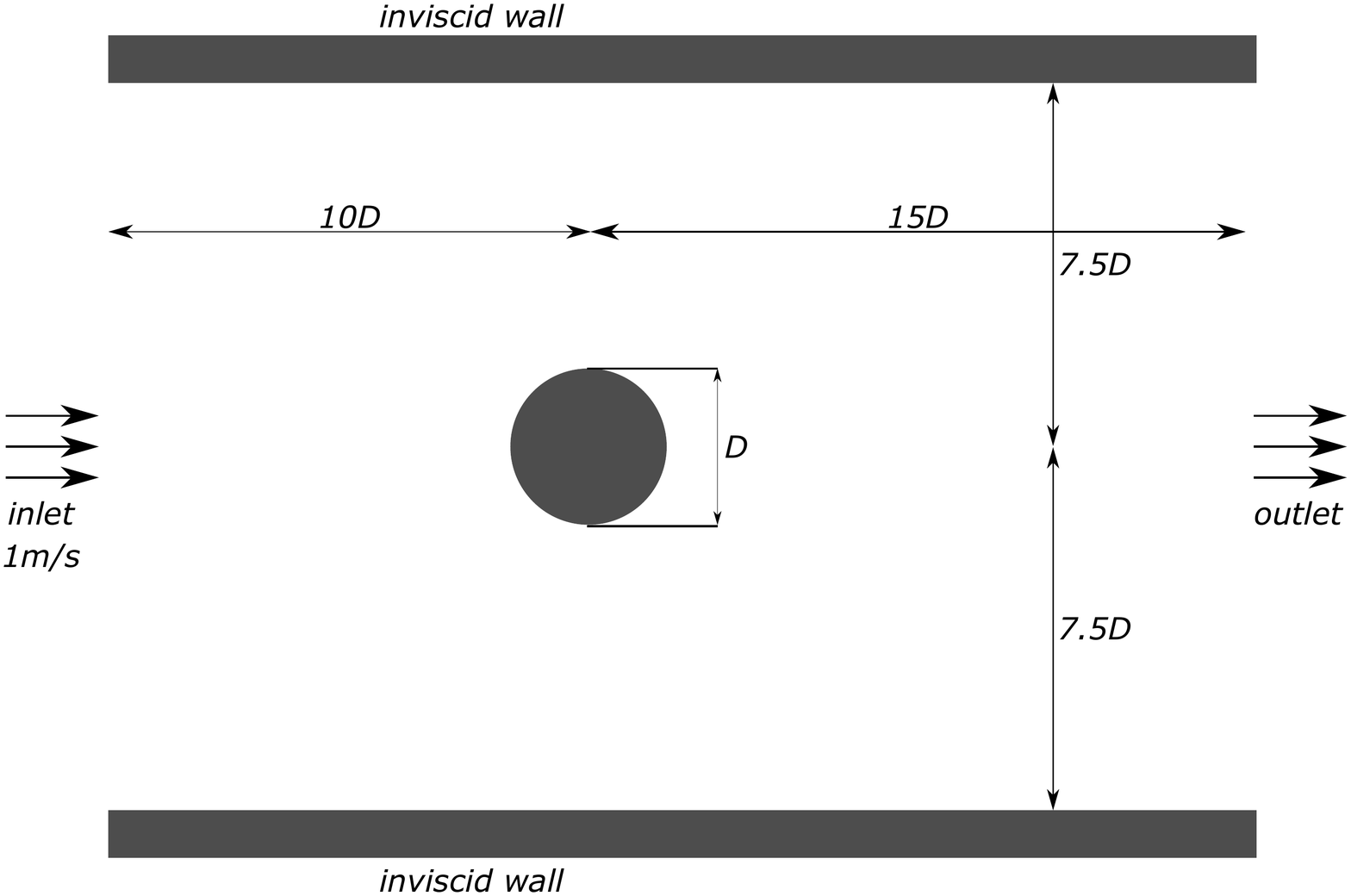}
  \caption{The domain dimensions for the flow past a circular cylinder
    problem.}%
  \label{fpc:setup}
\end{figure}

We study the flow past a circular cylinder problem for five different Reynolds
numbers ranging from 40 to 9500. We plot the coefficients of pressure drag and
skin friction as a function of time. We compare the results with the high
resolution vortex method of \citet{koumoutsakos1995}, and
\citet{ramachandran2004}. \Cref{fpc:setup} shows the domain setup, where the
diameter of the cylinder $D = 2 R = 2$ m. We use a non-dimensional time
$T = tU_{\infty}/ R$. We initialize the flow at $T=0$ with the potential flow
solution. The inlet velocity is 1 m/s, and the solid walls are inviscid.  In
order to minimize the reflection of the initial, undesirable, pressure waves
from the walls we employ the non-reflection boundary conditions
of~\citet{lastiwka2009} on the inviscid walls.

We simulate the problem with fixed refinement zones up to $T = 6$. For all the
simulations, the coarsest resolution in the domain is $D/\Delta x_{\max} = 4$.
We vary the finest resolution $D/\Delta x_{\min}$ from 100 to 500. For the
$Re = 9500$ case we use a finest resolution $D/\Delta x_{\min} = 1000$. Unless
explicitly mentioned we use a $C_r$ value of 1.08 for all the problems. The
parameters used in these simulations are summarized in~\cref{tab:fpc-params}.

\begin{table}[!ht]
  \centering
  \begin{tabular}[!ht]{ll}
    \toprule
    Quantity & Values \\
    \midrule
    $D$, Diameter & 2 m \\
    $\rho_0$, reference density & 1000 kg/m\textsuperscript{3} \\
    $c_s$ & 10 m/s \\
    $D/\Delta x_{\max}$, lowest resolution & 4 \\
    $D/\Delta x_{\min} $, highest resolution & 160, 250, 500\\
    $C_r$ & 1.08 \\
    Reynolds number & 40, 550, 1000, 3000, and 9500 \\
    Time of simulation & 6 \\
    Smoothing length factor, $h/\Delta x$ & 1.2\\
    \bottomrule
  \end{tabular}
  \caption{Parameters used for the flow past a circular cylinder problem.}%
  \label{tab:fpc-params}
\end{table}

\begin{figure}[!ht]
  \centering
  \includegraphics[width=0.6\textwidth]{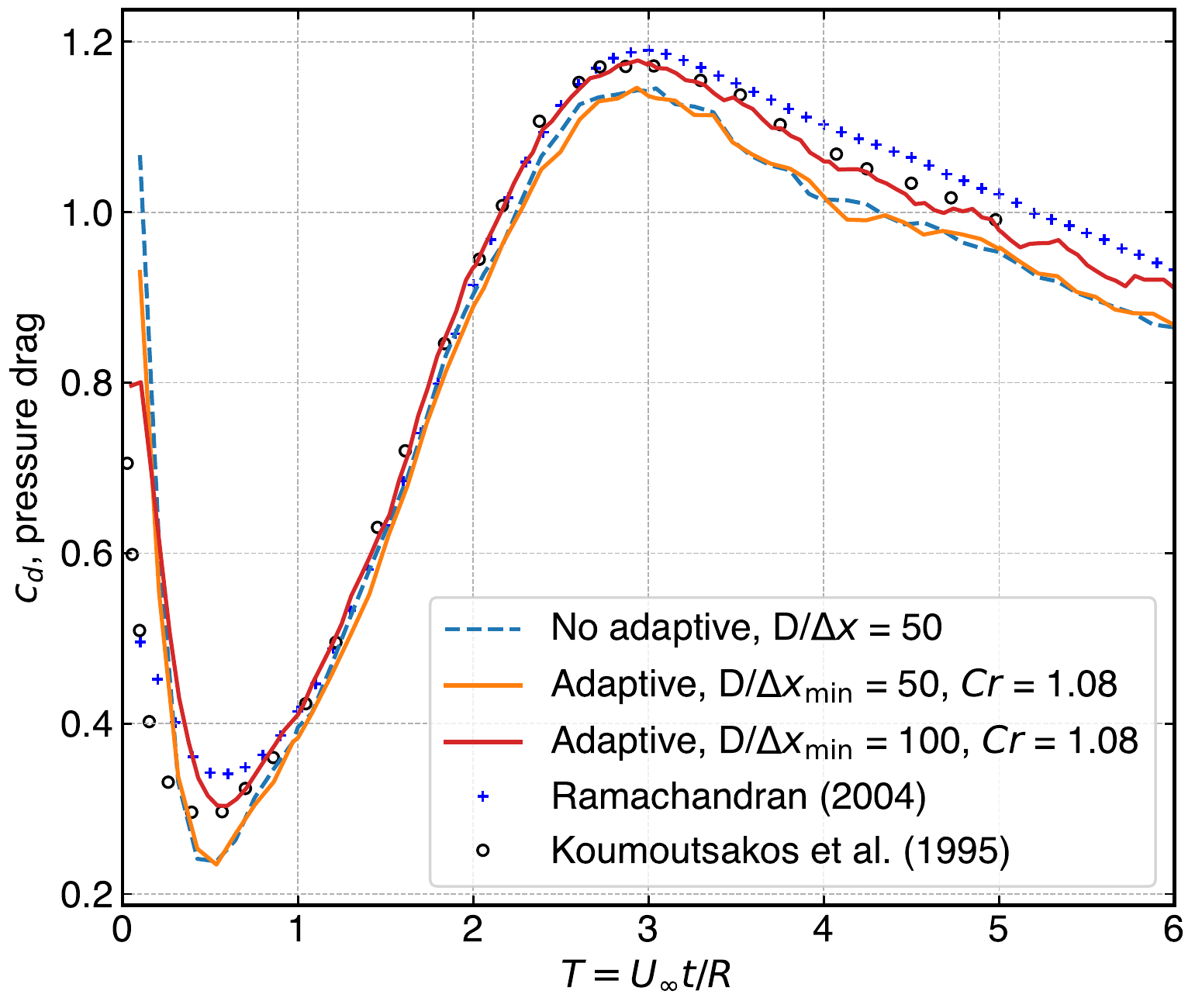}
  \caption{Time history of the coefficient of pressure drag for $Re = 1000$. We
    compare the adaptive cases with two different resolutions to the
    non-adaptive case with a fixed resolution of $D/\Delta x = 50$.}%
  \label{fpc:no-adapt-1000}
\end{figure}

\begin{figure}[!ht]
  \centering
  \includegraphics[width=0.6\textwidth]{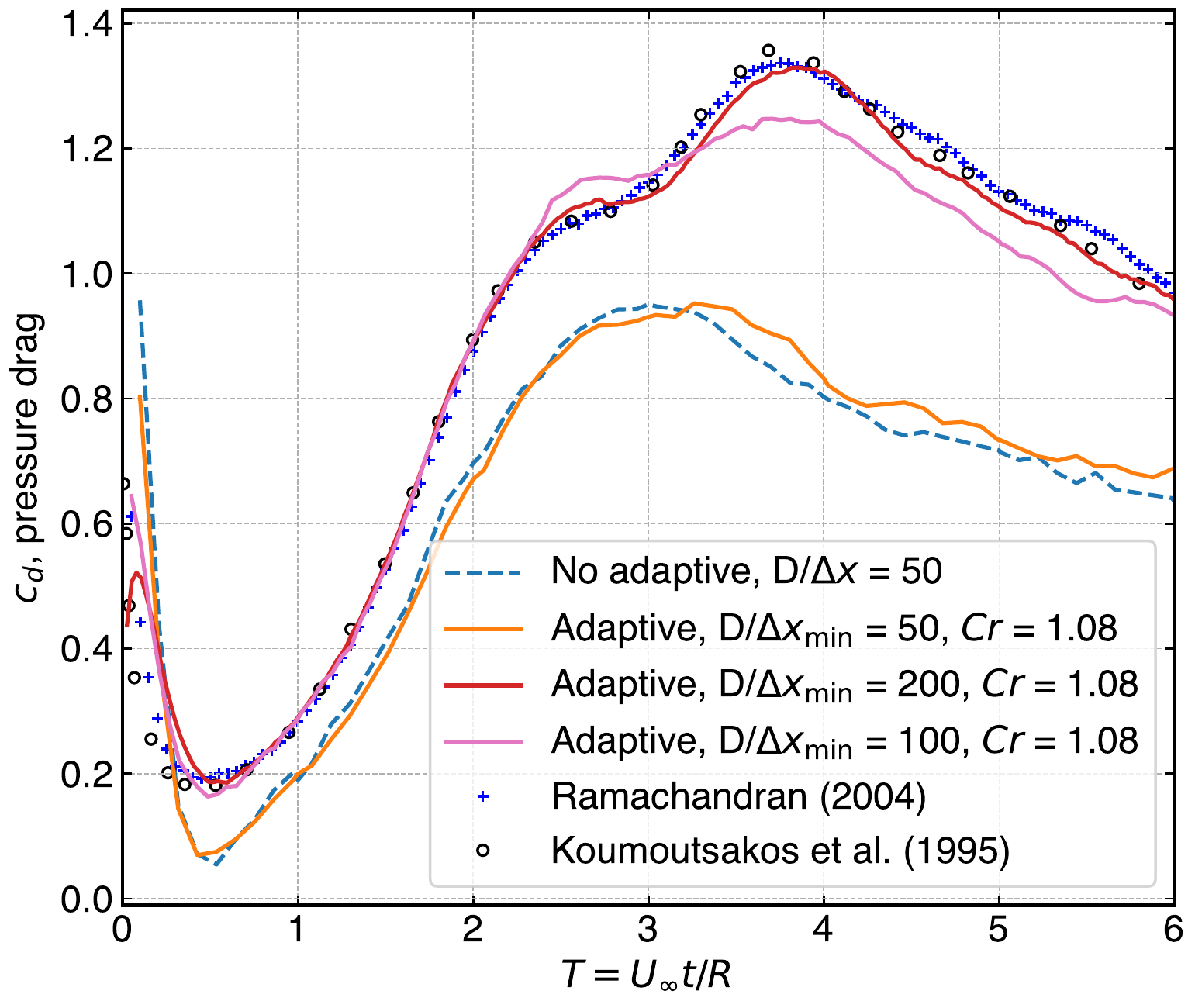}
  \caption{Time history of the coefficient of pressure drag for $Re = 3000$. We
    compare the adaptive cases with three different resolutions to the
    non-adaptive case with a fixed resolution of $D/\Delta x = 50$.}%
  \label{fpc:no-adapt-3000}
\end{figure}

First, we demonstrate the advantages of using adaptive particle refinement
over the non-adaptive case. We use two Reynolds numbers 1000 and 3000. We
simulate the problem up to $T = 6$ with both the non-adaptive case, using a
resolution $D/\Delta x = 50$, and the adaptive case, using two different
maximum resolutions $D/\Delta x_{\min} = 50$ and 100. We use solution
adaptivity based on the vorticity and this aspect is explored in greater
detail in the next section. \Cref{fpc:no-adapt-1000} shows the coefficients of
pressure drag for the $Re = 1000$ case. The coefficients match closely for
both the non-adaptive and the adaptive cases, the adaptive case with $D/\Delta
x_{\min} = 100$ is slightly better in comparison. The differences between the
adaptive and non-adaptive are not easy to assess in this case. On the other
hand, for $Re = 3000$ the advantage of using the adaptive resolution is clearly
seen in \cref{fpc:no-adapt-3000} and summarized in \cref{fpc-perf}. As can be
seen, the adaptive and non-adaptive cases match at $D/\Delta x_{\min} = 50$.
Although, at $D/\Delta x_{\min} = 50$ the adaptive case uses 35 times fewer
particles and is 22 times faster than the non-adaptive case. Given the
efficiency of the adaptive particle refinement, we increase the resolution in
the adaptive case to $D/\Delta x_{\min} = 100$ and further to $200$ and
observe significantly better results in \cref{fpc:no-adapt-3000}. The results
match those of the vortex method quite closely. For the $D/\Delta x_{\min} =
200$ case the number of particles is 2.6 times that of the $D/\Delta x_{\min}
= 50$ case with the adaptive particle refinement. This simulation still
requires 13 times fewer particles than the non-adaptive case at a much lower
resolution of $D/\Delta x_{\min}$ of 50. These results clearly indicate the
importance and performance of the adaptive particle resolution.

\begin{table}[ht]
  \centering
  \begin{tabular}[ht]{lllll}
    \toprule
    Parameter & & & \\
    \midrule
    Adaptive & Yes & Yes & Yes & No\\
    $D/\Delta x_{\min}$, Highest resolution & 50 & 100 & 200 & 50 \\
    $D/\Delta x_{\max}$, Lowest resolution & 4 & 4 & 4 & 50 \\
    time step (non-dimensional) & 0.0011 & 0.00055& 0.00027 & 0.0011 \\
    No.\,of particles & $44,091$ & $70,459$ & $114,082$ & $1,557,970$ \\
    CPU time taken (in mins) & 8.56 & 27.5 & 96.63 & 192.96 \\
    \bottomrule
  \end{tabular}
  \caption{Performance comparison of the adaptive cases with the non-adaptive
    cases for $Re = 3000$ at $T = 6$.}%
  \label{fpc-perf}
\end{table}

\begin{figure}[!ht]
  \centering
  \includegraphics[width=0.6\textwidth]{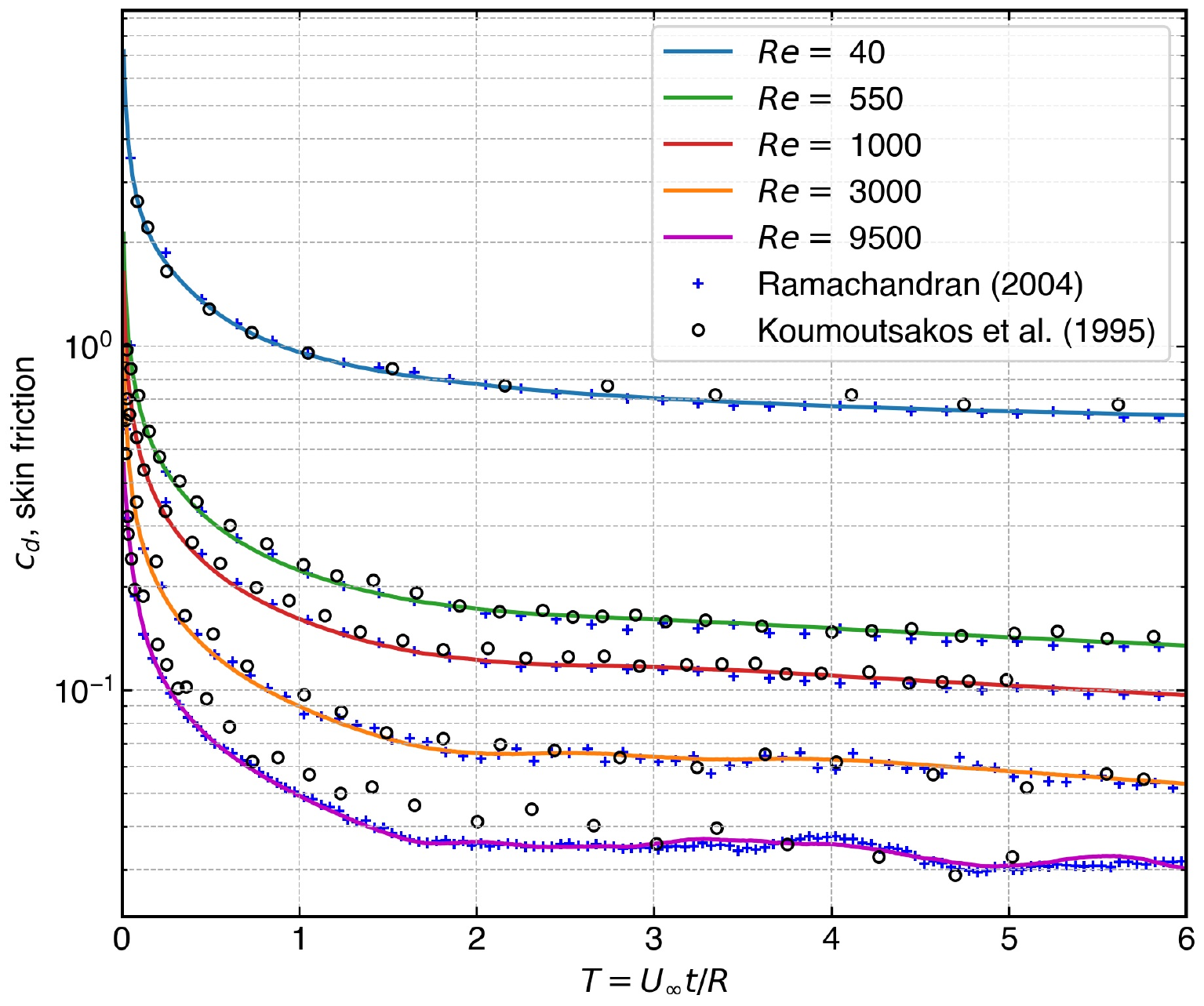}
  \caption{Coefficient of skin friction drag for $Re =$ 40, 550, 1000, 3000, and
    9500. We show only the results where the finest resolution,
    $D/\Delta x_{\min}$, is 500.}%
  \label{fpc:sf}
\end{figure}

\Cref{fpc:sf} shows the coefficient of skin-friction drag at different Reynolds
numbers. We only show the results where the finest resolution is 500. The
results are in good agreement with that
of~\cite{koumoutsakos1995,ramachandran2004}. For the case of $Re = 9500$ the
results of \cite{koumoutsakos1995} predicts slightly (note the logarithmic
scale) higher skin-friction drag, but our results match closely to that of
\cite{ramachandran2004}.

\begin{figure}[!ht]
\begin{subfigure}{0.5\textwidth}
  \centering
  \includegraphics[width=\textwidth]{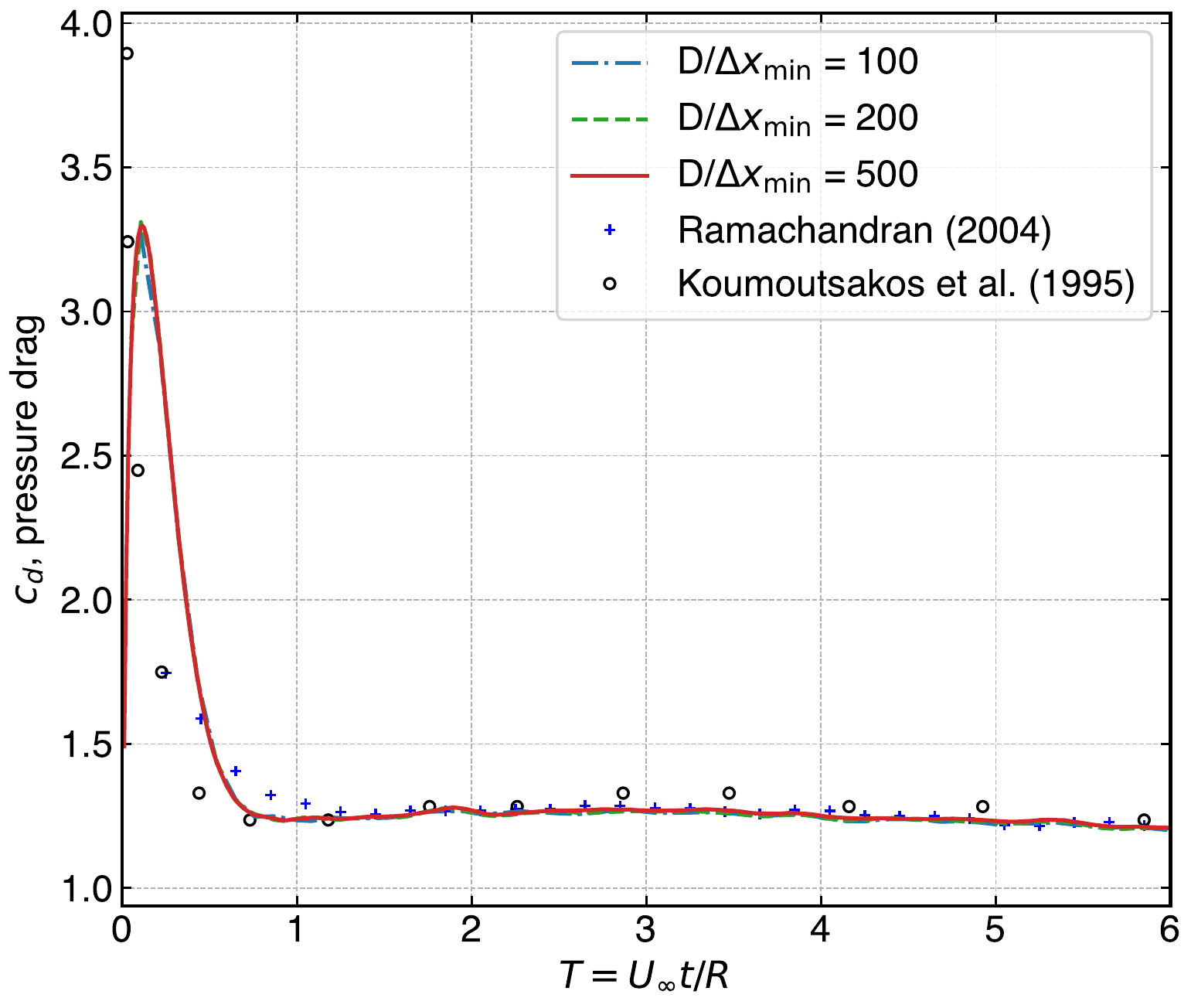}
\end{subfigure}
\begin{subfigure}{0.5\textwidth}
  \centering
  \includegraphics[width=\textwidth]{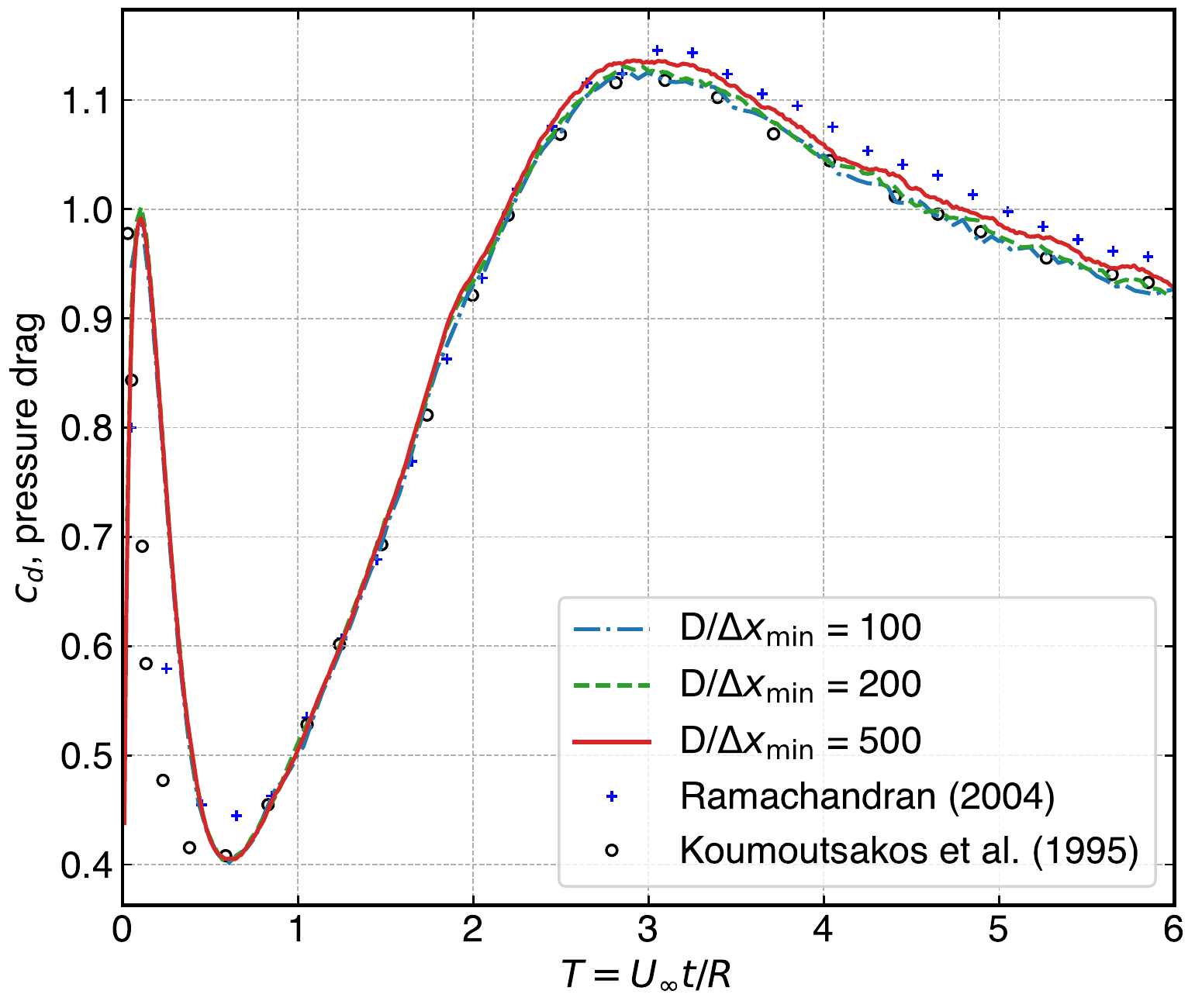}
\end{subfigure}
\caption{Coefficients of pressure drag at $Re = 40$ (left) and $Re = 550$
  (right) as a function of time while varying the finest resolution.}%
  \label{fpc:cd:40-550}
\end{figure}

\begin{figure}[!ht]
\begin{subfigure}{0.5\textwidth}
  \centering
  \includegraphics[width=\textwidth]{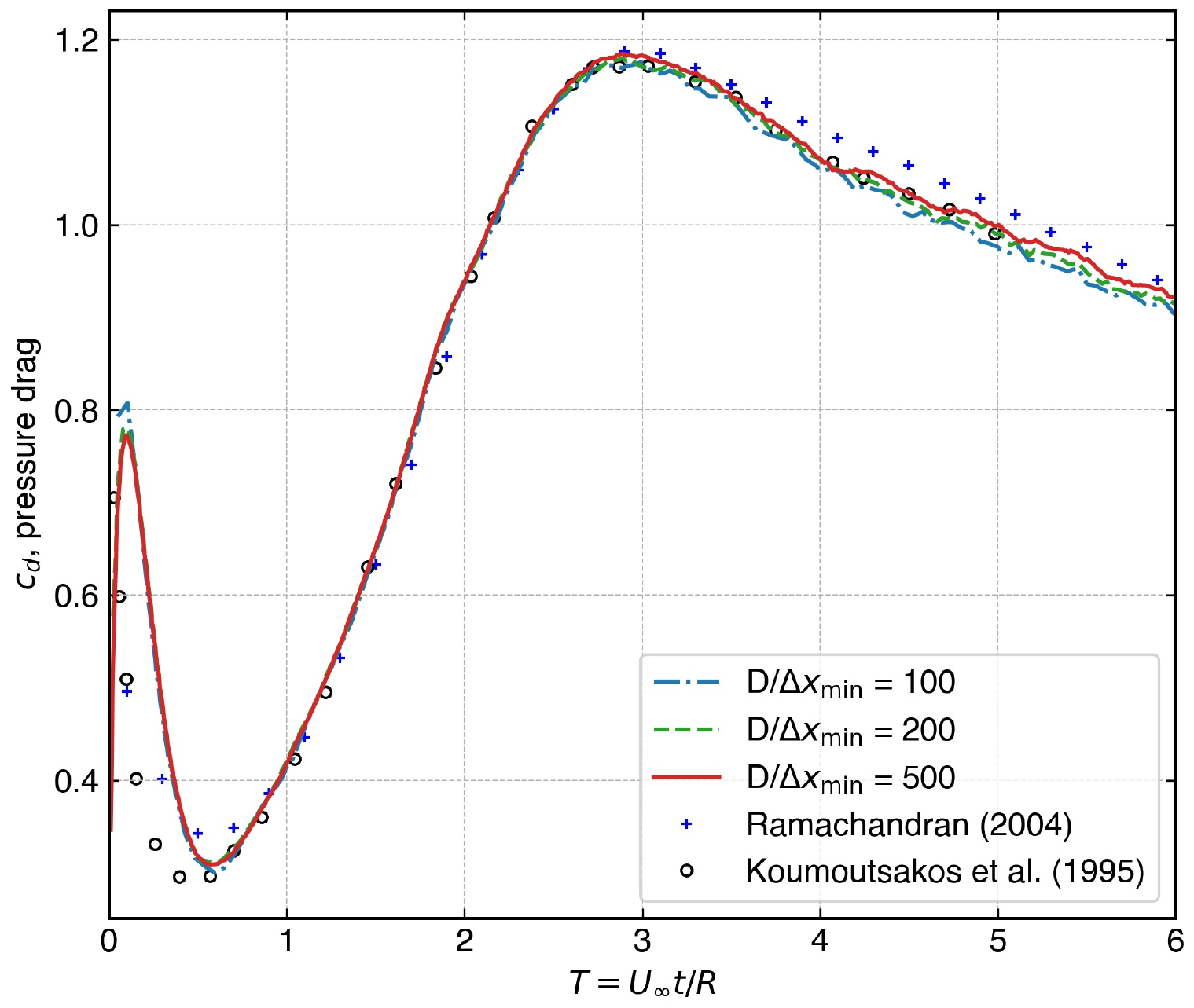}
\end{subfigure}
\begin{subfigure}{0.5\textwidth}
  \centering
  \includegraphics[width=\textwidth]{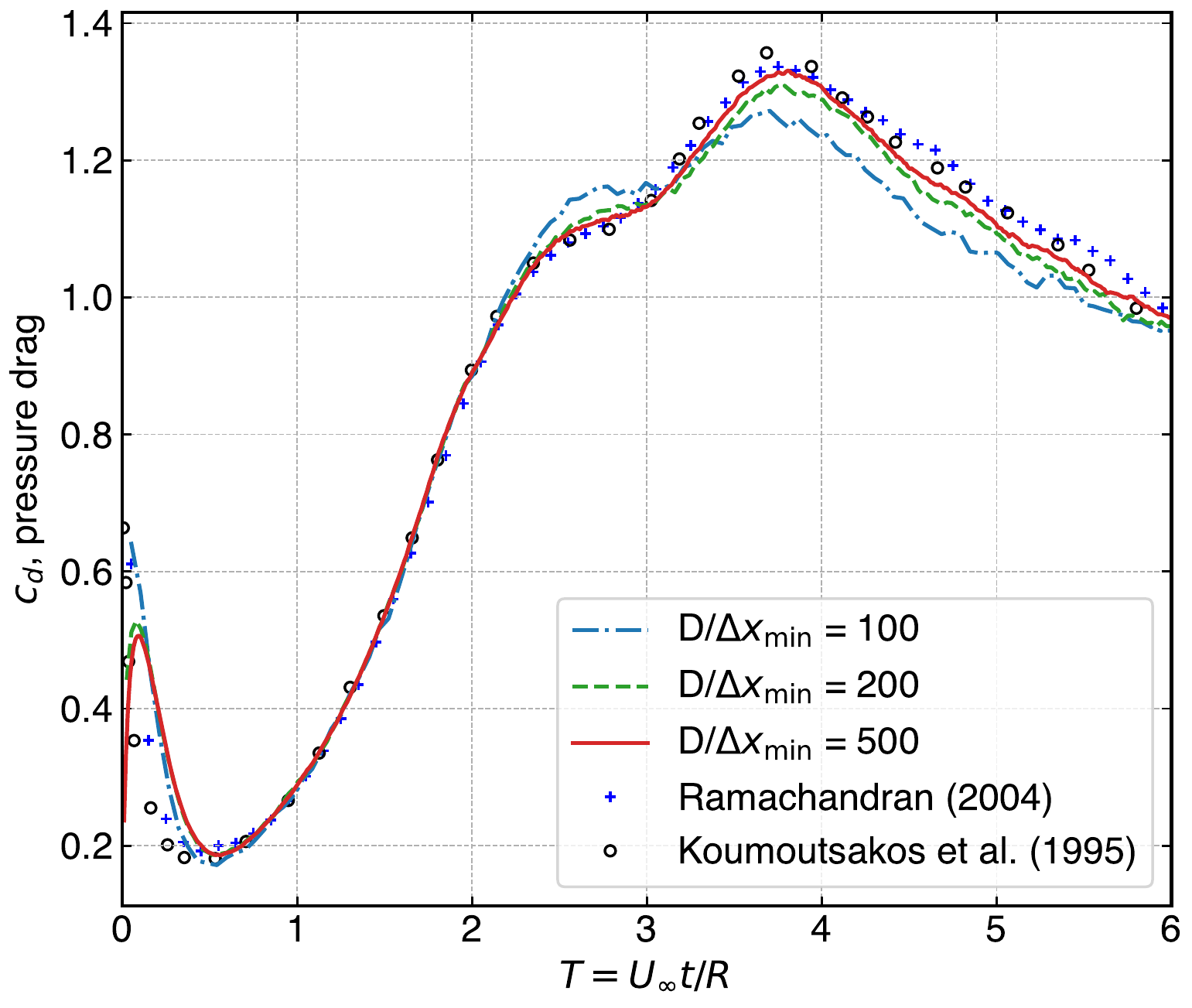}
\end{subfigure}
\caption{Coefficients of pressure drag at $Re = 1000$ (left) and $Re = 3000$
  (right) as a function of time while varying the finest resolution.}%
  \label{fpc:cd:1000-3000}
\end{figure}

In \cref{fpc:cd:40-550,fpc:cd:1000-3000} we plot the coefficient of pressure
drag for the Reynolds numbers, 40, 500, and 1000, 3000 respectively. Our results
differ from the established results at the start for up to $T = 0.5$.  This is
due to the weakly-compressible nature of our flow for where an initial pressure
wave is required to set the velocity from the potential start to the viscous
profile, whereas the established results use incompressible flow. Thereafter,
our results match closely with increase in the maximum-resolution.


\begin{figure}[!ht]
  \centering
  \includegraphics[width=0.6\textwidth]{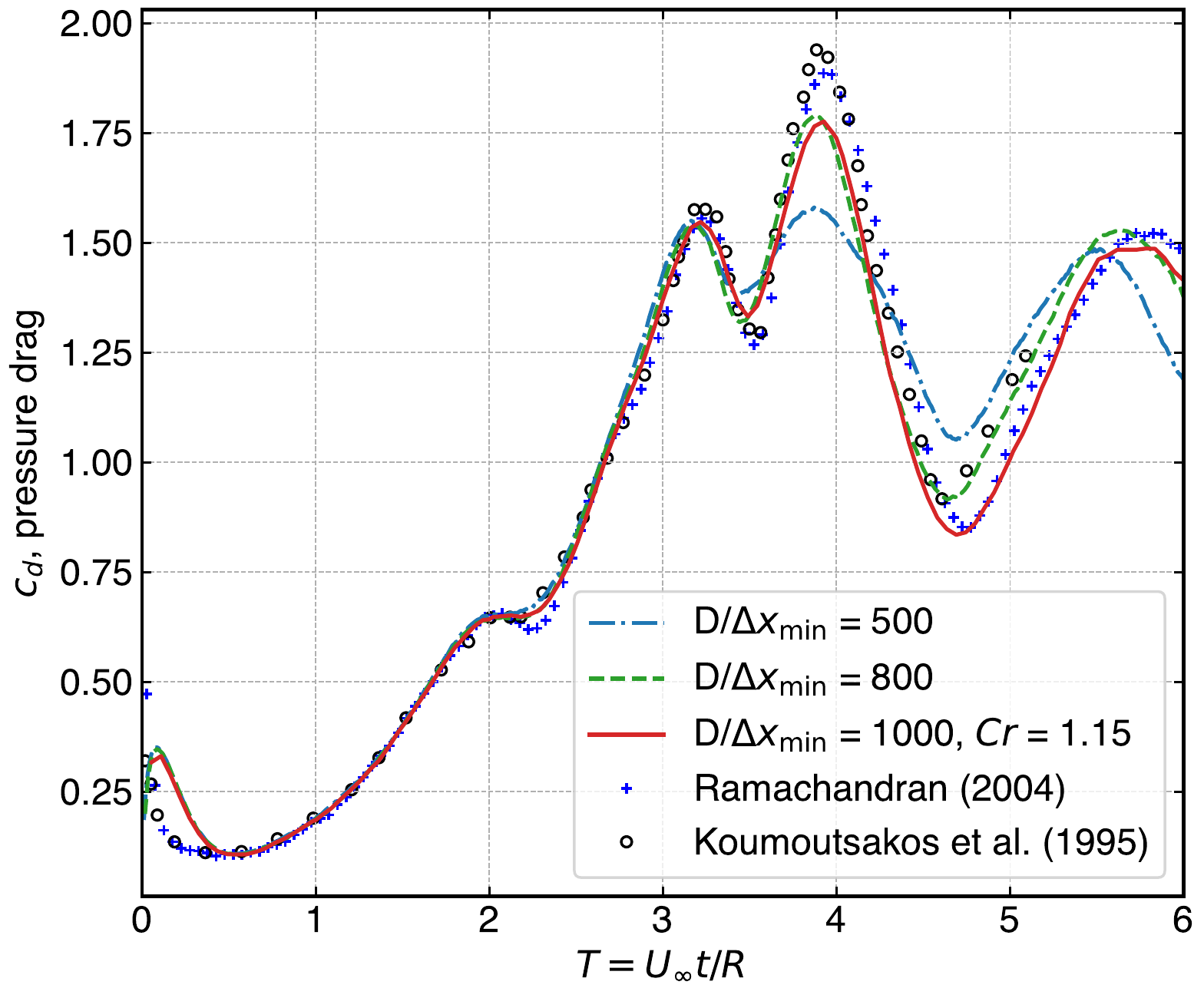}
  \caption{Coefficients of pressure drag at $Re = 9500$ as a function of time
    while varying the finest resolution.}%
  \label{fpc:cd:9500}
\end{figure}

For the $Re = 9500$ case shown in \cref{fpc:cd:9500} we use the finest
resolution $D/\Delta x_{\min} = 1000$ with a $Cr = 1.15$. The change in $C_r$ is
due to time and computational constraints. This is the
highest resolution used in our simulations. Even though the characteristic
features of the drag coefficient profile match with the established results the
curve does not reach the maximum, the trends are consistent with the established
results.  We note that the resolution used by \cite{ramachandran2004}
corresponds to a finest resolution of $D/\Delta x_{\min} = 1250$,
\cite{koumoutsakos1995} use a million vortices for their simulation, and the
present simulations employ around 200,000 fluid particles in the entire
domain. We would also like to note that after three seconds maintaining symmetry
is difficult and even with the DVH results of~\citet{dvh2015} there are
significant differences in the drag force.  The present results are clearly in
good agreement given the variation in literature.

\begin{figure}[!ht]
\begin{subfigure}{0.5\textwidth}
  \centering
  \includegraphics[width=\textwidth]{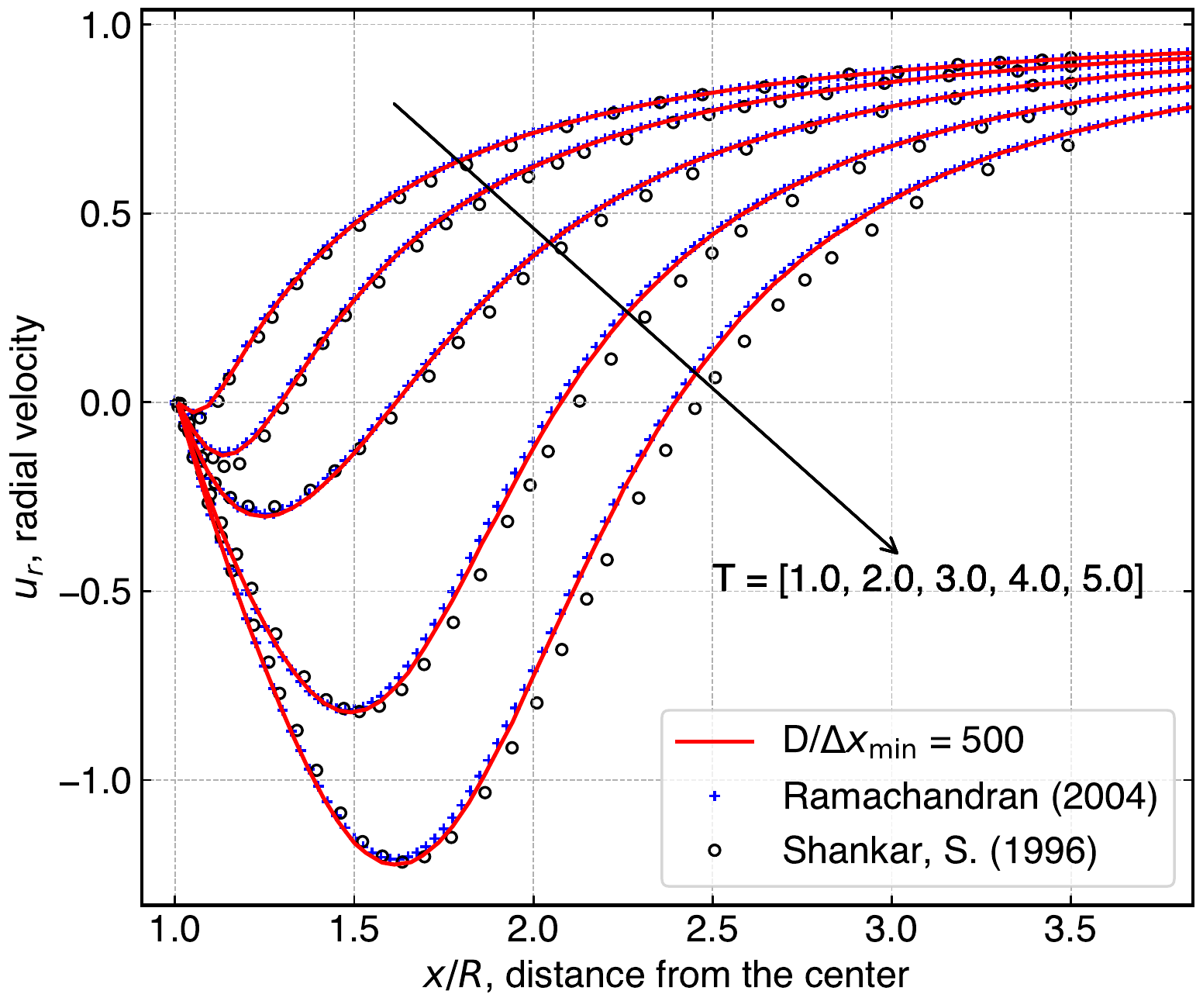}
\end{subfigure}
\begin{subfigure}{0.5\textwidth}
  \centering
  \includegraphics[width=\textwidth]{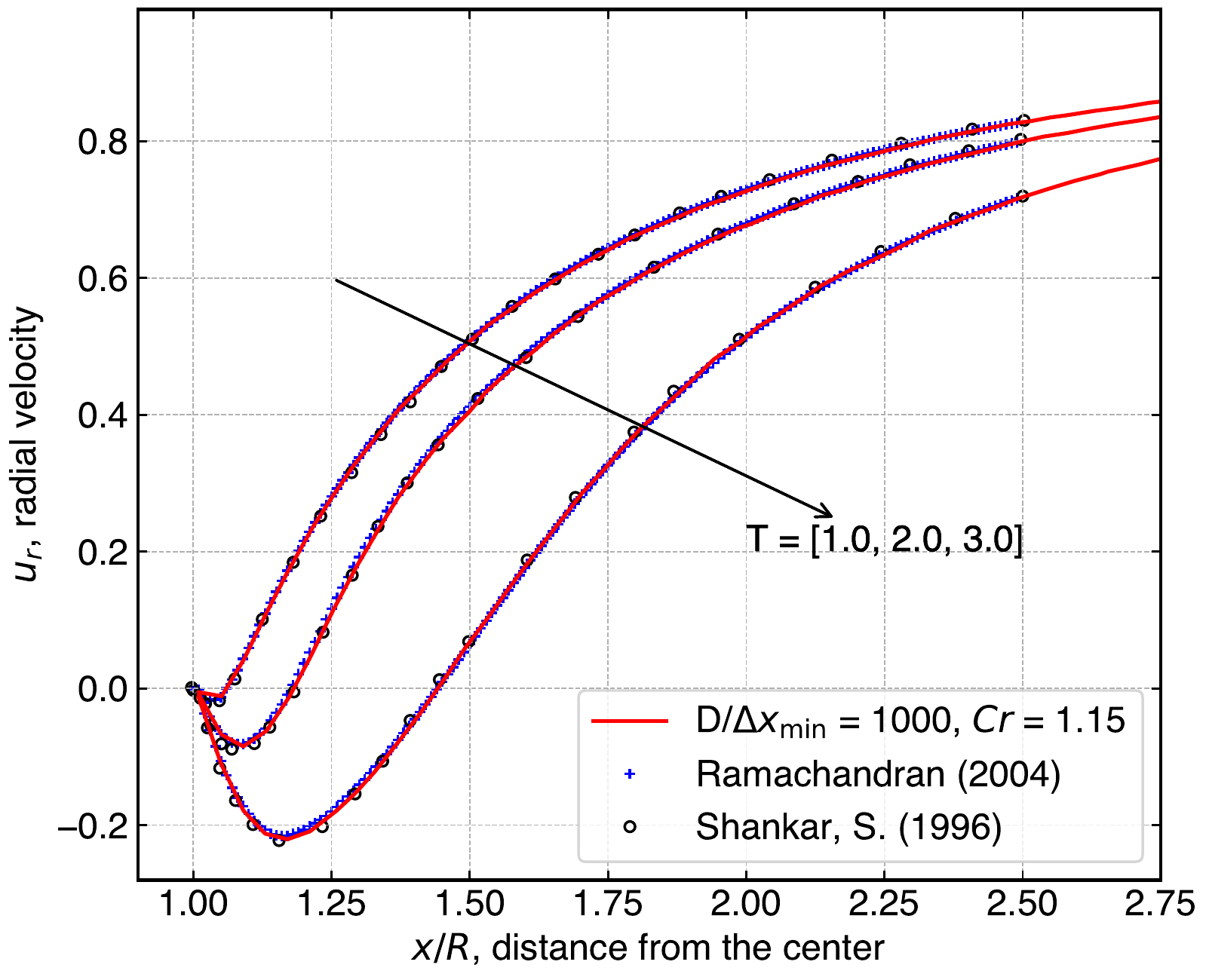}
\end{subfigure}
\caption{The radial velocity along the axis of symmetry in the rear of the
  cylinder for $Re = 3000$ (left) and $Re = 9500$ (right). The results are
  compared with~\citet{ramachandran2004} and~\citet{shankar1996}.}%
  \label{fpc:centerline}
\end{figure}

\Cref{fpc:centerline} show the radial velocity along the axis of symmetry on the
rear side of the cylinder for the Reynolds numbers 3000, with
$D/\Delta x_{\min} = 500$, and 9500, with $D/\Delta x_{\min} = 1000$ and
$Cr = 1.15$, at different times. In both the Reynolds numbers the results are in
good agreement with~\citet{ramachandran2004} while the results
of~\citet{shankar1996} show slight difference for larger time in the $Re = 3000$
case.

\begin{figure}[htp]
  \centering
\resizebox{.95\textwidth}{!}{%
\includegraphics[height=10cm]{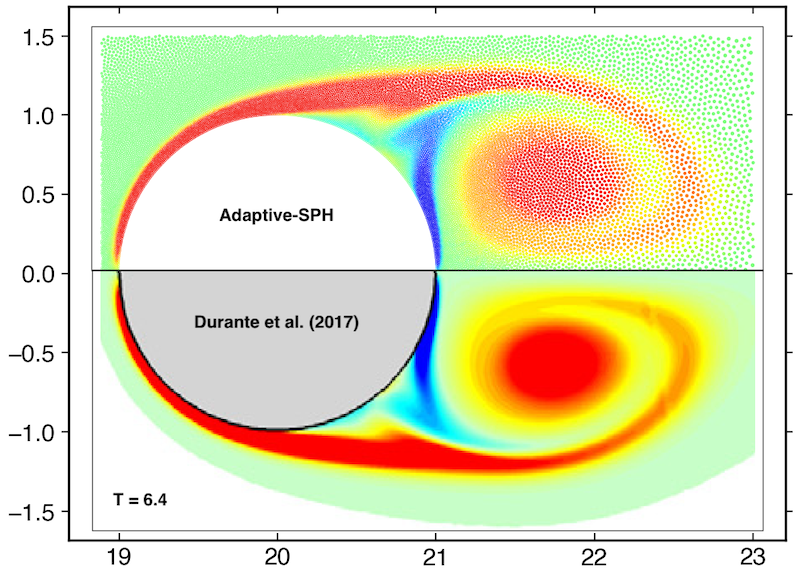}%
\includegraphics[height=10cm]{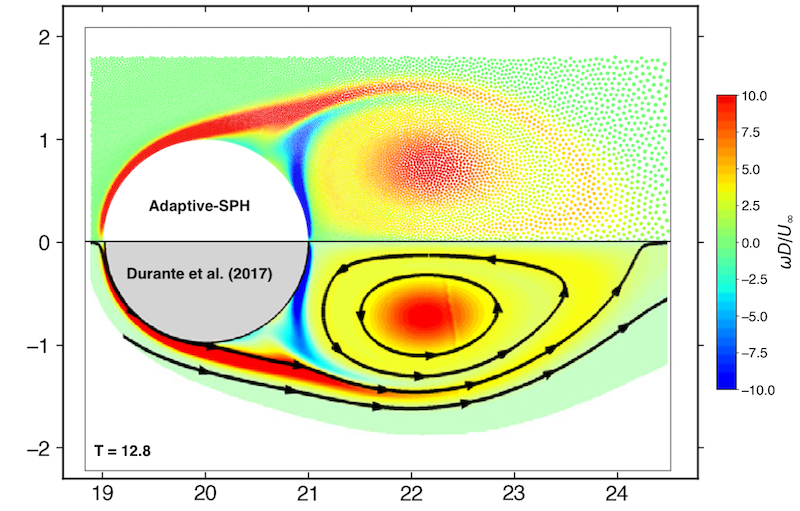}%
}
\caption{Comparison of vorticity distribution at $T = 6.4$ (left), and 12.8
  (right) for the Reynolds number 1000 with the vorticity distribution of
  \citet{durante2017}. Adaptive-SPH particles' size is proportional to their
  mass. The finest resolution, $D/\Delta x_{\min}$, is 200, and the coarsest
  resolution, $D/\Delta x_{\max}$, is 40. The value of $C_r$ is 1.08.}%
  \label{fpc:vor:1k}
\end{figure}

\Cref{fpc:vor:1k} compares the proposed formulation vorticity distribution with
\citet{durante2017}. The simulation is run at the Reynolds number 1000, and
times $T = 6.4,$ and 12.8 are shown. In the adaptive SPH figure (left) the
particles are sized proportional to the mass. There are some differences in the
color as a slightly different color map was used. The distribution show a good
similarity.

\begin{figure}[!ht]
  \centering
\resizebox{.95\textwidth}{!}{%
  \raisebox{-0.5\height}{ \includegraphics[width=0.28\textwidth]{%
      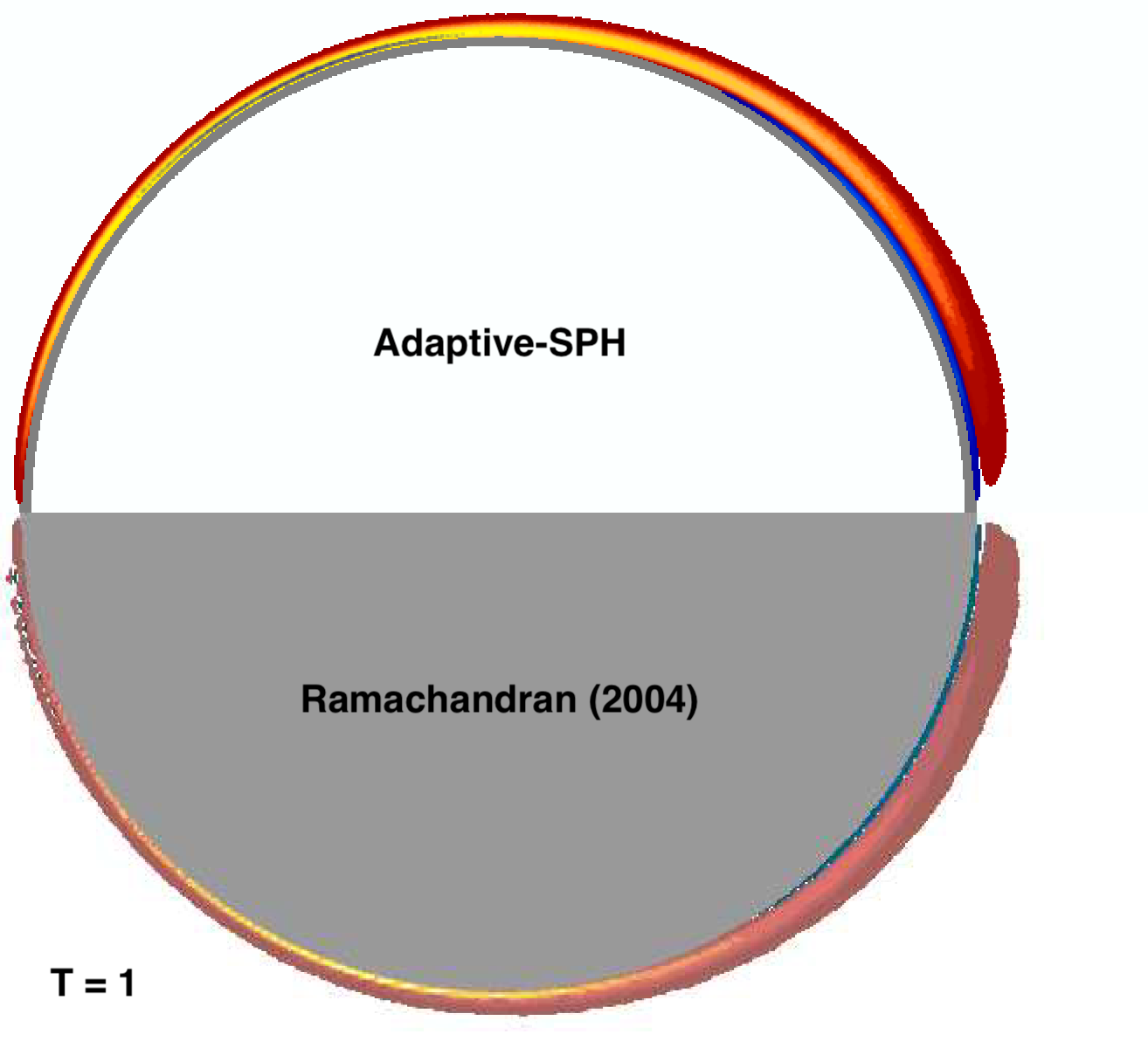}}
  \raisebox{-0.5\height}{ \includegraphics[width=0.28\textwidth]{%
      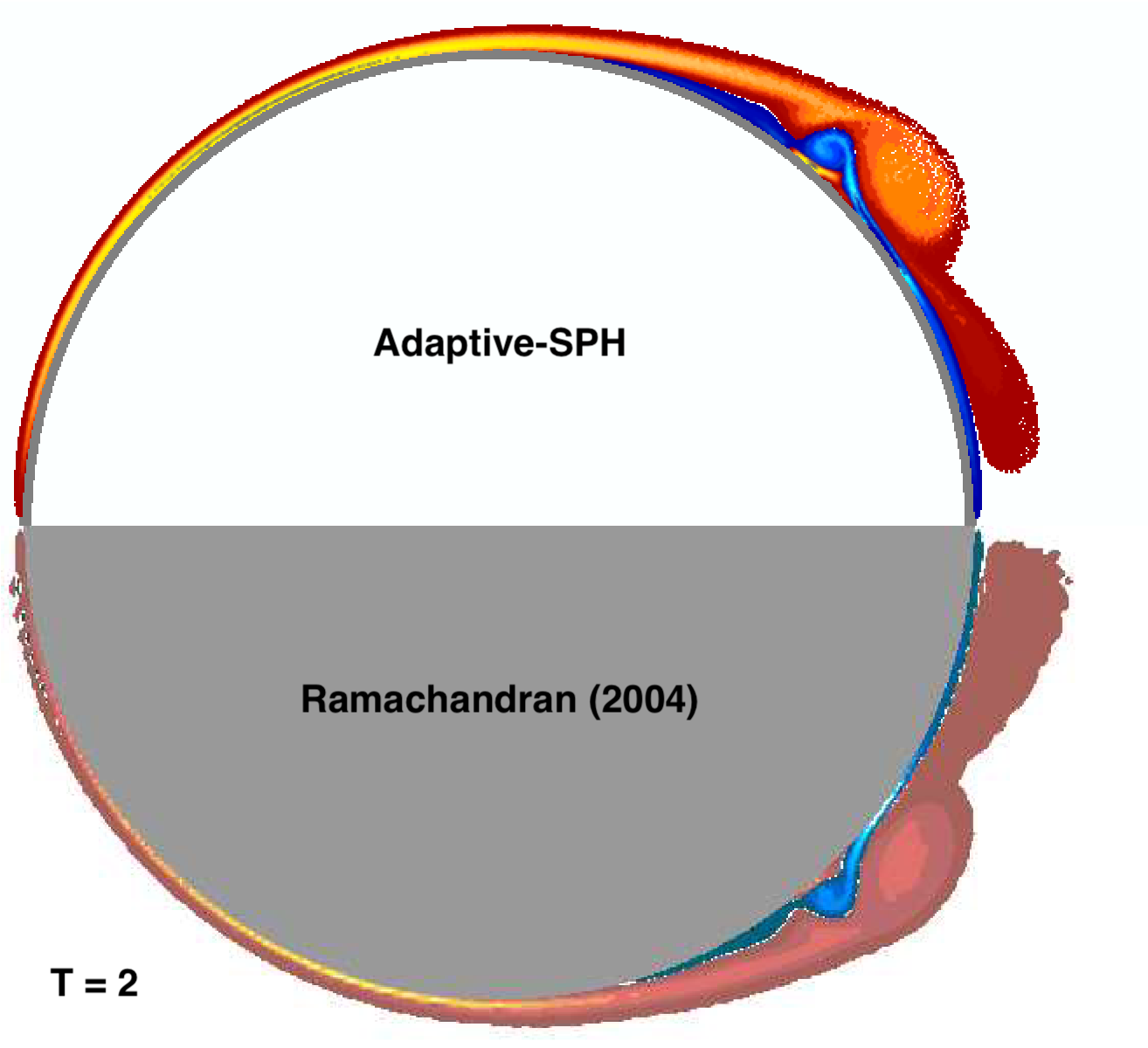}}
  \raisebox{-0.5\height}{ \includegraphics[width=0.28\textwidth]{%
      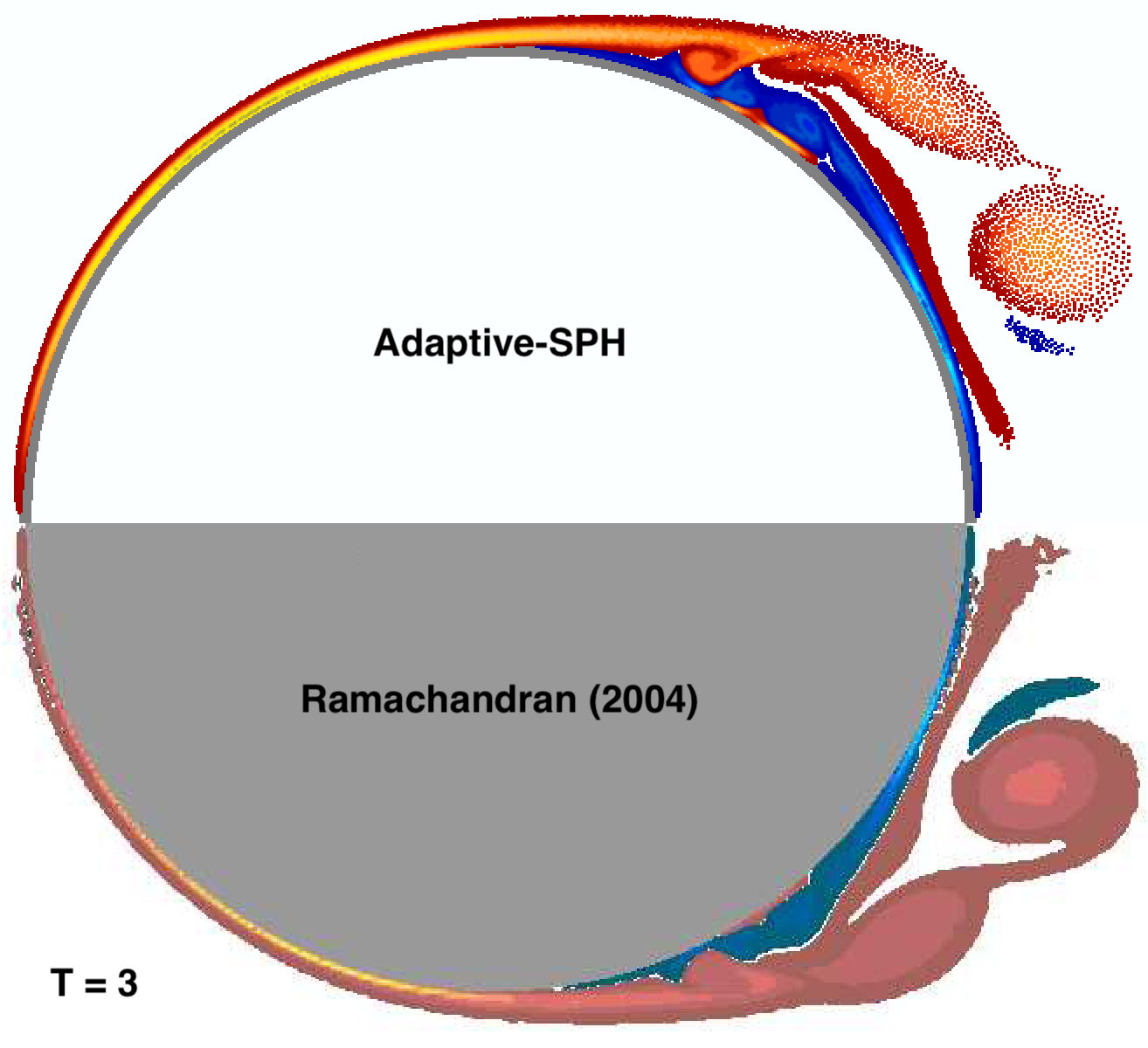}}
  \raisebox{-0.5\height}{ \includegraphics[width=0.04\textwidth]{%
      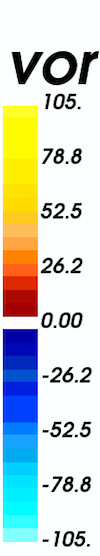}}
  }
  \caption{Comparison of vorticity distribution at $T = 1$ (left), 2 (center),
    and 3 (right) for the Reynolds number 9500 with the vorticity distribution
    of \citet{ramachandran2004}. The finest resolution, $D/\Delta x_{\min}$, is
    1000, and the coarsest resolution, $D/\Delta x_{\max}$, is 40. The value of
    $C_r$ is 1.15.}%
  \label{fpc:vor:9.5k}
\end{figure}

In \cref{fpc:vor:9.5k} we compare the vorticity distribution at the Reynolds
number 9500 with \citet{ramachandran2004}; times $T = 1, 2,$ and 3 are
shown. There are some differences in the color as a slightly different color map
was used in~\cite{ramachandran2004}. The vortices appear to maintain the
symmetry, and the secondary, tertiary, and further vortices generated at the
boundary layer are captured well. The boundary layer at the leading edge of the
cylinder is clearly observed. As the simulation progresses, the vortices grow
big and move across different layers having different smoothing lengths since
there is no solution adaptivity used in this case. At $T=3$, the plot clearly
shows the primary vortex at a different resolution than the boundary layer.
\begin{figure}[!ht]
  \centering
  \includegraphics[scale=0.38]{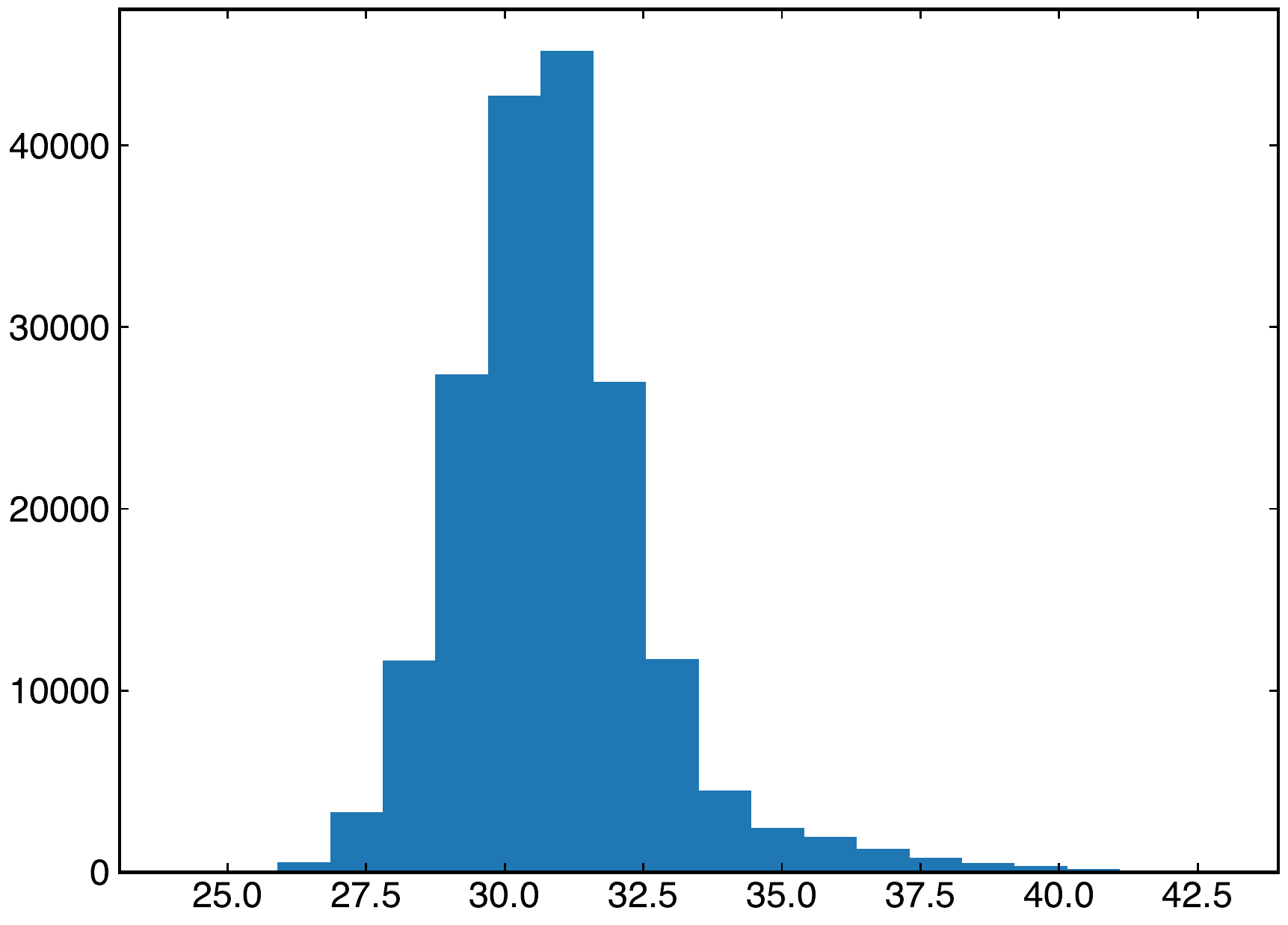}%
  \caption{Histogram showing the number of neighbors in the simulation of
    $Re = 9500$ case at $T = 6$. The highest resolution, $D/\Delta x_{\min}$,
    of particles in this figure is 1000.}%
  \label{fpc:neighbors}
\end{figure}

\begin{figure}[!ht]
  \centering
  \includegraphics[width=\textwidth]{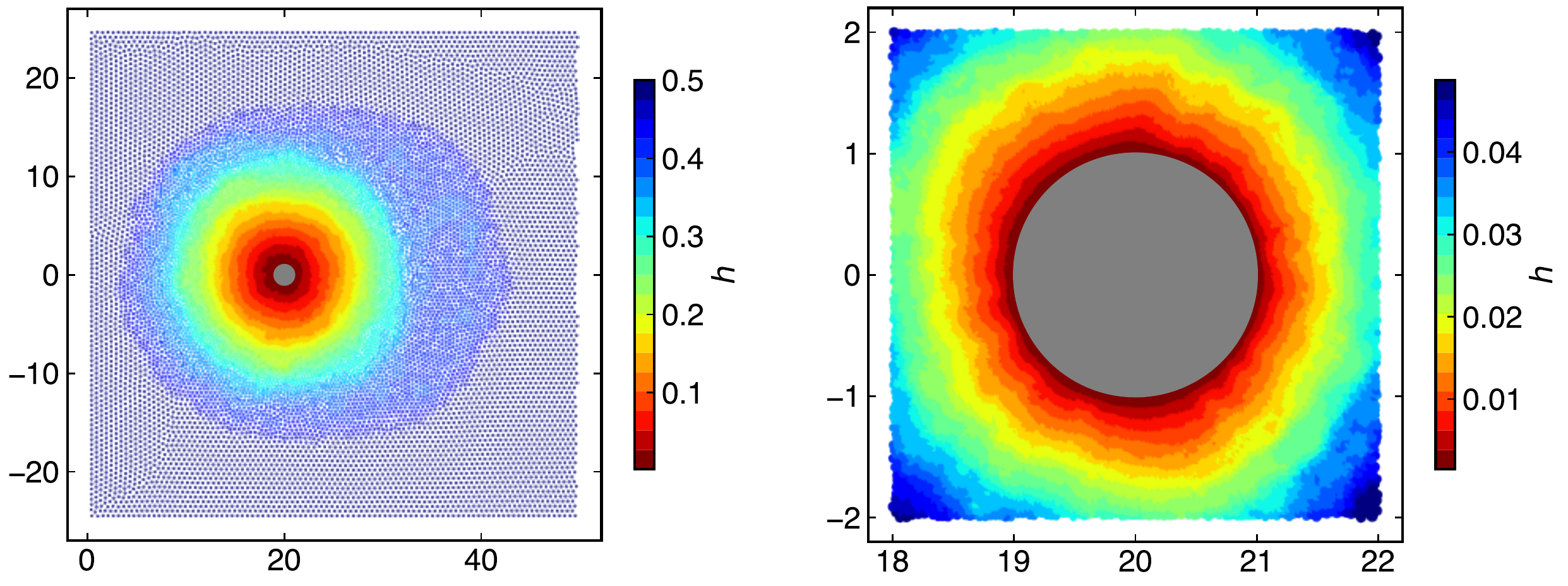}%
  \caption{Distribution of smoothing length of each particle at $T = 6$. The
    highest resolution $D/\Delta x_{\min}$ is 1000.}%
  \label{fpc:h}
\end{figure}

\Cref{fpc:neighbors} shows a histogram of the number of neighbors in the overall
simulation at $T = 6$ for the resolution $D/\Delta x_{\min} = 1000$. This shows
that for a majority of particles the number of neighbors are at 30. The highest
number of neighbors in the simulation is at 42. This shows the optimal neighbor
distribution further maximizing the performance. In \cref{fpc:h} we show the
smoothing length distribution for the same case. The left side shows the whole
domain and the right is a zoom-in near the cylinder. The smoothing length
varies across a large number of scales by a factor of 250. Even near the
cylinder it varies by about a factor of 20.

\FloatBarrier%

\subsection{Solution adaptivity}%
\label{sec:sol-adapt}

\begin{figure}[ht]
  \centering
  \includegraphics[width=0.6\textwidth]{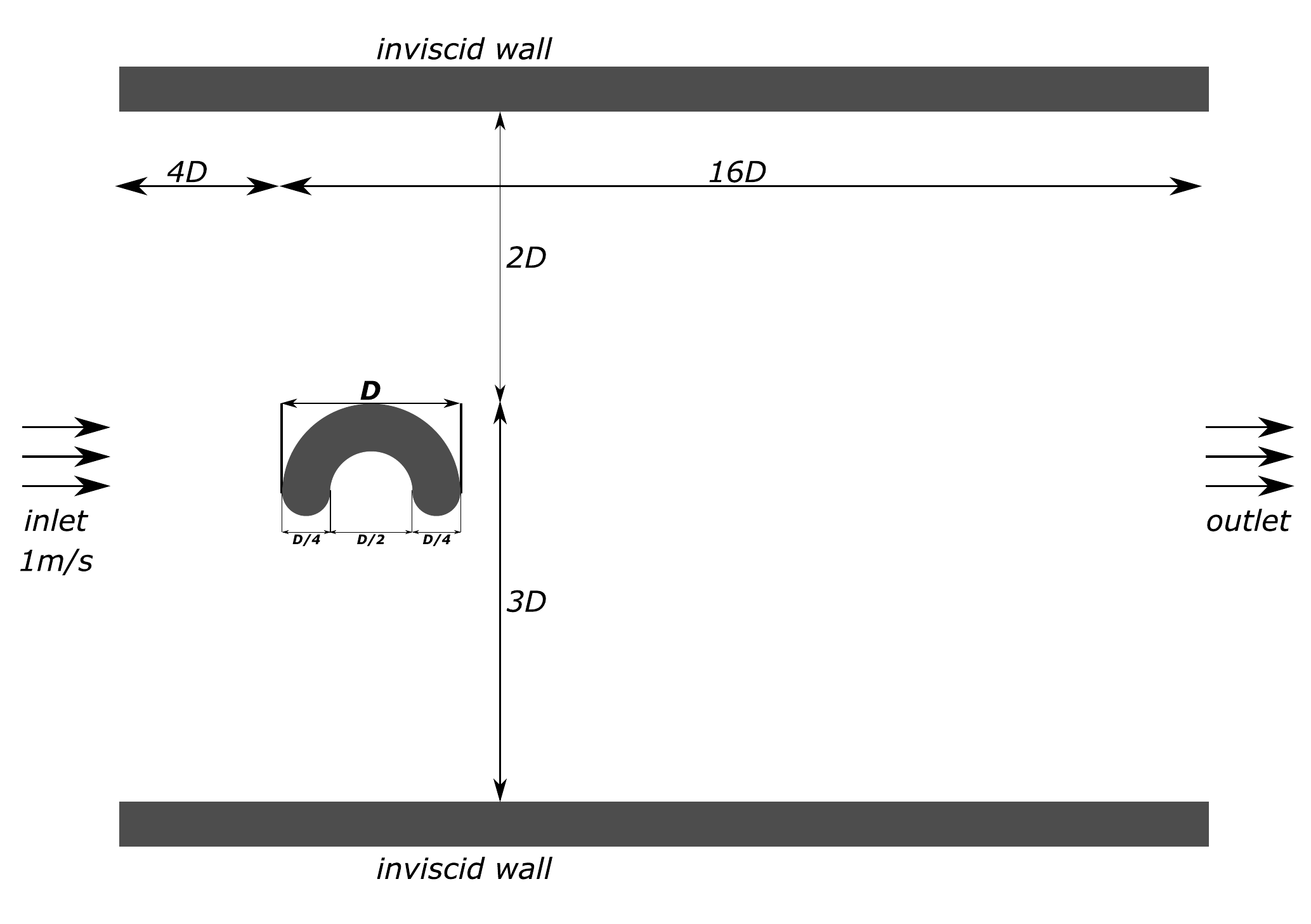}
  \caption{The domain dimensions for the flow past C-shape problem.}%
  \label{cshape:setup}
\end{figure}

In this section we demonstrate the solution-based adaptivity. We consider a flow
past C-shaped body at $Re = 2000$ and compare our results with \citet{dvh2015},
and \citet{sun_multi-resolution_2018}. The domain dimensions are given in
\cref{cshape:setup}, and the smoothing length factor, $h/\Delta x = 1.2$. The
outer diameter $D$ is 1 m. We first perform the simulation without adaptivity and
compare our results. The minimum-resolution $D/\Delta x_{\min}$ is 25 and the
maximum-resolution $D/\Delta x_{\min}$ is 200.  The minimum and
maximum-resolution match the respective resolutions of
\citet{sun_multi-resolution_2018}. We simulate the problem for $T = 30$.

\begin{figure}[!ht]
\begin{subfigure}{0.5\textwidth}
  \centering
  \includegraphics[width=\textwidth]{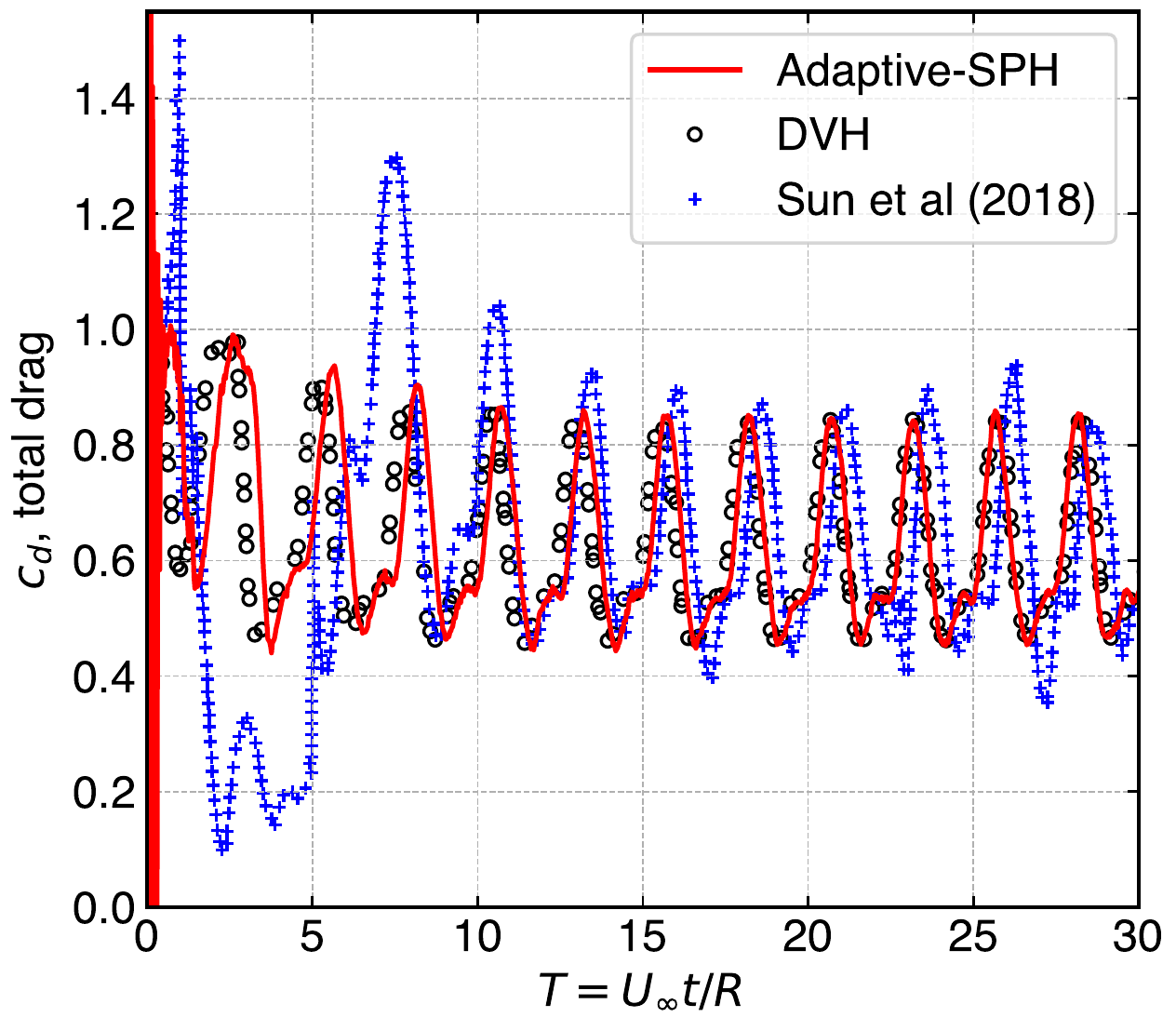}
\end{subfigure}
\begin{subfigure}{0.5\textwidth}
  \centering
  \includegraphics[width=\textwidth]{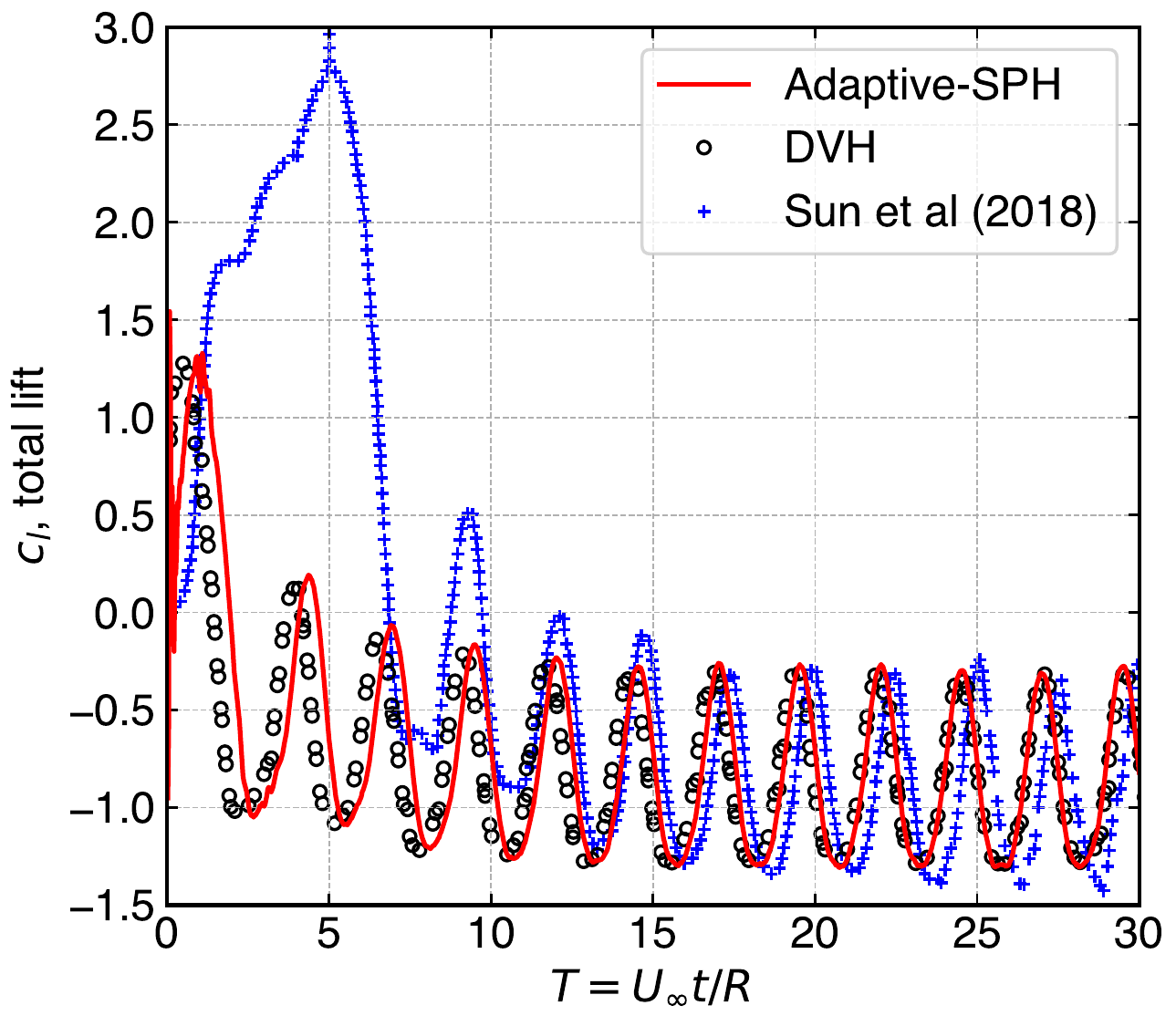}
\end{subfigure}
\caption{The coefficient of drag (left) and lift (right) for the flow past
  C-shape simulation at $Re = 2000$. The results are compared with
  \citet{dvh2015} and~\citet{sun_multi-resolution_2018}.}%
  \label{cshape:cd:cl}
\end{figure}

In \cref{cshape:cd:cl} we show the coefficients of total drag and lift. The
results are in good agreement with \citet{dvh2015}. The initial noise within
$T = 1$ is due to the weakly-compressible nature of our formulation.  We compare
the number of particles used in our simulation with the simulation of
\citet{sun_multi-resolution_2018}. \citet{sun_multi-resolution_2018} does not
mention the total length of the domain instead provides the dimensions
$[2.75, 5.75] \times [-0.8, 1.2]$ of an inner rectangular domain containing the
C-shape body, where the minimum-resolution is 50. We estimate the number of
particles inside this domain to be approximately 98,000, whereas for our
simulation the number of particles in this domain is about 38,735. This shows
that we use 2.53 times lower number of particles and achieve significantly
better results. This further demonstrates the efficiency and the accuracy of our
method.

\begin{figure}[!ht]
\begin{subfigure}{0.5\textwidth}
  \centering
  \includegraphics[width=\textwidth]{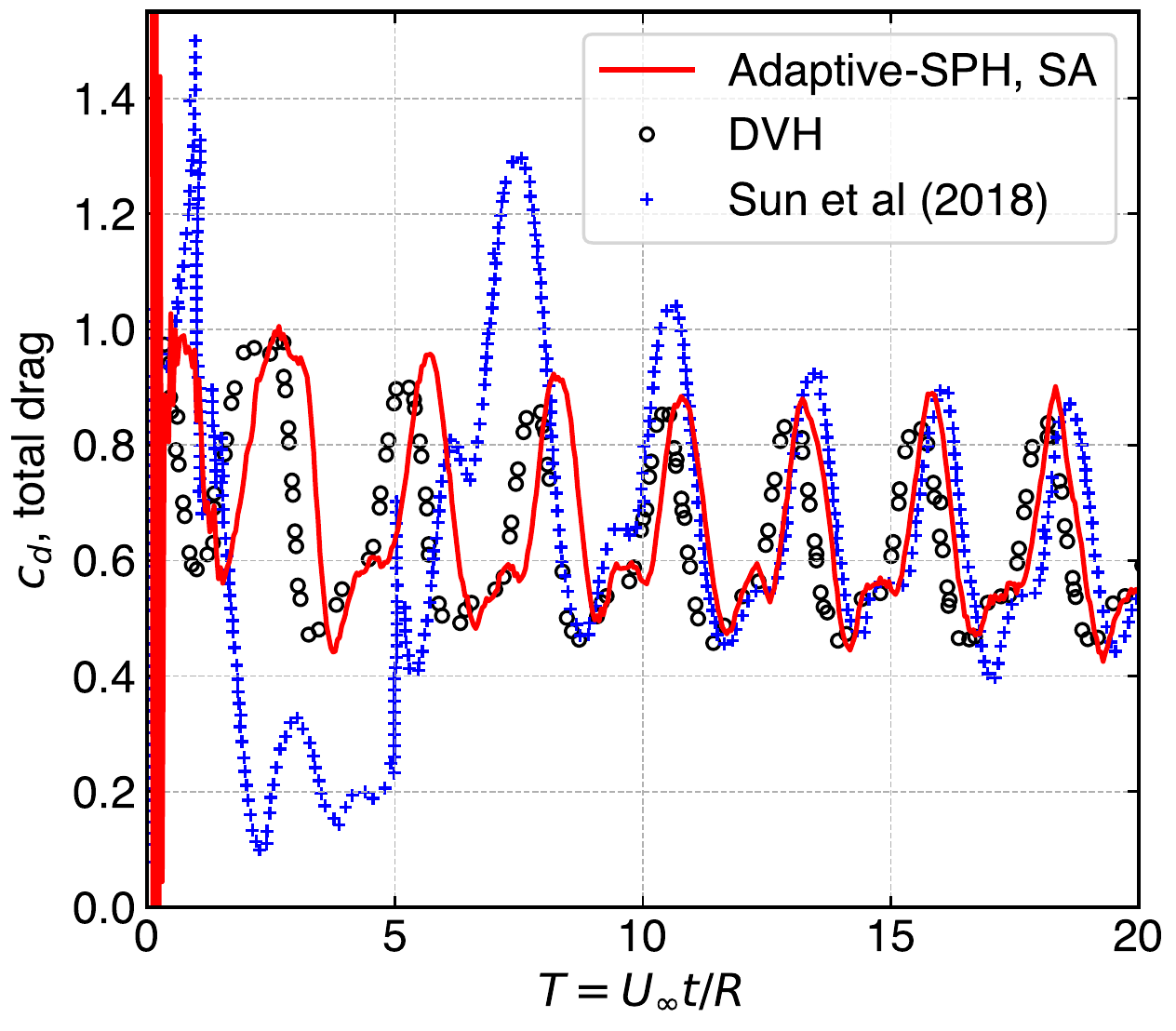}
\end{subfigure}
\begin{subfigure}{0.5\textwidth}
  \centering
  \includegraphics[width=\textwidth]{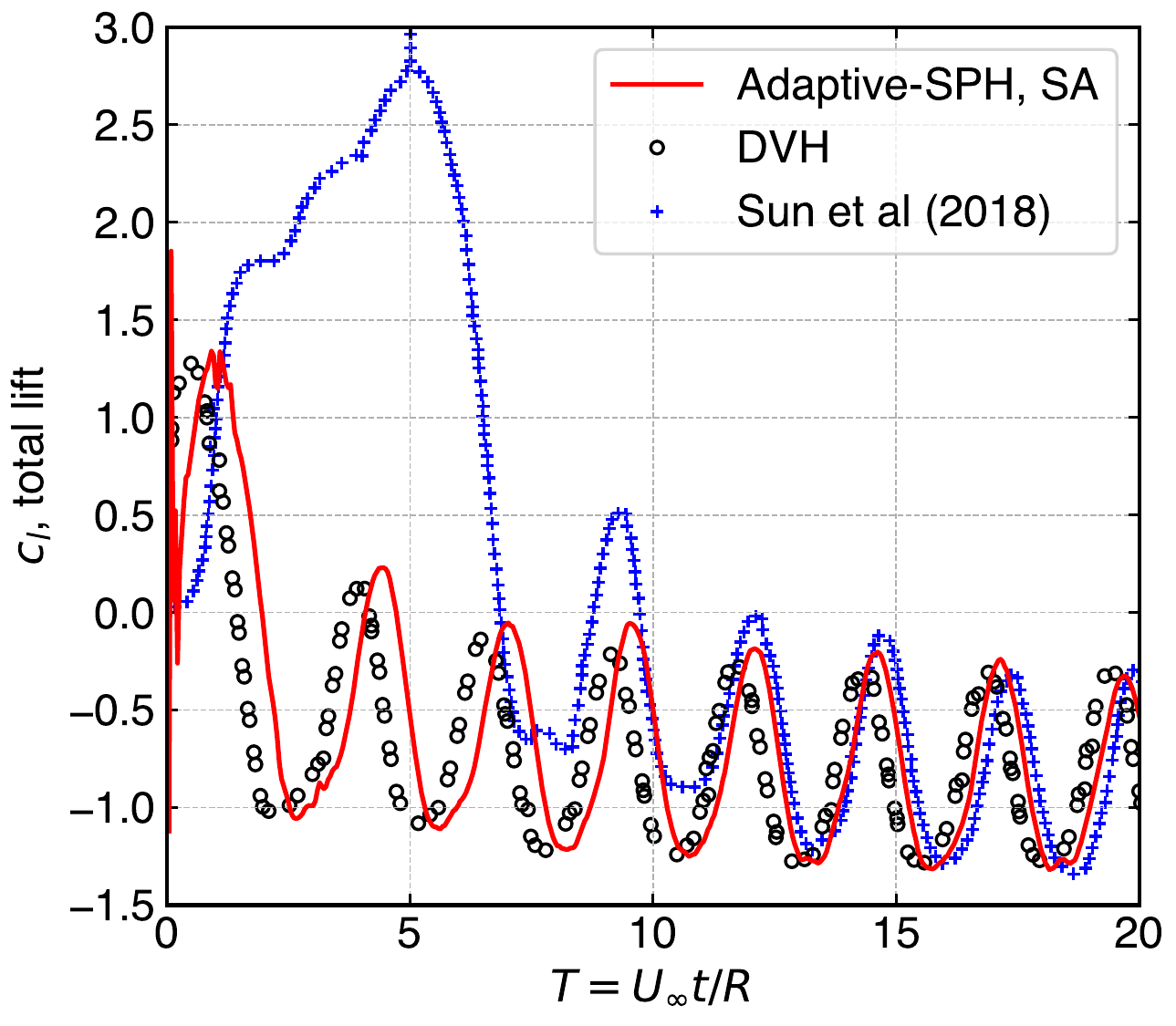}
\end{subfigure}
\caption{The coefficient of drag (left) and lift (right) for the flow past
  C-shape simulation at $Re = 2000$ with solution adaptivity are compared with
  \citet{dvh2015} and~\citet{sun_multi-resolution_2018}.}%
  \label{cshape:sol:cd:cl}
\end{figure}

Now, we simulate the flow past C-shape using solution-adaptivity, where the
vorticity in the flow is monitored and particles with absolute vorticity value
above 5\% of the maximum vorticity are resolved to the highest resolution. In
this simulation we use the maximum-resolution $D/\Delta x_{\min}$ of 125. The
simulation is performed for $T = 20$. In \cref{cshape:sol:cd:cl} the
coefficients of total drag and lift are shown which are in good agreement with
the results of \cite{dvh2015}.

\begin{figure}[!ht]
  \centering
  \includegraphics[width=\textwidth]{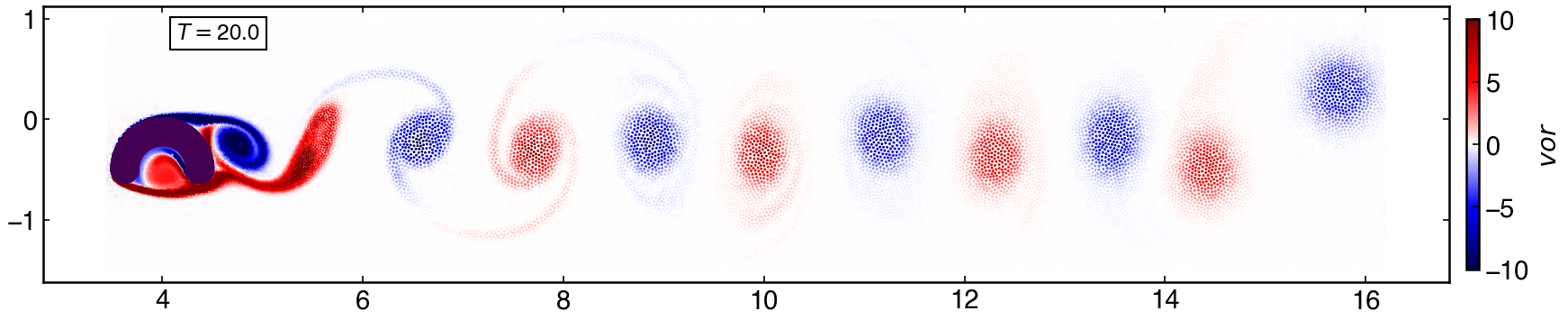}
  \includegraphics[width=\textwidth]{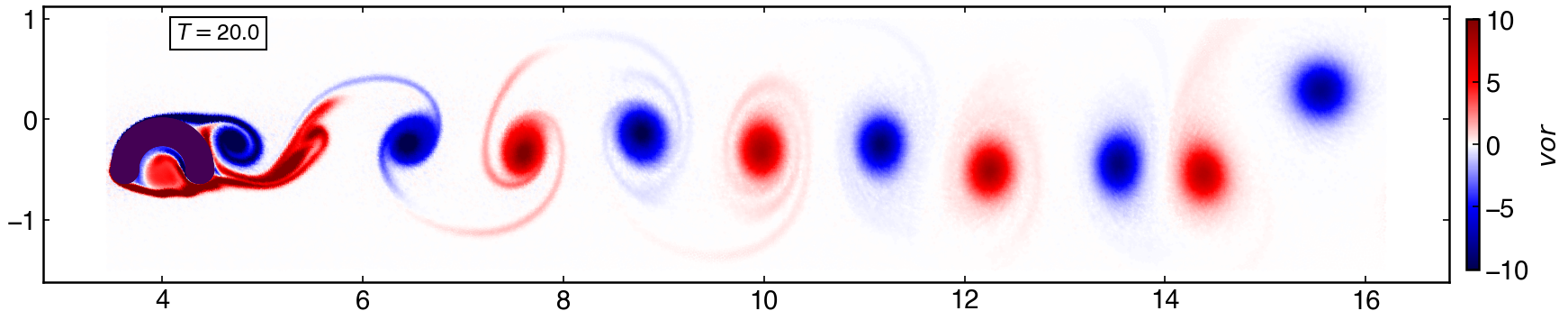}
  \caption{Vorticity distribution around the C-shape body simulated using
    adaptive-SPH, without solution-adaptivity (top) and with solution-adaptivity
    (bottom) at $T = 20$, respectively.}%
  \label{cshape:vor}
\end{figure}

\begin{figure}[!ht]
\begin{subfigure}{0.5\textwidth}
  \centering
  \includegraphics[width=\textwidth]{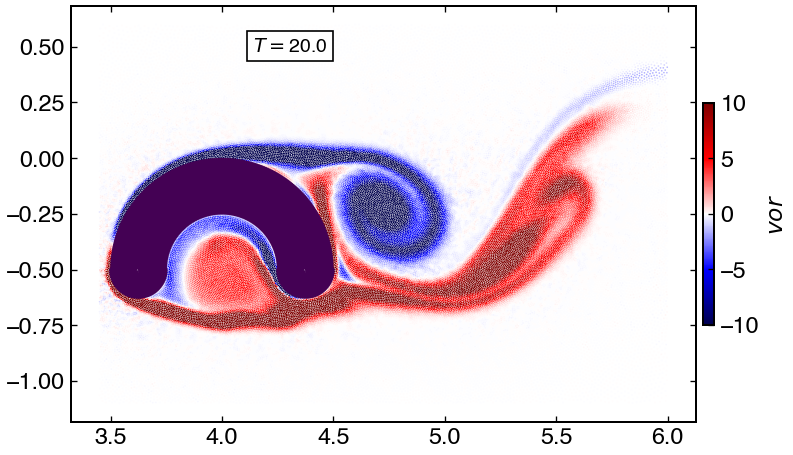}
  \subcaption{}%
  \label{cshape:vor:zoom}
\end{subfigure}
\begin{subfigure}{0.5\textwidth}
  \centering
  \includegraphics[width=\textwidth]{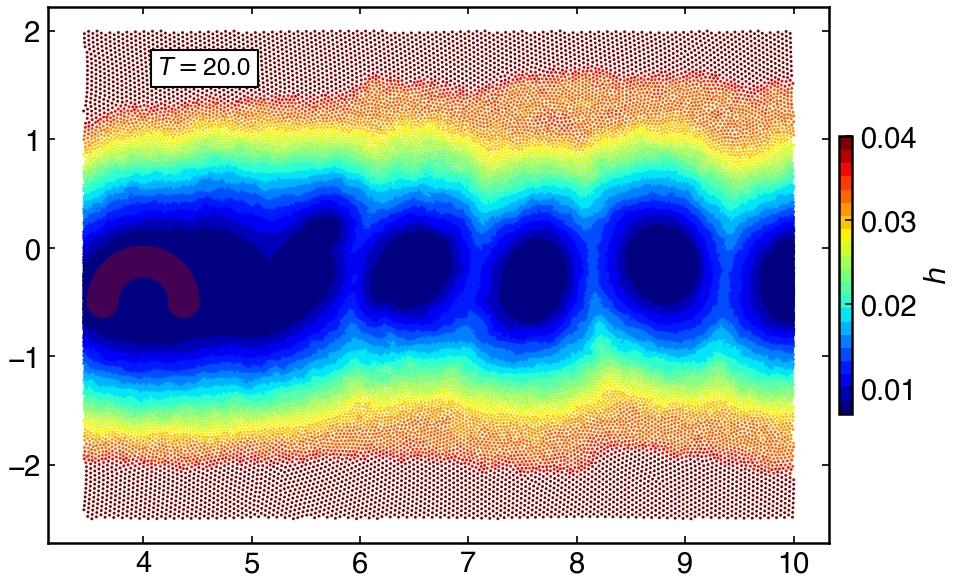}
  \subcaption{}%
  \label{cshape:h}
\end{subfigure}
\caption{(a) Zoomed-in view of the vorticity, (b) smoothing length distribution
  around the C-shape body simulated using the adaptive-SPH, with
  solution-adaptivity at $T = 20$.}%
\end{figure}

\Cref{cshape:vor} shows the vorticity distribution of the particles for the
non-solution adaptive and the solution adaptive cases at $T = 20$. It can be
seen that for the non-solution adaptivity case the trailing vortices are not
refined after they move certain distance away from the C-shape body, whereas in
the solution-based adaptivity the trailing vortices are resolved to the highest
resolution, based on the cut-off criteria stated above. In
\cref{cshape:vor:zoom} we show the zoomed-in view near the C-shape body for the
solution-based adaptive case, and in \cref{cshape:h} we show the smoothing
length distribution demonstrating the effect of solution adaptivity.

The results of this section demonstrate the accuracy and efficiency of the
proposed method even when there is a large change in the resolution. The method
shows little dissipation and it is capable of performing a high-resolution
simulation to capture all the features of a $Re = 9500$ flow. The results show
good accuracy with the solution-based adaptivity.

%

\section{Conclusions}%
\label{sec:conclusions}

In this work we have proposed an accurate and efficient method to handle
adaptive resolution in the context of weakly-compressible SPH. This is
achieved using (i) an accurate EDAC scheme~\cite{edac-sph:cf:2019} along with
the recent corrections of~\cite{adepu2021}, the use of variable-$h$
corrections of \cite{vacondio_accurate_2012}, and particle
shifting~\cite{diff_smoothing_sph:lind:jcp:2009}; (ii) adaptive splitting and
merging of particles where care is taken to ensure that the number of
particles is minimum and the number of neighbors is optimal.  We employ
background particles to specify the regions of refinement. Importantly, the
method allows for specifying fixed regions of refinement, automatic
geometry-based refinement, and automatic solution-based adaptivity elegantly
in the same framework. The algorithms employed are parallel. We provide an
open-source implementation of the entire algorithm along with complete
automation of all the results presented in this work.

We demonstrate the accuracy of the method using several benchmarks. The
Taylor-Green and Gresho-Chan benchmark problems clearly demonstrate that the
method is not diffusive and is more accurate than other recent adaptive
refinement techniques.

We perform simulations at unprecedented resolution for the flow past a
circular cylinder for a variety of Reynolds numbers in the range 40 to
9500. For example, at $Re = 9500$ we use a resolution of
$D/\Delta x_{\min} = 1000$ and $D/\Delta x_{\max} = 4$ giving a ratio of
length scales of 250. This requires 16 levels of refinement with a domain size
of $25D \times 15D$, requiring only 200,000 particles. The results are in good
comparison with that of \cite{koumoutsakos1995,ramachandran2004} who also
employ similar number of particles. This shows the effectiveness and accuracy
of the adaptive resolution method.

For a Reynolds number of 3000 with a $D/\Delta x_{\min} = 50$ we are able to
obtain similar accuracy with the adaptive refinement using 30 times less
particles, with a 25-fold speed improvement when compared with that using a
fixed resolution.

The current work has focused on the weakly-compressible SPH method. We have
demonstrated the method for two-dimensional problems without a
free-surface. The method in principle should work with a few modifications for
free-surface problems, as well as three-dimensional problems. We plan to
explore the adaptive particle refinement applied to incompressible SPH and
three-dimensional problems in the future.

\section*{Acknowledgements}

We would like to thank the Aerospace Computational Engine (ACE) at the
Department of Aerospace Engineering, Indian Institute of Technology Bombay for
providing computational resources.

\bibliographystyle{model6-num-names}
\bibliography{references}

\end{document}